\documentclass[12pt,letterpaper]{article}
\usepackage[usenames, dvipsnames]{xcolor}
\usepackage{jheppub}

\usepackage{tocloft}
\usepackage[]{todonotes}
\usepackage{bbold}
\usepackage{graphicx}
\usepackage{verbatim}
\usepackage[mathscr]{euscript}
\usepackage{slashed}
\usepackage{graphicx}
\usepackage{mathdots}
\usepackage{caption}
\usepackage[boxsize=0.5em,aligntableaux=center]{ytableau}

\usepackage[labelformat=simple]{subcaption}

\usepackage{enumerate}
\usepackage{bbm}
\usepackage{psfrag}
\usepackage{subfiles}
\usepackage{psfrag}
\usepackage{relsize}
\usepackage[T1]{fontenc}
\usepackage[utf8]{inputenc}
\usepackage{pgfplots}
\usepackage{tikzscale}

\usepackage{bbm} 					
\usepackage{slashed} 				
\usepackage{graphicx}				
\usepackage{subcaption}			
\usepackage{psfrag}				
\usepackage{tensor}				
\usepackage{fouridx}				
\usepackage{bm}					
\usepackage{mdframed}				
\usepackage{multirow}				
\usepackage{soul}					
\usepackage{bbold}				
\usepackage{multicol}				

\usepackage{amsmath}
\usepackage{amssymb}
\usepackage{amsthm}

\usepackage{mathtools}

\usepackage{feynmf}
\usepackage{marvosym}

\usepackage{import}

\newtheoremstyle{named}{}{}{\itshape}{}{\bfseries}{.}{.5em}{#3}
\theoremstyle{named}
\newtheorem*{namedconjecture}{Conjecture}

\newcommand{\e}{{\mathrm{e}}}
\newcommand{\eKK}{e_{\textrm{KK}}}
\newcommand{\gKK}{g_{\textrm{KK}}}
\newcommand{\LUV}{\Lambda_{\textrm{UV}}}

\newcommand{\cF}{\mathcal{F}}

\newcommand{\cM}{\mathcal{M}}

\newcommand{\Mpl}{M_{\textrm{Pl}}}
\newcommand{\Mpld}{M_{\textrm{Pl};d}}

\newcommand{\df}{\mathrel{:=}}
\newcommand{\noeq}{\mathrel{\phantom{=}}}

\newcommand{\diag}{\mathrm{diag}}

\DeclareMathOperator{\Tr}{Tr}

\definecolor{cobalt}{RGB}{44, 98, 120}
\definecolor{celadon}{rgb}{0.67, 0.88, 0.69}
\definecolor{dm}{cmyk}{.20, 0, .30, 0}
\definecolor{burgundy}{rgb}{0.5, 0.0, 0.13}
\definecolor{plotBlue}{RGB}{94, 130, 181}

\newcommand{\rmd}{\textrm{d}}
\def\be{\begin{equation}}
\def\ee{\end{equation}}
\def\bea{\begin{eqnarray}}
\def\eea{\end{eqnarray}}

\hypersetup{
  colorlinks,
  citecolor=Violet,
  linkcolor=cobalt,
  urlcolor=Blue}

\newif\iffastcompile

\fastcompilefalse

\iffastcompile
\newcommand{\js}[1]{}
\newcommand{\jsi}[1]{}
\newcommand{\cl}[1]{}
\newcommand{\lm}[1]{}
\else
\newcommand{\js}[1]{\todo[color=cobalt!30,size=\scriptsize, bordercolor=cobalt!30]{JS: #1}}
\newcommand{\jsi}[1]{\todo[color=cobalt!30,size=\scriptsize, bordercolor=cobalt!30, inline]{JS: #1}}
\newcommand{\cl}[1]{\todo[color=burgundy!30, size=\scriptsize, bordercolor=burgundy!30]{CL: #1}}
\newcommand{\lm}[1]{\todo[color=dm!90, size=\scriptsize, bordercolor=dm!90]{LM: #1}}
\fi

\ProvideTextCommandDefault{\Dbar}{%
\leavevmode\lower.5ex\rlap{\hskip-.07em\accent"16}D%
}

\begin{document}
	\newcommand{\main}{.}
\begin{titlepage}

\setcounter{page}{1} \baselineskip=15.5pt \thispagestyle{empty}

{\flushright ACFI-T22-07 \\ }

\bigskip\

\begin{center}
{\fontsize{20}{28} \bfseries Sharpening the Distance Conjecture  \\ \vspace{0.3cm} in Diverse Dimensions}

 \end{center}
\vspace{0.5cm}

\begin{center}
{\fontsize{14}{30}\selectfont Muldrow Etheredge$^a$, Ben Heidenreich$^a$, Sami Kaya$^b$, Yue Qiu$^a$, \\[4pt] and Tom Rudelius$^b$}
\end{center}

\begin{center}
\vspace{0.25 cm}
\textsl{$^{a}$Department of Physics, University of Massachusetts, Amherst, MA 01003 USA}\\
\textsl{$^{b}$Department of Physics, University of California, Berkeley, CA 94720 USA}\\

\vspace{0.25cm}

\end{center}

\vspace{.9cm}
\noindent 
The Distance Conjecture holds that any infinite-distance limit in the scalar field moduli space of a consistent theory of quantum gravity must be accompanied by a tower of light particles whose masses scale exponentially with proper field distance $\Vert\phi\Vert$ as $m \sim \exp(- \lambda \Vert\phi\Vert)$, where $\lambda$ is order-one in Planck units. While the evidence for this conjecture is formidable, there is at present no consensus on which values of $\lambda$ are allowed. In this paper, we propose a sharp lower bound for the lightest tower in a given infinite-distance limit in $d$ dimensions: $\lambda \geq 1/\sqrt{d-2}$. In support of this proposal, we show that (1) it is exactly preserved under dimensional reduction, (2) it is saturated in many examples of string/M-theory compactifications, including maximal supergravity in $d= \text{4 -- 10}$ dimensions, and (3) it is saturated in many examples of minimal supergravity in $d= \text{4 -- 10}$ dimensions, assuming appropriate versions of the Weak Gravity Conjecture. We argue that towers with $\lambda < 1/\sqrt{d-2}$ discussed previously in the literature are always accompanied by even lighter towers with $\lambda \geq 1/\sqrt{d-2}$, thereby satisfying our proposed bound. We discuss connections with and implications for the Emergent String Conjecture, the Scalar Weak Gravity Conjecture, the Repulsive Force Conjecture, large-field inflation, and scalar field potentials in quantum gravity. In particular, we argue that if our proposed bound applies beyond massless moduli spaces to scalar fields with potentials, then accelerated cosmological expansion cannot occur in asymptotic regimes of scalar field space in quantum gravity.

 \vspace{.9cm}

\bigskip
\noindent\today

\end{titlepage}
\setcounter{tocdepth}{2}
\tableofcontents

\section{Introduction}\label{INTRO}

In recent years, the search for universal features of theories of quantum gravity has produced an enormous body of research. This search is hindered significantly by the complicated nature of quantum gravity, and as a result our understanding of quantum gravity beyond the weakly coupled, supersymmetric context is very limited.

Within this world of darkness and confusion, the Distance Conjecture shines as a beacon of clarity and hope. This conjecture, originally referred to as ``Conjecture 2'' by Ooguri and Vafa in their seminal work \cite{Ooguri:2006in}, deals with asymptotic limits of moduli spaces of quantum gravity theories, which are parametrized by vacuum expectation values of massless scalar fields. Exactly massless scalar fields ordinarily require an infinite degree of fine-tuning, so they are expected to appear only in theories with eight or more supercharges, where their masses are protected by supersymmetry. Furthermore, asymptotic limits of moduli spaces represent weak coupling limits of quantum gravity. This means that the claims of the Distance Conjecture, in its most conservative formulation, are restricted to the weakly coupled, supersymmetric regime of quantum gravity, where our understanding is extensive and claims can be tested with a relatively high level of rigor.\footnote{Already in their original paper \cite{Ooguri:2006in}, Ooguri and Vafa conjectured that the Distance Conjecture should apply beyond massless moduli spaces to include scalar fields with potentials as well. We will discuss scalar fields with potentials further in \S\ref{DISC}.} So far, the Distance Conjecture has passed all of these tests.

Morally speaking, the Distance Conjecture stipulates the existence of a tower of exponentially light states in every infinite-distance limit of scalar field moduli space. A more precise statement is as follows:
\vspace{.2cm}
           \begin{namedconjecture}[The Distance Conjecture]
Let $\cM$ be the moduli space of a quantum gravity theory in $d \geq 4$ dimensions, parametrized by vacuum expectation values of massless scalar fields. Compared to the theory at some point $p_0 \in \mathcal{M}$, the theory at a point $p \in \mathcal{M}$ has an infinite tower of particles, each with mass scaling as
\be
m \sim \exp( -\lambda \Vert p - p_0\Vert )\,,
\label{DCdef}
\ee 
where $\Vert p - p_0\Vert$ is the geodesic distance in $\mathcal{M}$ between $p$ and $p_0$, and $\lambda$ is some order-one number in Planck units $(8 \pi G = \kappa_d^2 = 1)$.
            \end{namedconjecture}
    \vspace{.1cm}
\noindent
This conjecture has been confirmed in a vast array of top-down examples in string theory, and strong bottom-up arguments for its validity have been given in the context of effective field theory. However, despite this enormous body of research into the Distance Conjecture, a simple but crucial question remains: which values of $\lambda$ are allowed? ``Order-one'' carries a wide range of possible interpretations, but if $\lambda$ is too small, the constraints from the Distance Conjecture on low-energy effective field theory will be very weak. As a result, the key open question facing us is to place a lower bound on the coefficient $\lambda$ appearing in the definition of the Distance Conjecture \eqref{DCdef}.

In this paper, we propose a simple answer to this question: in quantum gravity in $d$ dimensions, any infinite-distance limit in moduli space features at least one tower which satisfies the Distance Conjecture with 
\be
\boxed{
\lambda \geq \frac{1}{\sqrt{d-2}}\,.
}
\label{prop}
\ee 
Note that this does not preclude the existence of additional, heavier towers in this infinite-distance limit with $\lambda < 1/\sqrt{d-2}$.

This proposed bound may come as a surprise to Distance Conjecture connoisseurs, since there are a number of examples of towers in 4d theories that scale as $m \sim \exp( -  \Vert p-p_0\Vert/\sqrt{6} )$, naively satisfying the Distance Conjecture with $\lambda = 1/ \sqrt{6}$ yet violating our proposed bound \eqref{prop}. Indeed, this has led a number of authors to single out $\lambda_{\textrm{min}} = 1/\sqrt{(d-1)(d-2)}$ as the minimal value for $\lambda$ in $d$ dimensions \cite{Grimm:2018ohb, Andriot:2020lea, Gendler:2020dfp, Lanza:2020qmt, Bedroya:2020rmd, Lanza:2021qsu}. However, in what follows, we will not only explain why towers with $\lambda = 1/\sqrt{(d-1)(d-2)}$ are ubiquitous in supergravity theories and string compactifications (namely, they arise upon dimensional reduction as Kaluza-Klein zero modes of towers of particles in the parent $(d+1)$-dimensional theory), but we will also argue that such towers are always accompanied in their appropriate limits in moduli space by even lighter towers (namely, Kaluza-Klein towers) that satisfy the Distance Conjecture with $\lambda = \sqrt{(d-1)/(d-2)}$ and in turn satisfy our proposed bound \eqref{prop}.
As mentioned above, the bound \eqref{prop} must therefore be understood as a bound on the \emph{lightest} tower of exponentially light states in a given infinite-distance limit of moduli space, not as a bound on \emph{all} towers of exponentially light particles in this limit. Indeed, there is no possible nontrivial lower bound on the coefficient $\lambda$ which applies to all such exponentially light particles in a given infinite-distance limit.\footnote{For example, consider a theory with two canonically normalized massless scalar fields $\rho$, $\phi$ and a tower of massive particles whose masses scale as $m \sim \exp ( - \lambda_\rho \rho)$. Take $\epsilon > 0$ to be an arbitrarily small positive number. We then find an arbitrarily small coefficient $\lambda <  \epsilon$ by considering the $s \rightarrow \infty$ infinite-distance limit of a geodesic $(\rho(s), \phi(s)) = (\delta s, s)$ for $\delta \equiv \epsilon / (2\lambda_\rho)$ (i.e., the limit $\rho, \phi \rightarrow \infty$ with $\rho/\phi \equiv \delta$ fixed). This scenario readily occurs, for instance, in circle compactification of Type II string theory to nine dimensions, where $\phi$ is the dilaton and $\rho$ is the radion. Our bound \eqref{prop} will nonetheless be satisfied in this scenario provided there is a second tower whose masses decay at a faster rate in this limit.} 

Our proposed bound \eqref{prop} is closely related to the Scalar Weak Gravity Conjecture \cite{Palti:2017elp, Lee:2018spm, Andriot:2020lea}, which we define as follows:\footnote{
In its original formulation \cite{Palti:2017elp}, the Scalar Weak Gravity Conjecture merely requires the existence of one particle of mass $m$ satisfying the bound
\be
\frac{g^{ij} \partial_i m \partial_j m }{ m^2} \geq \gamma \kappa_d^2\,,
\ee 
where $g^{ij}$ is the inverse metric on moduli space and $\gamma$ is some order-one coefficient. For $\gamma \geq (d-3)/(d-2)$, this bound implies that the attractive force mediated by massless scalar fields is greater in magnitude that the force of gravity. However, there is no compelling argument as to why the force mediated by massless scalar fields should be greater than the gravitational force, and this proposal cannot be true in non-supersymmetric systems like the world we live in.}
\vspace{.2cm}
           \begin{namedconjecture}[The Scalar Weak Gravity Conjecture]
           Given a massless scalar field modulus $\phi$ in a quantum gravity theory in $d$ spacetime dimensions, there necessarily exists a particle of mass $m$ satisfying
\be
\frac{ (\partial_\phi m )^2}{g_{\phi\phi} m^2} \geq  {\lambda_{\text{min}}^2}{\kappa_d^2}  \equiv \frac{\kappa_d^2}{d-2}\,,~~~ \text{   with }  \partial_\phi m < 0 \,,
\label{sWGC}
\ee 
where $g_{\phi\phi}$ is the $\phi\phi$ component of the metric on scalar moduli space, so that $\frac{1}{2} g_{\phi\phi} \rmd \phi \wedge \star \rmd \phi$ is the kinetic term for $\phi$ in the action.
            \end{namedconjecture}
    \vspace{.1cm}
\noindent
This is a natural extension of the ordinary Weak Gravity Conjecture \cite{Arkanihamed:2006dz}, which holds that given a 1-form gauge field $A$ with coupling constant $g_A$, there must exist a particle of mass $m$ and quantized charge $q_A \in \mathbb{Z}$ satisfying
\be
\frac{g_A^2 q_A^2  }{ m^2} \geq \gamma_d^2 \kappa_d^2\,, \label{WGCeq}
\ee 
where $\gamma_d$ is an order-one number fixed by the black hole extremality bound. Note that the condition $\partial_\phi m < 0$ is needed to ensure that the mass of the particle is decreasing as $\phi$ increases, since there is no analog of charge conjugation for scalar charges $\mu_\phi  = \partial_\phi m$ the way there is for gauge charges.

\begin{figure}
\centering
\includegraphics[width=70mm]{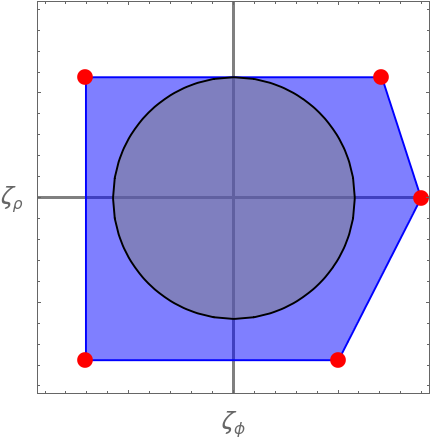}
\caption{The convex hull condition. The Scalar Weak Gravity Conjecture requires the convex hull (blue, shaded) generated by the scalar charge-to-mass vectors $\zeta_i$ of the particles of the theory (red dots) to contain a ball of radius $1/\sqrt{d-2}$ centered at the origin (gray, shaded).}
\label{fakefig}
\end{figure}

By varying over all possible 1-form gauge fields $A$ in the theory, one finds that the Weak Gravity Conjecture as stated in \eqref{WGCeq} is equivalent to the convex hull condition of \cite{Cheung:2014vva}. Similarly, by varying over all massless scalar fields $\phi$ in the theory, \eqref{sWGC} may be viewed as an analog of the convex hull condition for scalar charges rather than gauge charges, as proposed previously in \cite{Calderon-Infante:2020dhm}. More precisely, 
given a particle of mass $m$, we define the scalar charge-to-mass vector as
\be
\zeta_i\equiv\frac 1{\kappa_d}\frac\partial{\partial \phi^i}\log m\,,
\label{zetavec}
\ee
where the differentiation is performed with the $d$-dimensional Planck mass held fixed. The length of a scalar charge-to-mass vector is determined by contracting with the inverse of the scalar kinetic matrix, $|\vec{\zeta}| \equiv \sqrt{g^{ij} \zeta_i \zeta_j}$.
With these definitions, the Scalar Weak Gravity Conjecture defined above is equivalent to the statement that the convex hull generated by all of the $\vec\zeta$-vectors contains a ball of radius $\lambda_{\textrm{min}} = 1/\sqrt{d-2}$ centered at the origin of the scalar charge-to-mass vector space, as illustrated in Figure \ref{fakefig}.

The coefficient $1/(d-2)$ appearing in \eqref{sWGC} has been carefully chosen to coincide with the value of $\lambda_{\textrm{min}}^2$ we have proposed in \eqref{prop}, since for $ \phi \rightarrow \infty$, a tower of particles of mass $m = m_0 \exp( - \lambda \phi)$ will have $g^{\phi \phi} (\partial_\phi m)^2 / m^2 = \lambda^2$. In light of this, there is a simple analogy between the various conjectures highlighted in this paper: just as the ordinary Weak Gravity Conjecture requires a single particle satisfying 
the bound \eqref{WGCeq} and the tower Weak Gravity Conjecture requires a whole tower of particles satisfying that bound, so too does the Scalar Weak Gravity Conjecture require a single particle satisfying the bound \eqref{sWGC}, while the Distance Conjecture requires a whole tower of particles satisfying this latter bound. Our paper could therefore be viewed equally well as a sharpening of the Distance Conjecture or as a sharpening of the Scalar Weak Gravity Conjecture.

There is one important difference between the Distance Conjecture and the Scalar Weak Gravity Conjecture, however: the Scalar Weak Gravity Conjecture is a bound on the scalar charges with respect to \emph{all} massless scalars in the theory, including compact scalar fields known as axions. The Distance Conjecture, on the other hand, says very little about couplings to compact scalar fields, since it constrains infinite displacements in moduli space, and a compact scalar field can have only a finite displacement. Thus, despite significant overlap between the two conjectures, neither the Distance Conjecture nor the Scalar Weak Gravity Conjecture implies the other: the former requires an infinite tower of exponentially light particles while the latter only requires a finite number, but the latter requires \eqref{sWGC} to be satisfied even if $\phi$ is an axion, while the former does not constrain couplings to axions. Nonetheless, in what follows, we will see strong evidence that both conjectures are satisfied in supergravity theories with a coefficient $\lambda_{\textrm{min}} =  1/\sqrt{d-2}$.

We provide several lines of evidence in support of our proposal \eqref{prop}. To begin, in \S\ref{DIMRED} we argue that the bound is exactly preserved under dimensional reduction: a simple theory that saturates the Distance Conjecture with $\lambda \geq 1/\sqrt{D-2}$ in $D$ spacetime dimensions will, after dimensional reduction, saturate the Distance Conjecture with $\lambda = 1/ \sqrt{d-2}$ in $d=D-1$ dimensions. This feature of preservation under dimensional reduction is not a necessity: there is nothing wrong in principle with a bound that is saturated in $D$ dimensions yet satisfied comfortably after dimensional reduction to $d$ dimensions. However, many of the most rigorously tested quantum gravity conjectures--including the absence of global symmetries \cite{Hawking:1974sw, Banks:2010zn}, the Weak Gravity Conjecture \cite{Arkanihamed:2006dz} and the Repulsive Force Conjecture \cite{Palti:2017elp, Heidenreich:2019zkl}--are exactly preserved under dimensional reduction \cite{Banks:2010zn, Heidenreich:2015nta, Heidenreich:2019zkl}, and a number of more speculative conjectures may be sharpened by demanding such preservation as well \cite{Rudelius:2021oaz, Montero:2021otb}. The preservation of \eqref{prop} under dimensional reduction therefore offers a tantalizing hint that $\lambda_{\text{min}} = 1/\sqrt{d-2}$ is indeed the correct value for a lower bound on $\lambda$, but this hypothesis merits further testing.

In \S\ref{4D}-\ref{BOTTOMUP}, we thus carry out tests of our proposal in supergravity theories and string/M-theory compactifications to 4 -- 10 dimensions. These tests may be carried out in one of two ways. The ``top-down'' approach begins with a particular infinite-distance limit of an explicit string/M-theory compactification, computes the exponential decay of masses of towers of light particles with increasing field distance, and compares the coefficient to $\lambda_{\textrm{min}} = 1 / \sqrt{d-2}$. This method is difficult for general Calabi-Yau compactifications, though a number of examples have already been considered in the four-dimensional context. In \S\ref{4D}, we review these 4d examples and argue that they are consistent with our proposed bound $\lambda \geq 1/\sqrt{d-2}$.

In \S\ref{TOPDOWN}, we carry out further top-down checks of our proposed bound by considering maximal supergravities in $d= \text{4 -- 10}$ dimensions, which arise from M-theory compactified on $T^n$. Using U-duality \cite{Hull:1994ys}, we show that our bound $\lambda \geq 1/\sqrt{d-2}$ is saturated in all of these examples. We further argue that this bound is saturated in type I and heterotic string theory in 10 dimensions. In all of these cases where we have checked, whenever the bound is saturated, there is a string scale with an associated tower of string oscillator modes saturating the bound.

In \S\ref{BOTTOMUP}, we employ a ``bottom-up'' approach (which was previously used in \cite{Gendler:2020dfp}) to determine the coefficient $\lambda_{\textrm{min}}$ in minimal supergravity in 5 -- 9 dimensions. This approach proceeds by examining the behavior of the gauge couplings for the 1-form and 2-form gauge fields in the theory and then invoking appropriate versions of the Weak Gravity Conjecture. More precisely, the tower Weak Gravity Conjecture \cite{Heidenreich:2015nta, Heidenreich:2016aqi, Andriolo:2018lvp} holds that as the gauge coupling $g_A$ for some 1-form gauge field $A_1$ tends to zero, there will be a tower of light states with masses bounded in Planck units as
$m \lesssim g_A $.
If $g_A$ decays exponentially in some asymptotic limit of moduli space as $g_A \sim \exp(- \lambda \Vert p -p_0\Vert)$, therefore, the tower Weak Gravity Conjecture will immediately imply that the Distance Conjecture is satisfied with that same coefficient $\lambda$.
Similarly, the Weak Gravity Conjecture for a 2-form gauge field $B_2$ says that as the gauge coupling $g_B$ tends to zero, there will be a charged string whose tension in Planck units is bounded as $
T_{\textrm{string}} \lesssim g_B $.
If this string is a fundamental string, meaning that its core probes quantum gravity physics in the deep ultraviolet \cite{Reece:2018zvv} (see also \cite{Dolan:2017vmn}), then it will give rise to a tower of string oscillator modes beginning at the string scale $M_{\textrm{string}} \equiv \sqrt{2 \pi T_{\textrm{string}}} \lesssim \sqrt{g_B}$. If $g_B$ decays exponentially in some asymptotic limit of moduli space as $g_B \sim \exp(-2 \lambda \Vert p -p_0\Vert)$, therefore, the Weak Gravity Conjecture will immediately imply that the Distance Conjecture is satisfied with the coefficient $\lambda$.

Using this bottom-up approach, we find that the bound $\lambda \geq 1/\sqrt{d-2}$ is saturated in certain infinite-distance limits in moduli space in $d= \text{5 -- 9}$ dimensions. All of these limits are emergent string limits, meaning that some 2-form gauge coupling vanishes, and the Weak Gravity Conjecture implies a charged string whose oscillator modes satisfy the Distance Conjecture with $\lambda = 1/\sqrt{d-2}$. In contrast, the other infinite-distance limits we consider in these theories involve a 1-form gauge field whose gauge coupling vanishes in the limit, and the tower implied by the tower Weak Gravity Conjecture instead satisfies $\lambda > 1/\sqrt{d-2}$ with room to spare. In fact, in all the cases we encounter, these towers have the scaling behavior expected of Kaluza-Klein towers. Thus, both our top-down and bottom-up analyses lend strong support to the Emergent String Conjecture \cite{Lee:2019wij, Lee:2019xtm}, which holds that every infinite-distance limit must be either an emergent string limit or a decompactification limit, and they further suggest that only the emergent string limits may saturate our proposed bound $\lambda \geq 1/\sqrt{d-2}$. 

In \S\ref{DISC}, we conclude our analysis with brief discussion of open questions and applications of our proposed bound \eqref{prop}. Notably, we point out in \S\ref{SFP} that the lower bound $\lambda \geq 1/\sqrt{d-2}$ implies a low UV cutoff on effective field theory in any infinite-distance limit. If this bound applies to scalar fields with a potential and not merely massless moduli (as conjectured in \cite{Ooguri:2006in, Klaewer:2016kiy}), then this low cutoff leads to an upper bound on scalar potentials in asymptotic limits of scalar field space \cite{Hebecker:2018vxz, Andriot:2020lea,  Bedroya:2020rmd}. Assuming the Emergent String Conjecture applies to such limits, we show that the resulting bound forbids accelerated expansion of the universe in asymptotic regions of scalar field space \cite{Obied:2018sgi}, which agrees with the strong asymptotic de Sitter Conjecture of \cite{Rudelius:2021oaz, Rudelius:2021azq} and suggests that quintessence, like de Sitter, can persist for only a finite period of time in quantum gravity.\footnote{Note that our results do not forbid eternal inflation in the interior of scalar field space, which may occur even if every de Sitter vacuum is metastable and every period of quintessence ends after a finite period of time. For further discussion on this point, see \S\ref{SFP}.}

This low cutoff may also present a problem for large-field inflation models with $|\Delta \phi| \gtrsim 10 \Mpl$, as we discuss in \S\ref{LFI}. 
In \S\ref{AXIONS}, we comment on an extension to periodic scalar fields, also known as axions, and
 in \S\ref{BHRF} we consider implications of our bound for black holes in supergravity and the Repulsive Force Conjecture \cite{Palti:2017elp, Heidenreich:2019zkl}. In \S\ref{ESC}, we elaborate on connections to the Emergent String Conjecture and the possibility of an upper bound on the coefficient $\lambda$.

\section{Dimensional Reduction}\label{DIMRED}

In this section, we use dimensional reduction to isolate three special values of the Distance Conjecture parameter $\lambda$.

We begin with an Einstein-dilaton action in $D= d+1$ dimensions,
\be
S =  \int \rmd^D x \sqrt{-g} \left( \frac{1}{2\kappa_D^2} {\cal R}_D - \frac{1}{2} (\nabla \hat\phi)^2 \right)  \,. \label{eq:generalaction}
\ee
Here and in what follows, we often use $\hat{\cdot}$ to indicate that the scalar field $\cdot$ is canonically normalized.
We then consider the dimensional reduction ansatz:
\be
ds^2 = \e^{-\frac{\rho(x)}{d-2}} d{\hat s}^2(x) + \e^{\rho(x)} dy^2,
   \label{eq:dimredansatz}
\ee
where $y \cong y + 2 \pi R$.


Suppose that there is a tower of particles in $D$ dimensions with masses that scale in the limit $\hat \phi \rightarrow \infty$ as
\be
m_{\text{part}}^{(D)}   \sim \exp ( - \kappa_D \lambda_D  \hat \phi  )\,,
\ee 
Upon reduction, the tower of particles reduces to a tower of particles with masses that scale as 
\be
m_{\text{part}}^{(d)} \sim \exp \left(  -\kappa_d  \lambda_D  \hat{\phi} -  \frac{ \kappa_d}{\sqrt{(d-1)(d-2)} }\hat{\rho}  \right)\,,
\label{eq28}
\ee 
where we have defined the canonically normalized radion field $\hat \rho = \frac{2}{\kappa_d} \sqrt{\frac{d-2}{d-1}} \rho$. Defining another canonically normalized field
\be
\hat{\phi}' = \frac{1}{  \lambda_D^2 + \frac{1}{(d-1)(d-2)  } } \left( \lambda_D  \hat \phi + \frac{1}{\sqrt{(d-1)(d-2)} } \hat \rho \right)\,, 
\ee 
we may also write this as
\be
m_{\text{part}}^{(d)} \sim \exp \left(  - \kappa_d \left( \lambda_D^2 + \frac{1}{(d-1)(d-2)}  \right)^{1/2}  \hat \phi'    \right) \equiv \exp \left(   -\kappa_d \lambda_d  \hat \phi'    \right) \,.
\label{lambdaddef}
\ee 
There is also a tower of Kaluza Klein modes for the graviton with masses that scale as 
\be
m_{\text{KK}}^{(d)} \sim \exp \left(-  \kappa_d   \sqrt{\frac{ {d-1} }{{d-2} } }\hat{\rho}  \right)\,.
\label{KK}
\ee 
From the above analysis, we note three special values of the parameter $\lambda_d$:
\begin{enumerate}
\item $\lambda_d = 1 / \sqrt{(d-1)(d-2)}  $. This is the coefficient of the radion $\hat \rho$ for the dimensionally reduced tower in \eqref{eq28}. In other words, as $\hat \rho \rightarrow \infty$, this tower of states will become massless, with masses decaying exponentially as $m \sim \exp ( - \lambda_d  \kappa_d \hat \rho)$.
\item  $\lambda_d =  \sqrt{(d-1)/(d-2 )} $. This is the coefficient of the radion $\hat \rho$ for the Kaluza Klein modes in \eqref{KK}. Note that this value is always larger than the first value of $\lambda_d$, so these Kaluza Klein modes will become massless more quickly than the dimensionally reduced modes 
\item $\lambda_d = 1 / \sqrt{d-2}$. This value is distinguished by the fact that for $\lambda_{D} = 1 / \sqrt{D-2}$, \eqref{lambdaddef} gives $\lambda_d = 1 / \sqrt{d-2}$. In other words, this value of $\lambda_d$ is \emph{exactly preserved} under dimensional reduction.\footnote{More generally, $\lambda_d^2 = \frac{1}{d-2} + \beta$ is exactly preserved under dimensional reduction, but we will see many examples below that saturate this bound with $\beta = 0$, leading us to single out this particular value from the rest.} Preservation under dimensional reduction has proven to be a useful tool for sharpening various quantum gravity conjectures--see \cite{Heidenreich:2015nta, Heidenreich:2019zkl, Rudelius:2021oaz, Montero:2021otb} for examples.
\end{enumerate}

Of the three distinguished values, the first is the smallest. It is therefore tempting to conjecture that this is the 	``correct'' minimum value of $\lambda$ in the Distance Conjecture, i.e.,
\be
\lambda \geq \frac{1}{\sqrt{(d-1)(d-2)}}\,.
\label{att1}
\ee 
This bound was, in fact, proposed in \cite{Andriot:2020lea, Gendler:2020dfp}, motivated by the work of e.g. \cite{Grimm:2018ohb}. 

However, we note something interesting in the example above: although towers of states saturating the value $\lambda = {1}/{\sqrt{(d-1)(d-2)}}$ appear naturally in dimensional reduction, they are accompanied in this context by Kaluza Klein towers, which saturate the stronger bound
\be
\lambda \geq \sqrt{(d-1)/(d-2)} \,.
\label{att2}
\ee 
 Thus, every decompactification limit from $d$ to $D =d+1$ dimensions seems to introduce a tower of particles satisfying the Distance Conjecture with a coefficient $\lambda \geq \sqrt{(d-1)/(d-2)}$.

However, not every infinite-distance limit is a decompactification limit, and the bound $\lambda \geq \sqrt{(d-1)/(d-2)}$ is not satisfied in general. According to the Emergent String Conjecture \cite{Lee:2019xtm}, every infinite-distance limit that is not a decompactification limit is an emergent string limit, in which a charged fundamental string becomes tensionless asymptotically and a tower of string states become light. As we will see below, the tension of a fundamental string scales as $T_{\textrm{string}} \sim \exp ( -2  \kappa_d  \hat{\phi} / \sqrt{d-2} ) $ for $\hat \phi$ the canonically normalized dilaton in $d$ dimensions, which means that the tower of light string states satisfies the Distance Conjecture with coefficient $\lambda_d = 1 /\sqrt{d-2}$. This is nothing but the third distinguished value of $\lambda$, which we saw was exactly preserved under dimensional reduction.
This leads us finally to conjecture
\be
\lambda \geq \frac{1}{\sqrt{d-2}}
\label{sb}
\ee 
as the correct, sharpened version of the Distance Conjecture, as stated previously in \eqref{prop}. 

It is instructive to see how \eqref{sb} is satisfied in the dimensional reduction example considered above if we set $\lambda_D \geq 1/\sqrt{D-2}$. Upon dimensional reduction, we find a theory with two towers of states: one tower of Kaluza-Klein modes, and one that descends from the tower of particles in $D$ dimensions. By the definition in \eqref{zetavec}, these have $\vec{\zeta}$-vectors in the $(\hat \phi, \hat \rho)$ basis given by
\be
\vec{\zeta}_{\textrm{KK}}^{(d)} =  \begin{pmatrix}0\\ \sqrt{\frac{d - 1}{d -2}} \end{pmatrix}
\qquad\text{and}\qquad
\vec\zeta_\text{part}^{(d)} = \begin{pmatrix}\lambda_D \\\frac{1}{\sqrt{(d - 1) (d - 2)}} \label{eq:endpoints}\end{pmatrix},
\ee
which may be computed from the mass formulas in \eqref{eq28}, \eqref{KK}. For $\lambda_D = 1/\sqrt{D-2}$, the vector $\vec\zeta_\text{part}^{(d)}$ has magnitude $\lambda_d = 1/\sqrt{d-2}$ and therefore saturates our proposed bound in the limit $\phi' \rightarrow \infty$. The vector $\vec\zeta_\text{KK}^{(d)}$ has magnitude $\lambda_{\textrm{KK}}  = \sqrt{(d-1)/(d-2)}$ and satisfy our bound comfortably in the limit $\rho \rightarrow \infty$. In the intermediate regime $\rho, \phi \rightarrow \infty$ with fixed $\rho/\phi \geq 1/\sqrt{d-2}$, these Kaluza-Klein modes will still satisfy our bound $\lambda  \geq 1/\sqrt{d-2}$, saturating this bound for $\lambda_D = 1/\sqrt{D-2}$ when $\rho, \phi \rightarrow \infty$ with fixed $\rho/\phi = 1/\sqrt{d-2}$.

\begin{figure}
\begin{center}
\center
\includegraphics[width=60mm]{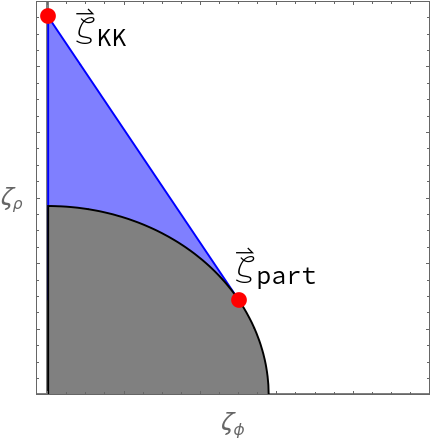}
\caption{Dimensional reduction and the convex hull condition. The gray region is the ball of radius $1/\sqrt{d-2}$ centered at the origin. The vector $\vec\zeta_\text{KK}=(0,\sqrt{(d-1)/(d-2)})$ comes from the Kaluza Klein modes, and the vector $\vec\zeta_\text{part}=(1/\sqrt{d-1},1/\sqrt{(d-1)(d-2)})$ comes from particles with scalar charges in the $D$-dimensional theory that saturate the proposed bound $\lambda_\text{min}^{(D)}=1/\sqrt{D-2}$. Since the convex hull of these vectors contains the ball of $1/\sqrt{d-2}$, the Distance Conjecture will be satisfied with a coefficient $\lambda \geq 1/\sqrt{d-2}$ in every direction $\rho, \phi \rightarrow \infty$ with $\rho/\phi \geq 1/\sqrt{d-2}$. \label{fig:KKCHC}}
\end{center}
\end{figure}

This is illustrated pictorially in Figure \ref{fig:KKCHC}. The part of the convex hull generated by $\vec{\zeta}_{\textrm{KK}}^{(d)}$ and $\vec{\zeta}_{\textrm{part}}^{(d)}$ remains outside the ball of radius $1/\sqrt{d-2}$ and is tangent to the ball at the point $\vec{\zeta}_{\textrm{part}}^{(d)}$ provided $\lambda_D = 1/\sqrt{D-2}$. This means that the Distance Conjecture will be satisfied with $\lambda \geq 1/\sqrt{d-2}$ everywhere along this line segment, i.e., in any infinite-distance limit with $\rho, \phi \rightarrow \infty$, $\rho/\phi \geq 1/\sqrt{d-2}$.

One additional feature of this line segment is worth noting, as it further distinguishes the value $\lambda_d = 1/\sqrt{d-2}$. Namely, when $\lambda_{D} \leq 1/\sqrt{D -2}$, the point on this line segment closest to the origin is simply the endpoint $\vec\zeta_\text{part}^{(d)}$. In contrast, when $\lambda_D > 1/\sqrt{D - 2}$, the point on this line segment closest to the origin lies in the interior of the line segment.

A very similar analysis applies to a more general compactification from $D=d+n$ dimensions to $d$ dimensions, for $n \geq 1$. Such a compactification has \cite{MReece}
\begin{align}
m_{\textrm{KK}}^{(d)}& \sim \exp \left(-  \kappa_d   \sqrt{\frac{ {n+d-2} }{{n(d-2)} } }\hat{\rho}  \right) \,,\label{nKK} \\
m_{\textrm{part}}^{(d)}& \sim \exp \left( - \lambda_D \phi -  \kappa_d   \sqrt{\frac{ n}{{(n+d-2)(d-2)} } }\hat{\rho}  \right)  \,.
\end{align}
This leads to
\be
\vec{\zeta}_{\textrm{KK}}^{(d)} =  \begin{pmatrix}0\\  \left( \frac{n+d-2}{n(d -2)}\right)^{1/2} \end{pmatrix}
\qquad\text{and}\qquad
\vec\zeta_\text{part}^{(d)} = \begin{pmatrix}\lambda_D \\ \left( \frac{n}{(n+d - 2) (d - 2)} \right)^{1/2} \end{pmatrix}.
\label{ndimred}
\ee
As in the $n=1$ case above, the vector $\vec\zeta_\text{part}^{(d)}$ lies on the ball of radius $\lambda_{\textrm{min}}= 1/\sqrt{d-2}$ for $\lambda_D = 1/\sqrt{D-2}$, and the line between this vector and $\vec\zeta_\text{KK}^{(d)}$ is tangent to this ball, as depicted in Figure \ref{fig:KKCHC}. In the limit $n \rightarrow \infty$, the vectors $\vec\zeta_\text{part}^{(d)}$ and $\vec\zeta_\text{KK}^{(d)}$ coalesce, and $\lambda_{\textrm{KK}} \rightarrow 1/\sqrt{d-2}$.

\subsection{Winding Modes and Kaluza-Klein Monopoles}

Although the picture in Figure \ref{fig:KKCHC} is suggestive, it is not complete. In order satisfy our bound \eqref{prop} in all infinite-distance limits, we must ensure that the entire ball of radius $\lambda_{\textrm{min}} = 1/\sqrt{d-2}$ is contained in the convex hull of the $\vec{\zeta}$-vectors. So far in our discussion of dimensional reduction, we have only considered limits in which the dilaton $\phi$ and the radion $\rho$ tend to $+ \infty$. What about the opposite limits, in which one or both of these fields tend to $- \infty$?

In these cases, satisfying the Distance Conjecture typically requires new ingredients beyond what we have considered so far in this section. In string theory, the strong coupling limit $\phi \rightarrow -\infty$, in which the dilaton diverges, is typically equivalent to a weak coupling limit in a dual frame. This is seen most clearly in the case of Type IIB string theory in 10 dimensions or Type II string theory compactified on a circle to 9 dimensions, which we will review below.

In the limit $\rho \rightarrow - \infty$, the radius of the dimensional reduction circle vanishes. Here, the Kaluza-Klein tower becomes heavy, but a tower of light charged particles appears from string winding modes. In particular, wrapping a string of tension $T$ in $D$ dimensions around the dimensional reduction circle will produce a tower of charged particles in $d$ dimensions whose masses scale with the canonically-normalized radion $\hat \rho$ as 
\be
m \sim \exp \left(   \frac{ d-3   }{   \sqrt{(d-1)(d-2)}  }   \kappa_d  \hat \rho  \right) \,.
\ee 
For $d \geq 5$, these winding modes will satisfy our proposed bound \eqref{prop}. For $d=4$, we have $\lambda = 1 /\sqrt{6}$, so these winding modes do not satisfy our proposed bound.

However, in $d=4$, we expect another tower of light particles to appear in the $\rho \rightarrow \infty$ limit: Kaluza-Klein monopoles. Our dimensional reduction ansatz produces a Kaluza-Klein gauge field $A^{\text{KK}}_1$ in $d$ dimensions with gauge coupling
\be
\frac{1}{\eKK^2} = \frac{R^2}{ 2 \kappa_d^2} \e^{  2  \sqrt{(d-1)/(d-2)} \kappa_d \hat \rho} \,.
\ee 
 In the limit $\hat \rho \rightarrow - \infty$, the magnetic gauge coupling $\gKK \equiv 2 \pi / \eKK$ vanishes as  $\gKK \sim \exp ( - \sqrt{(d-1)/(d-2)}  \kappa_d |\hat \rho|)$. In 4d, the tower Weak Gravity Conjecture applied to the electromagnetic dual gauge field therefore implies a tower of Kaluza-Klein monopoles with $m \lesssim \gKK / \kappa_4 \Rightarrow \lambda \geq \sqrt{3/2}$, so the tower of Kaluza-Klein monopoles satisfies our proposed bound \eqref{prop}.
 
In 5d, the Kaluza-Klein monopole is a string. This  string will be charged magnetically under the Kaluza-Klein gauge field, and its tension scales as
\be
T_{\textrm{string}} \lesssim  1/ (\eKK \kappa_5) \sim \exp\biggl( \frac{2}{\sqrt{3}} \kappa_5 \hat \rho\biggr) \,.
\ee 
The oscillator modes of this string will produce a tower of massive particles beginning at the scale $M_{\text{string}} = \sqrt{2 \pi T_{\text{string}}} \lesssim \exp( \kappa_5 \hat \rho /\sqrt{3} )$, satisfying the bound $\lambda \geq 1/\sqrt{d-2}$ in the small radius limit $\hat \rho \rightarrow - \infty$.
 
We expect, therefore, that in the small radius limit $\hat \rho  \rightarrow  -\infty$ of a dimensional reduction, the Distance Conjecture with $\lambda \geq 1/\sqrt{d-2}$ will be satisfied by winding string modes in $d \geq 5$ and by Kaluza-Klein monopoles in $d = 4, 5$. Indeed, T-dualities in string theory suggest that the small radius limit of a theory is likely equivalent to a large radius limit in another duality frame, so it is unsurprising that our proposed bound is satisfied in each of these two limits.

Let us pause here to emphasize a parallel between the large radius limit and the small radius limit of  a dimensional reduction from $D=5$ to $d=4$ dimensions. In the large radius limit $\hat \rho \rightarrow  \infty$, we found one tower with $\lambda = 1/\sqrt{6}$, which came from the Kaluza Klein zero modes of a tower of particles in the parent theory in $D=5$ dimensions. We also found another tower with $\lambda = \sqrt{3/2}$, which came from the Kaluza-Klein modes of the graviton. Similarly, in the small radius limit $\hat \rho \rightarrow  -\infty$, we found one tower with $\lambda = 1/\sqrt{6}$, which came from winding string modes, and another tower with $\lambda = \sqrt{3/2}$, which came from Kaluza-Klein monopoles. The Emergent String Conjecture suggests this is not an accident: the $\hat \rho \rightarrow - \infty$ limit should correspond to a decompactification limit in a dual frame, so the towers of winding modes and Kaluza-Klein monopoles are respectively identified with towers of 5d particles and Kaluza-Klein modes in this dual frame. More generally, while towers with $\lambda = 1/\sqrt{6}$ (which do not satisfy our proposed bound \eqref{prop}) seem to be common in infinite-distance limits in 4d, the Emergent String Conjecture strongly suggests that these limits will feature even lighter towers with $\lambda = \sqrt{3/2}$ (which do satisfy the bound \eqref{prop}). In the following section, we will see compelling evidence that this expectation is borne out in 4d supergravity theories and string compactifications.

To conclude this subsection, we consider winding modes and wrapped branes in more general compactifications. Consider a reduction from $D = d+n$ dimensions to $d$ dimensions, and suppose that a $(P-1)$-brane wraps $k$ dimensions of the internal geometry, yielding a $(p-1)$-brane in $d$ dimensions (here, $P=p+k$).
The tension of the resulting $(p-1)$-brane will then scale with the canonically normalized radion in the small volume limit $\hat \rho \rightarrow \infty$ as \cite{MReece}
\be
T_p \sim \exp \left( - \sqrt{ \frac{d-2}{n (n+d-2)}} \frac{p n - k (d-2) }{d-2}  \kappa_d \hat \rho \right) \,.
\ee 
Let us now examine several special cases of this formula. First, we consider the case $p =1$, which corresponds to a particle in $d$ dimensions, for which the tension $T$ is simply the mass. We further suppose that this particle arises from wrapping a brane over the entire $n$-dimensional compactification manifold. Plugging in $p=1$, $k = n$, we have
\be 
m \sim \exp \left(   \sqrt{ \frac{n (d-2)}{n+d-2}} \frac{ d-3 }{d-2}  \kappa_d \hat \rho \right)~~ \Rightarrow~~ \lambda =  \sqrt{ \frac{n (d-2)}{n+d-2}} \frac{ d-3 }{d-2} \,,
\ee 
which satisfies $\lambda \geq 1/\sqrt{d-2}$ for $n \geq 2$, $d \geq 4$. In other words, our proposed bound \eqref{prop} will be satisfied by $n$-branes wrapping a compactification $n$-manifold in the small volume limit for $n >1$.

Next, we consider the case $p=2$, $n=k=1$, which corresponds to a 2-brane wrapping a circle to produce a string in $d$ dimensions. 
This gives 
\be 
T_{\text{string}} \sim \exp \left( \frac{d-4 }{\sqrt{(d-1)(d-2)}}  \kappa_d \hat \rho \right) \,.
\ee 
This string will give rise to a tower of string oscillator modes at the mass scale $M_{\text{string}} = \sqrt{2 \pi T_{\text{string}}}$, which satisfy the Distance Conjecture with a coefficient of $\lambda = \frac{d-4 }{2\sqrt{(d-1)(d-2)}} $.
This saturates the bound $\lambda \geq 1/\sqrt{d-2}$ for $d=10$. Indeed, this describes the weak coupling limit of the Type IIA superstring, which may be realized as the small radius limit of an M2-brane wrapped on a circle.

\subsection{A Convex Hull Condition}\label{CHC}

We now combine the ingredients above in a simple yet illustrative toy model of dimensional reduction, showing how the Distance Conjecture and the Scalar Weak Gravity Conjecture may be satisfied with a coefficient $\lambda_\text{min}^{(d)} = 1/\sqrt{d-2}$ under the assumption that they are satisfied in $D=d+1$ dimensions with $\lambda_\text{min}^{(D)} = 1/\sqrt{D-2}$.

We consider a $D$-dimensional theory with a single modulus, the dilaton $\phi$, and we suppose that this theory has two kinds of strings, where one kind of strings has a tension which scales with the dilaton by $T_\text{string}^+\sim\exp\left(2 \lambda_\text{string}^{(D)}\kappa_D\phi\right)$, and the other kind of strings has a tension which scales with the dilaton by $T_\text{string}^-\sim \exp\left(- 2\lambda_\text{string}^{(D)}\kappa_D\phi\right)$, where $\lambda_\text{string}^{(D)}=1/\sqrt{D-2} = 1/\sqrt{d-1}$. This behavior occurs, for instance, in Type IIB string theory, where S-duality switches the strong and weak coupling limits $\phi \rightarrow \pm \infty$. The string oscillator modes of these respective strings form towers with masses which scale with the dilaton as $m_\text{string}^{(D)}(\phi)\sim \exp(\pm\lambda_\text{string}^{(D)}\kappa_D\phi)$, and thus they saturate the Distance Conjecture and the Scalar Weak Gravity Conjecture in $D$ dimensions.

We may then calculate the minimal radius $\lambda_\text{min}^{(d)}$ by examining the convex hull of all the $\vec\zeta^{(d)}$-vectors, as discussed in the introduction. This convex hull is generated by the $\vec\zeta^{(d)}$-vectors for (a) the Kaluza-Klein modes and (b) the string winding modes. By our discussion earlier in this section, these have $\vec\zeta^{(d)}$-vectors in the $(\hat\phi, \hat\rho)$ basis given by
\be
\vec \zeta_\text{KK}^{(d)}=\begin{pmatrix}0\\ \sqrt{\frac{d - 1}{d -2}} \end{pmatrix},\qquad\vec \zeta_\text{wind}^{(d)}=\begin{pmatrix}\pm \frac 2{\sqrt{d-1}}\\-\frac{d-3}{\sqrt{(d-1)(d-2)}}\end{pmatrix}.
\ee
When $d=4$, we must include an additional point corresponding to the Kaluza-Klein monopoles:
\be
\vec \zeta_\text{mon}^{(4)}=\begin{pmatrix}0\\-\sqrt{\frac 32}\end{pmatrix}\,.
\ee
The string oscillator modes lie on the boundary of the convex hull. They have
\begin{equation}  
\zeta_\text{string}^{(d)}=\begin{pmatrix}\pm \frac 1{\sqrt{d-1}}\\\frac{1}{\sqrt{(d - 1) (d - 2)}}  \end{pmatrix}\,.
 \end{equation}
When $d=5$, we may include an additional point corresponding to the string oscillator modes for the Kaluza-Klein monopole string:
\be
\vec \zeta_\text{mon. str.}^{(5)}=\begin{pmatrix}0\\-\frac{1}{\sqrt{3}}\end{pmatrix}\,.
\ee

\begin{figure}
\begin{center}
\begin{subfigure}{0.475\textwidth}
\center
\includegraphics[width=50mm]{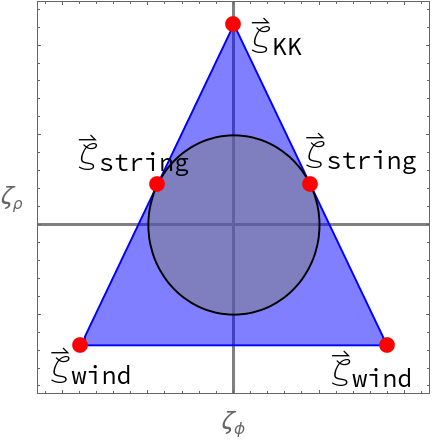}
\caption{$d>5$} \label{sfig:zetadgeq6}
\end{subfigure}
\begin{subfigure}{0.475\textwidth}
\center
\includegraphics[width=50mm]{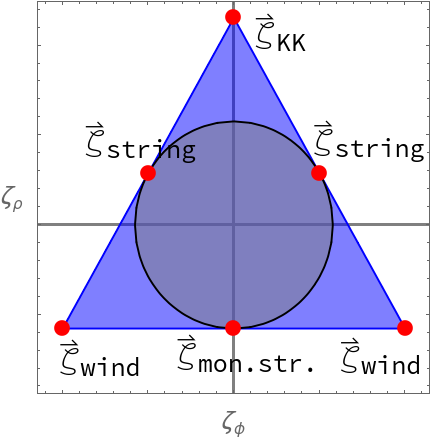}
\caption{$d=5$} \label{sfig:zeta5d}
\end{subfigure}
\hfill
\begin{subfigure}{0.475\textwidth}
\center
\includegraphics[width=50mm]{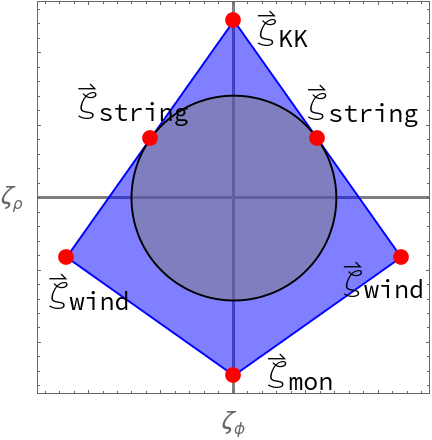}
\caption{$d=4$} \label{sfig:zeta4d}
\end{subfigure}
\caption{The convex hull condition in a toy model of dimensional reduction. The gray region in both figures is the ball of radius $1/\sqrt{d-2}$ centered at the origin. The vector $\vec\zeta_\text{KK}^{(d)}=(0,\sqrt{(d-1)/(d-2)})$ comes from the Kazula-Klein modes; the vectors $\vec\zeta_\text{string}=(\pm1/\sqrt{d-1},1/\sqrt{(d-1)(d-2)})$ come from string oscillation modes in the $D$-dimensional theory, and the vectors $\vec \zeta_\text{wind}=(\pm 2/\sqrt{d-1},-(d-3)/\sqrt{(d-1)(d-2)}$ come from winding states. In the $d=5$ case in \subref{sfig:zeta5d}, there is additionally the $\vec \zeta_\text{mon}^{(5)}$-vector coming from the string oscillator modes of the Kaluza-Klein monopole string. In the $d=4$ case in \subref{sfig:zeta4d}, the vector $\vec \zeta_\text{mon}^{(4)}=(0,-\sqrt{3/2})$ comes from the Kaluza-Klein monopole and is needed for the ball of radius $1/\sqrt{d-2}$ to be contained in the convex hull.\label{fig:zetach}}
\end{center}
\end{figure}

For $d \geq 4$, the convex hulls of the $\vec\zeta^{(d)}$-vectors have a minimal radius $\lambda_\text{min}^{(d)}$ of precisely
\be
\lambda_\text{min}^{(d)}=\frac 1{\sqrt{d-2}}\,,
\ee
as shown in Figure \ref{fig:zetach}. Thus, the bound $\lambda \geq 1/\sqrt{d-2}$ is saturated in $d$ dimensions, just as it was in $D$ dimensions. Whenever it is saturated, there is always a tower of string oscillator modes with $\lambda = 1/\sqrt{d-2}$, which either descend from string oscillator modes in $D$ dimensions or else come from the Kaluza-Klein monopole string in $d=5$.

It is interesting to note that the string oscillator modes are not generators of the convex hull in these examples, since they always saturate the bound $\lambda \geq 1/\sqrt{d-2}$. Instead, the convex hull is generated by a combination of Kaluza-Klein modes, winding modes, and Kaluza-Klein monopoles. More generally, if the Emergent String Conjecture is true, then any generator $\vec{\zeta}^{(d)}$ of the convex hull must either correspond to a tower of string oscillator modes or else a tower of Kaluza-Klein modes in some duality frame. By \eqref{nKK}, this means there is only a finite set of possibilities for the length of such a generator:
\begin{equation}
|\vec{\zeta}^{(d)}| = \frac{1}{\sqrt{d-2}} ~~\text{ or } ~~ |\vec{\zeta}^{(d)}| = \sqrt{ \frac{ n+d-2 }{n (d-2)} }\,,
\end{equation}
for some $n \geq 1$.

In our brief analysis in this section, we have notably ignored the possibility of axions. If the $D$-dimensional theory includes 1-form gauge fields, then the $d$-dimensional theory will have axions, and the electric charge-to-mass ratios in the $D$-dimensional theory determine the axion-charge components of the $\vec\zeta^{(d)}$-vectors of the $d$-dimensional theory. The Weak Gravity Conjecture and Repulsive Force Conjectures in the higher dimensional theory then have important consequences for the axion components of the $\vec\zeta^{(d)}$-vectors. We defer to future research a thorough investigation of these consequences.

\section{A Reexamination of Supergravity in Four Dimensions}\label{4D}

So far, nearly all studies of the Distance Conjecture coefficient $\lambda$ have taken place within the context of four-dimensional supergravity \cite{Grimm:2018ohb, Blumenhagen:2018nts, Joshi:2019nzi, Erkinger:2019umg, EnriquezRojo:2020hzi, Ashmore:2021qdf}. Multiple works have suggested $\lambda_{\textrm{min}} = 1/ \sqrt{6}$ as the minimal bound in four dimensions \cite{Grimm:2018ohb, Andriot:2020lea, Gendler:2020dfp}, and several examples have been previously claimed to saturate this bound. Such examples naively violate our proposed bound $\lambda \geq 1/\sqrt{d-2}$, so before we provide evidence for this bound in higher dimensions, we must first address this apparent contradiction between our proposal and previous claims in the literature.

As we saw in the previous section, the coefficient $\lambda = 1 /\sqrt{6}$ appears readily in decompactification limits: it is the scaling behavior expected for the Kaluza-Klein zero modes of a tower of particles in $D=5$ dimensions after compactification to $d=4$ dimensions. Such towers are always accompanied in these decompactification limits by towers of Kaluza-Klein modes with $\lambda = \sqrt{3/2}$, which ensure consistency with our proposed bound \eqref{prop} on the lightest tower of particles in any infinite-distance limit. In the remainder of this section, we argue that this situation is generic: explicit examples with $\lambda = 1/\sqrt{6}$ discussed previously in the literature always feature lighter towers with $\lambda = \sqrt{3/2}$, so they in fact represent examples in support of our bound, not counterexamples to it.
 
The first extensive discussion of the coefficient $\lambda = 1/\sqrt{6}$ appeared in \cite{Grimm:2018ohb}, which examined the behavior of towers of massive particles in asymptotic regions of scalar field space in four dimensions. In one-dimensional vector multiplet moduli spaces of Type IIB compactifications, they argued that asymptotic limits are characterized by one of three possible values for $\lambda$:
\be
\lambda = \frac{1}{\sqrt{6 }}\,,  \frac{1}{\sqrt{2}} \,, 1 \,.
 \ee 
These values can be determined from the type of singularity that occurs in the infinite-distance limit of complex structure moduli space, which can be classified using the theory of mixed Hodge structures \cite{Grimm:2018ohb, Corvilain:2018lgw, Grimm:2018cpv, Grimm:2019ixq}. This classification, however, only implies the existence of some exponentially light tower with coefficient $\lambda$ given by one of the values above. It does not imply that this is the only exponentially light tower, nor that it is the lightest such tower. Thus, while infinite-distance limits in this classification corresponding to $\lambda = 1/\sqrt{2}$ and $\lambda = 1$ necessarily satisfy our bound \eqref{prop}, limits with $\lambda = 1/\sqrt{6}$ require further analysis.

Such an analysis was recently carried out in the supergravity context in \cite{Gendler:2020dfp}, which pointed out that for $\mathcal{N} = 2$ theories with a single vector multiplet, there are three possible forms for the perturbative part of the prepotential, which in turn lead to two distinct infinite-distance limits. Reference \cite{Gendler:2020dfp} considered one such prepotential in detail:
\be
\cF =  - \frac{(X^1)^3}{X^0}\,.
\ee 
After setting the axion vevs to vanish, this prepotential leads to a gauge kinetic matrix of the form
\be
a_{IJ} = \diag( \e^{ \sqrt{6} \kappa_4 \hat \rho } , 3 \e^{\sqrt{2 /3} \kappa_4 \hat \rho} )\,,
\ee 
for $\hat \rho$ a canonically normalized scalar field. In the limit $\hat \rho \rightarrow \infty$, assuming the tower Weak Gravity Conjecture, we therefore expect two towers of charged particles, one for each gauge field, with
\be
\lambda = \frac{1}{\sqrt{6}} \,,~ \lambda = \sqrt{\frac{3}{2}}\,,
\label{4dvalues}
\ee 
respectively. This reflects precisely the behavior we expect: the existence of a tower with $\lambda = 1/\sqrt{6} < 1/\sqrt{2}$ is accompanied by another tower with $\lambda \geq 1/\sqrt{2}$.

Indeed, these values of $\lambda$ match precisely with what we expect from dimensional reduction of $\mathcal{N}=1$ supergravity with no massless vector multiplets in five dimensions, which has one 1-form gauge field but no scalar fields. Dimensionally reducing to 4d, a tower of superextremal charged particles with $\lambda_5 = 0$ will reduce to a tower of charged particles with $\lambda_4 = 1 / \sqrt{6}$ as in \eqref{eq28}, and there will be Kaluza Klein towers with $\lambda_4 = \sqrt{3/2}$ as in \eqref{KK}. These indeed match the values in \eqref{4dvalues}. We see here that although there is a tower of particles saturating the bound $\lambda \geq 1/ \sqrt{(d-1)(d-2)}$ in the limit $\hat \rho \rightarrow \infty$, there is an even lighter tower of particles that satisfies our proposed bound, $\lambda  \geq 1 / \sqrt{d-2}$.

This behavior also fits nicely with the results of \cite{Lanza:2020qmt, Lanza:2021qsu, Heidenreich:2021yda}, which considered the behavior of axion strings\footnote{Recall that an axion string is a string charged magnetically under an axion $\theta$, so that $\theta \rightarrow \theta + 2 \pi$ as one circles the core of the string.} in infinite-distance limits in moduli space. Reference \cite{Lanza:2021qsu} argued that any such limit corresponds to the tensionless limit of an axion string, and the mass of the lightest tower of particles scales (in Planck units) as either $m^2 \sim T_{\textrm{string}}$, $m^2 \sim T_{\textrm{string}}^2$, or $m^2 \sim T_{\textrm{string}}^3$ in this limit. According to \cite{Heidenreich:2021yda}, the large radius limit of pure 5d supergravity with a nontrivial Chern-Simons coupling on a circle features two gauge fields $A_1$ and $A_1^{\textrm{KK}}$ with gauge couplings $e$ and $\eKK$ satisfying the relation $e \sim \eKK^{1/3}$, and consistency with various forms of the Weak Gravity Conjecture implies an axion string whose tension scales as $T_{\textrm{string}} \sim e^2$. The tower Weak Gravity Conjecture for $A_1^{\textrm{KK}}$ then implies a tower of particles whose masses scale as $m_{\textrm{KK}}^2 \sim \eKK^2 \sim T_{\textrm{string}}^3$, in agreement with \cite{Lanza:2021qsu}. This tower is, of course, simply the Kaluza Klein tower with $\lambda = \sqrt{3/2}$, whereas the tower with $\lambda = 1 /\sqrt{6}$ is the tower of oscillator modes for the axion string.

The presence of a tower with $\lambda = \sqrt{3/2}$ accompanying a tower with $\lambda = 1/\sqrt{6}$ has been observed not only in supergravity and Kaluza-Klein theory, but also in UV complete string compactifications.
In particular, \cite{Joshi:2019nzi} found that near an ``M-point'' of a Calabi-Yau compactification of Type IIA string theory, there exists a tower of light D0-branes with $\lambda = \sqrt{3/2}$ accompanied by a tower of light D2-branes with $\lambda = 1/\sqrt{6}$.\footnote{Note that the conventions of \cite{Joshi:2019nzi} differ from ours by a factor of $\sqrt{2}$: $\lambda_{\textrm{them}} = \sqrt{2} \lambda_{\textrm{us}} $, as previously noted in \cite{Andriot:2020lea}.} This scaling behavior is unsurprising, as this limit may be viewed as a decompactification of the M-theory circle of Type IIA.

In that same paper, \cite{Joshi:2019nzi} studied towers of particles that occur near the ``s-point'' of a Calabi-Yau geometry they called $X_{2,2,2,2}$. In Section 6.2.3. of that paper, they found that the masses of the light particles are given in terms of a pair of integers $x$, $y$ by
\begin{equation}
m(x,y) = m_0 |y|  e^{-  \kappa_4 \hat \phi / \sqrt{6} } + m_1 |2 i y \log(4) - x| e^{-  \kappa_4 \hat \phi  \sqrt{3/2} } + O(e^{-  \kappa_4 \hat \phi  \sqrt{5/2} })\,,
\end{equation}
where $m_0$, $m_1$ are constants, and the s-point corresponds to the limit $\hat \phi \rightarrow \infty$. For $|y| \neq 0$, therefore, there is a tower of light particles indexed by $y \in \mathbb{Z}$ with $\lambda = 1/\sqrt{6}$. For $|y|=0$, however, the first term vanishes, and there is a tower of light particles indexed by $x \in \mathbb{Z}$ with $\lambda = \sqrt{3/2}$. Again, these towers have the expected scaling behavior for a decompactification limit to five dimensions.

Similarly, \cite{EnriquezRojo:2020hzi} found towers with $\lambda = \sqrt{3/2}$ and $1/\sqrt{6}$ in a Type IIB Calabi-Yau orientifold compactification. This example is especially interesting because it features $\mathcal{N}=1$ supersymmetry in four dimensions, so the scalar field in question is not even a massless modulus. We will elaborate on the application of the Distance Conjecture to massive scalar fields below in \S\ref{DISC}.

The authors of \cite{Lanza:2021qsu} studied the spectrum of charged particles and strings in a model of Type IIA string theory compactified on a Calabi-Yau threefold $X$ given by a particular $\mathbb{P}^1$ fibration over $\mathbb{P}^2$. In one limit, they found an asymptotically tensionless string with $T_{\textrm{string}} \sim \exp(-\sqrt{2/3} \kappa_4  \hat \rho)$ and a tower of Kaluza-Klein modes for the Calabi-Yau threefold $X$ with $m \sim \exp(-\kappa_4 \hat \rho/ \sqrt{6}   ) $. From our discussion in \S\ref{DIMRED}, this scaling behavior is precisely what is to be expected for M-theory compactified first on the Calabi-Yau $X$ and then on $S^1$, where $\hat\rho$ is the canonically normalized radion of the $S^1$. From an M-theory perspective, the tower of Kaluza-Klein modes for $X$ with $\lambda = 1/\sqrt{6}$ in the $\hat \rho \rightarrow \infty$ limit will be accompanied in four dimensions by a tower of light Kaluza-Klein modes for the M-theory circle with $\lambda = \sqrt{3/2}$. From a Type IIA perspective, these latter Kaluza-Klein modes will be D0-branes, which were not considered in the analysis of \cite{Lanza:2021qsu}.

Reference \cite{Gendler:2020dfp} also considered a 4d $\mathcal{N}=2$ theory with two vector multiplets and a prepotential of the form
\be
\cF = - \frac{1}{6 X^0} T S^2 \,. 
\ee 
They argued that every infinite-distance limit has a tower satisfying $\lambda \geq 1 / \sqrt{2}$, and this bound is in fact saturated in certain directions in scalar field space. This matches precisely with our proposed bound.

To our knowledge, our discussion in this section has now addressed every reference in the literature to a Distance Conjecture tower in four dimensions with an exact coefficient $\lambda < 1/\sqrt{2}$. We have seen that all such examples feature even lighter towers with $\lambda \geq 1/\sqrt{2}$, and a number of examples in fact saturate this bound. The only potential counterexamples left to discuss are numerical examples from e.g. \cite{Blumenhagen:2018nts, Erkinger:2019umg}, which are listed in Table 3 of \cite{Andriot:2020lea}. These examples involve an averaging procedure, so the associated values of $\lambda$ are known only within a range, $\lambda \in [\lambda_- , \lambda_+ ]$. Some of these ranges have $\lambda_- < 1 /\sqrt{2}$, indicating a possible violation of our proposed bound \eqref{prop}, but all of them have $\lambda_+ > 1/\sqrt{2}$, which is consistent with our bound. In fact, the central values $\lambda_0 = (\lambda_- + \lambda_+)/2$ across all the examples range from $\lambda_0 = 0.7668$ to $0.8638$, tantalizingly close to our proposed minimum value $\lambda_{\text{min}} = 1/\sqrt{2} \approx 0.7071$. Thus, more precise studies of these examples could lead to either a counterexample to our bound or remarkable evidence in favor of it, but we leave this to future study.

 \section{Top-Down Evidence in Maximal Supergravity}\label{TOPDOWN}
 In these sections, we explicitly compute $\lambda_\text{min}$ in maximal supergravity in four to ten dimensions. In each dimension, we find that the Distance Conjecture and Scalar Weak Gravity Conjecture are satisfied in all directions of moduli space by a tower of particles with $\lambda \geq \lambda_\text{min}=1/\sqrt{d-2}$. For each dimension, this bound is in fact saturated in one or more directions, and in all cases that we have checked (namely, for $d \geq 7$), these directions correspond to emergent string limits featuring a tower of string oscillator modes with $\lambda = 1/\sqrt{d-2}$.

\subsection{Ten Dimensions}\label{10D}

There are three types of 10d supergravity for us to consider: Type I, Type IIA, and Type IIB. The relevant parts of their actions (in Einstein frame) take the form \cite{Polchinski:1998rr}:
\begin{align}
S_{\textrm{IIA}} &=  \frac{1}{2 \kappa_{10}^2} \left( \int  - \frac{1}{2} \rmd \sigma \wedge \star \rmd \sigma - \frac{1}{2 } e^{\sigma} H_3 \wedge \star H_3  - \frac{1}{2}  e^{-3 \sigma/2} F_2 \wedge \star F_2   \right)  \\
S_{\textrm{IIB}} &=  \frac{1}{2 \kappa_{10}^2} \left( \int  - \frac{1}{2} \rmd \sigma \wedge \star \rmd \sigma - \frac{1}{2} e^{\sigma} H_3 \wedge \star H_3   \right)  \\
S_{\textrm{I}} &= \frac{1}{2 \kappa_{10}^2} \left( \int  - \frac{1}{2} \rmd \sigma \wedge \star \rmd \sigma - \frac{1}{2} e^{\sigma} H_3 \wedge \star H_3  - \frac{1}{g_A^2} e^{ \sigma/2}   \textrm{Tr} (F_2 \wedge \star F_2)  \right) \,. 
\end{align}
Let us consider each of these in turn, beginning with Type IIA. Here, in the limit $\sigma \rightarrow \infty$, the gauge coupling for the 2-form $B_2$ scales as $g_B \sim e^{-\sigma/2}$. Assuming that a fundamental string charged under $B_2$ satisfies the Weak Gravity Conjecture bound $T \lesssim g_B \kappa_{10} $, its string oscillator modes will scale as
\be
M_{\textrm{string}} \sim \sqrt{2 \pi T} \sim \exp ( \kappa_{10}  \hat \sigma  /\sqrt{8})\,,
\ee 
where we have defined $\hat \sigma = \sigma / (\sqrt{2} \kappa_{10} )$ to be the canonically normalized dilaton. Thus, this tower of string oscillator modes will satisfy the Distance Conjecture with
\be
\lambda = 1 / \sqrt{8}\,,
\ee 
saturating our proposed bound $\lambda \geq 1/ \sqrt{d-2}$. Of course, such a string is not merely a hypothetical entity: this is simply the IIA superstring.

In the other infinite-distance limit, $\sigma \rightarrow - \infty$, the gauge field $A_1$ will become weakly coupled as $g_A \sim \exp( 3 \sigma/4) \sim \exp(  \sqrt{9/8}\kappa_{10} \hat \sigma   )$. The tower Weak Gravity Conjecture then implies a tower of light particles with masses beginning at the scale $g_A \kappa_{10}$, which satisfy the Distance Conjecture with a coefficient of 
\be
\lambda = \sqrt{\frac{9}{8}} \,.
\ee 
This matches the value expected for Kaluza Klein modes \eqref{KK}, and indeed it points towards the well-known fact that this tower of charged particles in Type IIA string theory (D0-branes) is in fact a Kaluza Klein tower for M-theory on a circle.

Next, let us consider Type IIB supergravity. The limit $\sigma \rightarrow + \infty$ is identical to the Type IIA case, and the bound $\lambda \geq 1 / \sqrt{d-2}$ is saturated by oscillator modes of the Type IIB superstring. The limit $\sigma \rightarrow -  \infty$ is mysterious from the perspective of the supergravity action we have written above, as the gauge field $B_2$ becomes strongly coupled. However, here we invoke the well-known S-duality of the Type IIB superstring, which implies that the strong coupling limit of one Type IIB superstring is the weak coupling limit of a different Type IIB superstring. The bound $\lambda \geq 1 / \sqrt{d-2}$ will therefore be saturated in this limit also. More generally, taking into account the axion $C_0$, the convex hull condition for the $\vec{\zeta}$-vectors will be satisfied in every direction in scalar field space by towers of string oscillator modes for $(p, q)$-strings, as the corresponding vectors $\vec\zeta_{p,q}$ densely fill in the sphere of radius $\lambda = 1/\sqrt{d-2}$. 
We see that the duality web plays a crucial role here in satisfying our proposed bound.

Finally, we have Type I supergravity. Here, the limit $\sigma \rightarrow \infty$ once again introduces a tower of string oscillator modes saturating the bound $\lambda  \geq 1/ \sqrt{d-2}$, by the same calculation as in Type IIA. The strong coupling limit $\sigma \rightarrow  - \infty$ is more mysterious, and again we must use known details of the string duality web. Three different string theories have Type I supergravity as their low energy limit: Type I string theory, $SO(32)$ heterotic string theory, and $E_8 \times E_8$ heterotic string theory. The first two of these are S-dual, so the $\sigma \rightarrow - \infty$ limit of Type I string theory corresponds to the $\sigma \rightarrow + \infty$ limit of $SO(32)$ heterotic string theory, and vice versa. Thus, each of these limits will saturate the bound $\lambda \geq 1/\sqrt{d-2}$ as well. The $\sigma \rightarrow - \infty$ limit of $E_8 \times E_8$ heterotic string theory, on the other hand, corresponds to M-theory on an interval separating two Ho\v rava-Witten walls. Here, there is a tower of Kaluza Klein modes with $\lambda = \sqrt{9/8}$, as in the decompactification limit of Type IIA string theory to M-theory above. Once again, we see that string dualities ensure that our proposed Distance Conjecture bound is satisfied.

\subsection{Nine Dimensions}
In $d=9$ dimensional maximal supergravity, coming from M-theory on $T^2$, there are three moduli, all originating from the eleven-dimensional graviton. To determine their couplings, we reduce the $D=11$ dimensional Einstein-Hilbert action
with the ansatz
\be
ds^2_D= \Vert g \Vert^{-\frac{1}{d-2}} g_{\mu \nu}dx^\mu dx^\nu+g_{m n} d y^m d y^n\,, \label{eq:metricansatz}
\ee
where $m,n$ index the $k=D-d$ compact directions, $\mu, \nu$ index the $d$ noncompact directions, $y^m \cong y^m + 2\pi R$, and $\Vert g \Vert = \det g_{mn}$. It is convenient to decompose $g_{mn}$ in terms of volume and shape parameters $U$ and $\tau = \tau_1 + i \tau_2$, respectively, where
\be
g_{m n} = e^U \frac 1{\tau_2}\begin{pmatrix}1&\tau_1\\\tau_1&|\tau|^2\end{pmatrix}.
\ee
With this, the Einstein-moduli sector of the dimensionally reduced action is
\be
S_9=\frac 1{2\kappa_9^2}\int d^9x\sqrt{-g}\left(\mathcal R-\frac 9{14}(\partial U)^2-\frac{(\partial \tau_1)^2+(\partial \tau_2)^2}{2\tau_2^2}\right)\,.
\ee

Working in modified nine-dimensional Planck units where $2 \kappa_9^2 = (2\pi)^6$ for convenience, the spectrum of 1/4 BPS particles is
\be
m_{p,q,w}=\frac{|p+\tau q|}{\sqrt{\tau_2} R}e^{-\frac 9{14} U}+R^{4/3} |w| e^{\frac 67U}\,,
\ee
where $p, q\in \mathbb{Z}$ are the Kaluza-Klein charges and $w \in \mathbb{Z}$ is the M2 brane winding charge. These particles are $1/2$ BPS when either $w=0$ or $p=q=0$.

At a particular point in moduli space, the canonically normalized moduli are
\be
\hat \phi^a=(\hat U,\hat \tau_1,\hat \tau_2)= \frac{1}{\kappa_9} \Biggl(\sqrt{\frac{9}{14}}U,\frac{\tau_1}{\sqrt{2} \langle \tau_2\rangle},\frac{\tau_2}{\sqrt{2} \langle \tau_2\rangle}\Biggr)\,,
\ee
where $\langle \tau_2\rangle$ is the value of $\tau_2$ at the point in question (not including its fluctuations).
The canonically normalied scalar charge-to-mass vectors $\zeta_a=\frac{1}{\kappa_9} \frac\partial{\partial \hat \phi^a}\log m_{p,q,w}$ are then \\[-10pt]
\begin{subequations}
\begin{align}
\zeta_{\hat{U}} &= \frac{4  \sqrt{\tau_2} e^{\frac{3 U}{2}} R^{7/3} |w| -3  |p+\tau q|}{\sqrt{14} \left(|p+\tau q|+ \sqrt{\tau_2} e^{\frac{3 U}{2}} R^{7/3}  |w|\right)},\\
\zeta_{\hat{\tau}_1} &= \frac{\sqrt{2} q \tau_2 (p+\tau_1 q)}{|p+\tau q| \left(|p+\tau q|+\sqrt{\tau_2} e^{\frac{3 U}{2}} R^{7/3}  |w| \right)},\\
\zeta_{\hat{\tau}_2} &= \frac{q^2 \tau_2^2-(p+\tau_1 q)^2}{\sqrt{2} |p+\tau q| \left(|p+\tau q|+ \sqrt{\tau_2} e^{\frac{3 U}{2}} R^{7/3}  |w|  \right)}\,,
\end{align}
\end{subequations}
for the $1/4$ BPS particles.

\begin{figure}
\begin{center}
\center
\includegraphics[width=3.5in]{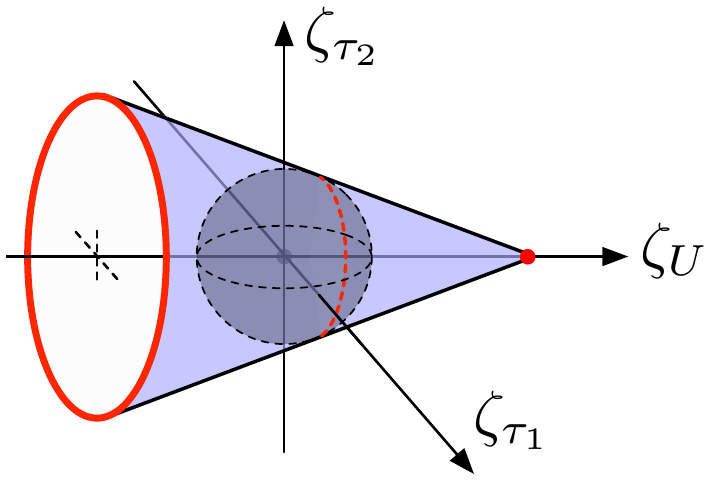}
\caption{The convex hull in 9d maximal supergravity. The $\vec{\zeta}$-vectors of BPS states densely fill the shaded cone, with $1/2$ BPS states at the base and apex and $1/4$ BPS states in between. The ball of radius $1/\sqrt{d-2} = 1/\sqrt{7}$ lies entirely within the cone, touching it along the dashed red circle, which is itself densely populated by the oscillator modes of $1/2$ BPS fundamental strings.
\label{fig:zetaKK}}
\end{center}
\end{figure}

As illustrated in Figure \ref{fig:zetaKK}, these $\vec\zeta$-vectors lie on a cone. At the tip of the cone lie the 1/2 BPS states with $p=q=0$ and $w\neq 0$, corresponding to an M2 brane wrapped $w$ times on $T^2$:
\be
\vec \zeta_9^\text{winding}=\left(\sqrt{\frac87},0,0\right).
\ee
 The base of the cone is populated by the 1/2 BPS Kaluza-Klein modes with $w=0$ but nonzero $p$ or $q$:
 \be
\vec\zeta_9^\text{KK}\in\left\{\left(-\frac {3}{\sqrt {14}},\zeta_{\tau_1},\zeta_{\tau_2}\right)\in \mathbb R^3:\zeta_{\tau_1}^2+\zeta_{\tau_2}^2=\frac12\right\}.
\ee
 The remaining 1/4 BPS states lie along the cone somewhere between its tip and circular base. From this example, we see that the convex hull generated by the 1/2 BPS states contains the convex hull generated by the 1/4 BPS states.

The purely winding $\vec \zeta_9^\text{winding}$-vectors are a distance $\sqrt{\frac87}$ from the origin, and so too are the Kaluza Klein mode $\vec \zeta_9^\text{KK}$-vectors. Thus, the points on the cone closest to the origin lie on a circle halfway between the base and apex. The radius of this circle is $\lambda_\text{min}=1/\sqrt{d-2}$, saturating our proposed bound.

Note that wrapping an M2 brane on the $(p,q)$ cycle of the torus gives rise to $1/2$ BPS fundamental strings of tension
\begin{equation}
T_{\textrm{string}} = \frac{R^{1/3}}{2 \pi}  \frac{|p+\tau q|}{\sqrt{\tau_2}} e^{\frac{3 U}{14}} \,.
\end{equation}
The corresponding string oscillator modes of mass $m \propto \sqrt{2 \pi T_{\textrm{string}}}$ densely populate the $\lambda = 1/\sqrt{d-2}$ circle where the bound is saturated. Thus, when our bound is saturated, there is an associated string scale, as before.

\subsection{Eight Dimensions}

Reducing the M-theory effective action
\be
S_{11}=\frac 1{2\kappa_{11}^2}\int d^{11}x\sqrt{-g} \left(\mathcal R-\frac{1}{2} |F_4|^2\right) - \frac{1}{12 \kappa_{11}^2} \int C_3 \wedge F_4 \wedge F_4 \,, \qquad F_4=dC_3 \,,
\ee
with the ansatz \eqref{eq:metricansatz}, we obtain the Einstein-moduli sector in $d=8$ dimensions,
\begin{align}
S_d &= \frac{1}{2 \kappa_d^2} \int d^d x \sqrt{- g} \bigg(\mathcal R - \frac{1}{4} \left( g^{m m'} g^{n n'} + \frac{1}{d - 2} g^{m n} g^{m' n'} \right) \partial g_{m n} \cdot \partial g_{m' n'} \nonumber\\
&\noeq - \frac{1}{2 \times 3!} g^{m m'} g^{n n'} g^{p p'} (\partial C_{m n p} \cdot \partial C_{m' n' p'}) \bigg)\,, \label{eqn:EinsteinModuliD}
\end{align}
where $g_{m n} = e^U \hat{g}_{m n}$, $g^{m n}$ is its matrix inverse, and $C_{m n p}$ are the components of $C_3$ along the compact directions.

The BPS states are characterized by the integral Kaluza-Klein momenta $N_m$ as well as the integral M2 brane wrapping numbers $W^{mn}=-W^{nm}$ around the various cycles of the three-torus, and 1/2 BPS states exist when $N_m W^{m n} = 0$. In terms of the rescaled quantities
\be
n_m = \frac{N_m}{R} \,, \qquad w^{m n} = R^{\frac{d-5}{3}} W^{m n} \,,
\ee
the 1/2 BPS mass formula is \cite{Obers:1998fb}
\be
m^2 = \Vert g\Vert^{- \frac{1}{d - 2}} \left(Z_m Z^m + \frac{1}{2} {w}_{mn} {w}^{mn} \right),  \quad
Z_m = {n}_m + \frac{1}{2} C_{m n p} {w}^{n p} , \quad
Z_n {w}^{m n} = 0, \label{eq:8dmass}
\ee
again in modified 8d Planck units $2 \kappa_d^2 = (2\pi)^{d-3}$,
where indices are raised and lowered using $g_{m n}$ and $\Vert g \Vert = \det g_{m n}$.

At this point, it is expedient to specialize more particularly to the case $d=8$. We define $w_m = \frac{1}{2} \varepsilon_{m n p} w^{n p}$ where $\varepsilon_{m n p} = \pm 1$ is the Levi-Civita symbol on the $k=3$ compact directions. Then in terms of $C$ such that
$C_{m n p} =  \varepsilon_{m n p} C$,
 the 1/2 BPS mass formula becomes
\be
m^2 = \Vert g\Vert^{- \frac{1}{6}} ((\vec n  + C \vec w )^2 + \Vert g\Vert \vec w ^2)\,, \qquad
\varepsilon^{m n p} n_n w_p = 0,
\ee
using the notation $\vec v^2=v_pv^p$. Furthermore, the shortening condition $\varepsilon^{m n p} n_n w_p=0$ implies that $\vec n$ and $\vec w$ are proportional, so $\vec w  = \frac{1}{r}\vec n$ for some $r \in \mathbb{R}$. Thus,
\be
m^2 = \Vert g\Vert^{- \frac{1}{6}} \bigl[(r+C)^2 + \Vert g\Vert\bigr] \frac{\vec{n}^2}{r^2} \,.
\ee
In terms of the moduli
\be
\phi^a = \frac{1}{2\kappa_8} \begin{pmatrix} g_{m n}\\ C \end{pmatrix},
\ee
we obtain
\begin{subequations}
\begin{align}
\vec\zeta (\theta, \hat n) &= \begin{pmatrix} \zeta^{m n}\\ \zeta \end{pmatrix} = \begin{pmatrix} \frac{1}{2} \cos (\theta) g^{m n} + \frac{1}{3} g^{m n} - \hat{n}^m \hat{n}^n\\ \sin (\theta)\end{pmatrix} ,\\
\vec \zeta \cdot \vec{\tilde{\zeta} } &= G^{a b} \zeta_a \tilde{\zeta}_b = \zeta^{m n} \tilde{\zeta}_{m n} - \frac{1}{9} \zeta^m_m \tilde{\zeta}^n_n + \frac{1}{2} \zeta \tilde{\zeta} ,
\end{align}
\end{subequations}
where $\theta = 2 \arctan(r+C)$, $\hat n=\vec n/\sqrt{\vec n^2}$, $G^{a b}$ is the inverse metric on moduli space read off from the effective action \eqref{eqn:EinsteinModuliD}, and we set $\langle \Vert g \Vert \rangle = 1$ (i.e., $\langle U \rangle = 0$) after taking the moduli derivatives.

Notice that $\vec \zeta(\theta,\hat n) = \vec N (\hat n) + \vec T (\theta)$ decomposes into pieces $\vec N(\hat n)$ and $\vec T(\theta)$ depending only on $\hat{n}$ and $\theta$, respectively, where
\begin{align}
\vec N (\hat n) &\df \begin{pmatrix} \frac{1}{3} g^{m n} - \hat{n}^m \hat{n}^n\\ 0 \end{pmatrix}, & \vec T (\theta) &\df \begin{pmatrix} \frac{1}{2} \cos (\theta) g^{m n}\\ \sin (\theta) \end{pmatrix}, \\
\intertext{and}
| \vec N (\hat{n}) |^2 &= \frac{2}{3}\,, \qquad
\vec N (\hat{n}) \cdot \vec T (\theta) = 0\,, &
| \vec T (\theta) |^2 &= \frac{1}{2}\,, \qquad \vec \zeta^2 (\theta, \hat{n}) = \frac{7}{6}\,.
\end{align}
Thus, the convex hull of 1/2 BPS $\vec \zeta$ vectors is the product of the radius $1/\sqrt{2}$ circle around the origin traced out by $\vec{T}(\theta)$ with the convex hull traced out by $\vec N (\hat{n})$ in the remaining five directions orthogonal to this circle. In particular, $\lambda_\text{min}$ for the overall convex hull is the smaller of $1/\sqrt{2}$ ($\lambda_\text{min}$ for the circle $\vec{T}(\theta)$) and $\lambda_\text{min}$ for the convex hull of the $\vec N (\hat{n})$.

To determine the latter, we first outline a general strategy for obtaining $\lambda_\text{min}$ from a set of $\vec \zeta$-vectors that we will use repeatedly. 
Suppose that $\Delta$ is the convex hull generated by a set of points $\mathcal B=\{\vec e_i\}$. To find $\lambda_\text{min}$ for $\Delta$, it is sufficient to find
\be
\lambda(\hat n)=\max_{\vec e\in \mathcal B}(\vec e\cdot \hat n),
\ee
and then minimize over different choices of $\hat n$. Alternatively, for each direction $\hat n$, we can choose $\vec \pi\propto \hat n$ such that
\be
\vec e_i\cdot \vec \pi\leq 1
\ee
for all $\vec e_i\in\mathcal B$, where \emph{at least one} $\vec e_i$ saturates the bound. Then $|\vec \pi|=\frac 1{\lambda(\hat n)}$ and we find $\lambda_\text{min}$ by \emph{maximizing} $|\vec \pi|$ as we vary $\hat n$.

\begin{figure}
\begin{center}
\center
\includegraphics[width=2.5in]{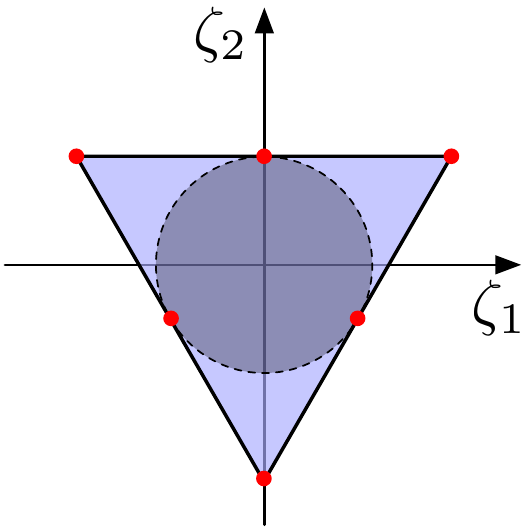}
\caption{A schematic representation of the convex hull of the $\vec{N}(\hat{n})$ vectors. Only the moduli arising from the diagonal components of $\hat{g}_{m n}$ are depicted and likewise only the points corresponding to $\hat{n} = (1,0,0)$, $\hat{n} = (0,1,0)$ and $\hat{n} = (0,0,1)$ are shown, 
\label{fig:zeta8d}}
\end{center}
\end{figure}

Applying the latter method to the case at hand, consider the vector space of traceless, symmetric $k\times k$ matrices with the inner product $\vec M\cdot \vec N=\text{Tr}(MN)$, with $\mathcal B$ consisting of those of the form
\be
E^{mn}=\hat e^m\hat e^n-\frac 1k\delta^{m n}
\ee
for any unit vector $\hat{e}^m$.
Choosing any direction $\hat \Pi^{mn}$, we use an $O(k)$ transformation to diagonalize $\hat \Pi^{mn}$, so that
\be
\Pi^{mn}=\diag(\lambda_1,\dots,\lambda_k),\qquad \sum_{i=1}^k \lambda_i=0.
\ee
We impose $\text{Tr}(\Pi^{mn}E_{mn})=\hat e_m \Pi^{mn}\hat e_n\leq 1$ with at least one unit vector giving equality. This is nothing but the condition $\lambda_i\leq 1$ with at least one equality, i.e.,
\be
\Pi^{mn}=\diag(1,\lambda_2,\dots,\lambda_k),\qquad \lambda_2,\dots,\lambda_k\leq 1,\qquad \sum_{i=2}^k \lambda_i=1.
\ee
Now we want to scan over directions to maximize $\Pi^2$, per the strategy explained above. Note that for two variables $x$ and $y$ with a fixed total $t$, $x^2+y^2=x^2+(t-x)^2$ is \emph{minimized} when $x=y=t/2$, and \emph{increases} as the difference between $x$ and $y$ increases (in either direction). Thus, for any pair of $\lambda_i$'s, we can increase $\Pi^2$ by increasing their difference while maintaining $\sum_{i=2}^k\lambda_i=-1$, until we saturate one of the inequalities $\lambda_2,\dots,\lambda_k\leq 1$. Thus, there is always a way to increase $\Pi^2$ unless \emph{all but one} of the $\lambda_i$'s is equal to 1, i.e., the longest $\Pi^{mn}$ will be of the form
\be
\Pi^{mn}_\text{max}=\diag(1,\dots,1,-(k-1)) \quad\Rightarrow\quad
\Pi^2_\text{max}=(k-1)+(k-1)^2=k(k-1),
\ee
and so
\be
\lambda_\text{min}=\frac 1{\sqrt{k(k-1)}} = \frac{1}{\sqrt{6}} ,
\ee
since $k=3$ in the case of interest. The convex hull of the $\vec{N}(\hat{n})$ vectors is illustrated schematically in Figure~\ref{fig:zeta8d}.

Since $1/\sqrt{6} < 1/\sqrt{2}$, we conclude that $\lambda_{\text{min}} = 1/\sqrt{6} = 1/\sqrt{d-2}$ for 8d maximal SUGRA, again saturating our bound. As before, there are fundamental strings arising from wrapping M2 branes on the various one-cycles of $T^3$, with tension
\be
T_{\text{string}} = \frac{\Vert g \Vert^{-1/6}}{2\pi} \sqrt{g_{m n} W^m W^n}\,,
\ee
where the integers $W^m$ describe the cycle in question. The corresponding string oscillator modes $m \propto \sqrt{2 \pi T_{\textrm{string}}}$ have $\vec{\zeta}$ vectors
\be
\vec\zeta (\hat W) = \begin{pmatrix} -\frac{1}{6} g^{m n} + \frac{1}{2} \hat{W}^m \hat{W}^n\\ 0\end{pmatrix} ,\\
\ee
and indeed these are precisely the directions in which our bound was saturated. Thus, there is a string scale associated to each such direction, as before.

\subsection{Seven Dimensions \label{sec:topdown7d}}

The 7d Einstein-moduli action is still given by~\eqref{eqn:EinsteinModuliD}, now with $d=7$. Likewise, the $1/2$ BPS mass formula~\eqref{eq:8dmass} is still valid, except that an additional shortening condition $w^{[m n} w^{p q]} = 0$ (trivial for $k<4$) now comes into play. Defining $v_{mn} = \frac{1}{2} \varepsilon_{mnpq}w^{pq}$ and $C^q$ such that $C_{m n p} = \epsilon_{mnpq} C^q$, we obtain
\be
m^2 = \Vert g\Vert^{- \frac{1}{5}} \left(Z_m Z^m + \frac{\Vert g \Vert}{2} {v}_{mn} {v}^{mn} \right),  \quad
Z_m = {n}_m + v_{mq} C^q ,
\ee
with the $1/2$ BPS shortening conditions
\be
\varepsilon^{m n p q} Z_n v_{p q} = 0, \qquad \varepsilon^{m n p q} v_{m n} v_{p q} = 0.
\ee
Thus, in terms of the moduli
\be
\phi^a = \frac{1}{2\kappa_7} \begin{pmatrix} g_{m n}\\ C^m \end{pmatrix},
\ee
we obtain
\begin{subequations}
\begin{align}
\vec\zeta &= \begin{pmatrix} \zeta^{m n}\\ \zeta_m \end{pmatrix} = \begin{pmatrix} - \frac{1}{5} \delta^{m n} + \frac{\delta^{m n} v^2 - Z^m Z^n - v^{m p} v^n\,_p}{Z^2 + v^2}\\ 2 \frac{Z^n v_{n m}}{Z^2 + v^2}
\end{pmatrix},\\
\vec \zeta \cdot \vec{\tilde{\zeta} } &= \zeta_{m n}  \tilde{\zeta}^{m n} - \frac{1}{9} \zeta^m_m  \tilde{\zeta}^n_n + \frac{1}{2} \zeta_m \tilde{\zeta}^m.
\end{align}
\end{subequations}
where $Z^2 \df Z_m Z^m$, $v^2 \df \frac{1}{2} v_{m n} v^{m n}$, and we choose a basis where $g_{m n} = \delta_{m n}$ for simplicity.
Note that for $Z_m \ne 0$ the $1/2$ BPS shortening conditions can be solved by $v_{m n} = Z_m u_n - Z_n u_m$ for some $u_m$ that can be chosen to be orthogonal to $Z_m$ without loss of generality. One can easily check that this leads to $\vec{\zeta}^2 = 6/5$, which also holds when $Z_m = 0$.

To find the convex hull of the $1/2$ BPS states, we simplify the above expressions by encapsulating $\zeta^{m n}$ and $\zeta_m$ into a single $5 \times 5$ symmetric traceless matrix
\be
  \zeta^{a b} = \left(\begin{array}{cc}
    \zeta^{m n} - \frac{1}{3} \zeta^p_p \delta^{m n} & \frac{1}{2} \zeta^n\\
    \frac{1}{2} \zeta^m & \frac{1}{3} \zeta^m_m
  \end{array}\right),
\ee 
so that
\be
  \zeta_{a b}  \tilde{\zeta}^{a b} = \zeta_{m n}  \tilde{\zeta}^{m n} -
  \frac{1}{9} \zeta^m_m \tilde{\zeta}^n_n + \frac{1}{2} \zeta_m \zeta^m = \vec{\zeta} \cdot \vec{\tilde{\zeta}}\,.
\ee 
For the $1/2$ BPS states, we find explicitly:
\be
\zeta^{a b} =\begin{pmatrix}
\frac{2}{5} \delta^{m n} - \frac{Z^m Z^n + v^{m p} v^n\,_p}{Z^2 + v^2} & \frac{Z^n v_{n m}}{Z^2 + v^2}\\
\frac{Z^n v_{n m}}{Z^2 + v^2} & \frac{2}{5} - \frac{Z^2}{Z^2 + v^2}
\end{pmatrix}.
\ee 
Defining the $5 \times 5$ antisymmetric matrix,
\be
  V_{a b} = \begin{pmatrix}
    v_{m n} & - Z_m\\
    Z_n & 0
  \end{pmatrix},
\ee 
we see that the above is the same as
\be
  \zeta^{a b} = \frac{2}{5} \delta^{a b} - \frac{V^{a c} V^b\,_c}{V^2}, \qquad V^2 \equiv \frac{1}{2} V_{a b} V^{a b},
\ee 
where the $1 / 2$ BPS shortening conditions are now
\be
  \varepsilon^{a b c d e} V_{b c} V_{d e} = 0 . \label{eqn:7dshortening}
\ee 
The simplicity of these expressions is a manifestation of U-duality, where the 7d U-duality group is $SL(5,\mathbb{Z})$, which enhances to $SL(5,\mathbb{R})$ in the low-energy effective action (ignoring charge quantization).

The shortening condition~\eqref{eqn:7dshortening} implies that $V_{a b} = X_a Y_b - X_b Y_a$ has rank two. In particular,
choosing $X_a$ and $Y_a$ to be orthogonal without loss of generality,
\be
  \Pi^a_b = \frac{V^{a c} V_{b c}}{V^2} = \hat{X}^a \hat{X}_b + \hat{Y}^a
  \hat{Y}_b,
\ee 
is a rank-two projection matrix, i.e., satisfying $\Pi^2 = \Pi$ and $\Tr
\Pi = 2$. Thus,
\be
  \zeta^{a b} = \frac{2}{5} \delta^{a b} - \Pi^{a b} \qquad \Rightarrow \qquad
  \zeta_{a b} \zeta^{a b} = \text{Tr} \left[ \frac{4}{25} \mathbf{1}+
  \frac{1}{5} \Pi \right] = \frac{4}{5} + \frac{2}{5} = \frac{6}{5} .
\ee 
Now consider an arbitrary symmetric traceless matrix $P^{a b}$, and diagonalize
\be
  P^{a b} = \left(\begin{array}{ccc}
    \lambda_1 &  & \\
    & \ddots & \\
    &  & \lambda_5
  \end{array}\right), \qquad \sum_{i = 1}^5 \lambda_i = 0, \qquad \lambda_1
  \geq \lambda_2 \geq \cdots \geq \lambda_5 .
\ee 
We have
\be
  P \cdot \zeta = - \sum_a \lambda_a \Pi_{a a} .
\ee 
The rank-two projector is constrained by $\Pi_{a a} \leq 1$ (no sum) and
$\sum_a \Pi_{a a} = 2$, so
\be
  P \cdot \zeta \geq - \lambda_1 - \lambda_2,
\ee 
and the minimum value is achieved for $\Pi_{a b} = \text{diag} (1, 1, 0, 0,
0)$. Thus, we set $\lambda_1 + \lambda_2 = 1$ to obtain $P \cdot \zeta
\geq - 1$ with the bound saturated in at least one direction. We then
have:
\begin{align}
\begin{aligned}
P^{a b} &= \text{diag} (\lambda_1, \lambda_2, \lambda_3, \lambda_4, \lambda_5), \qquad\lambda_1 + \lambda_2 = 1,\\
\lambda_3 + \lambda_4 + \lambda_5 &= - 1,\qquad \min (\lambda_{1, 2}) \geq \max (\lambda_{3, 4, 5}) .
\end{aligned}
\end{align}
We want to maximize $P^{a b} P_{a b} = \lambda_1^2 + \lambda_2^2 + \lambda_3^2
+ \lambda_4^2 + \lambda_5^2$ subject to these constraints. Recall that for
fixed $x + y = t$, $x^2 + y^2$ \emph{increases} as the difference between
$x$ and $y$ increases. Thus, for fixed $\lambda_{1, 2}$ it is optimal to
saturate the bound $\min (\lambda_{1, 2}) \geq \max (\lambda_{3, 4, 5})$
with two out of three of $\lambda_3$, $\lambda_4$ and $\lambda_5$, so that
(again taking $\lambda_1 \geq \lambda_2 \geq \cdots \geq
\lambda_5$ without loss of generality)
\be
  P^{a b} = \text{diag} (1 - \lambda_2, \lambda_2, \lambda_2 , \lambda_2,
  - 1 - 2 \lambda_2), \qquad \frac{1}{2} \geq \lambda_2 \geq -
  \frac{1}{3} .
\ee 
Now we just have to maximize
\be
  P_{a b} P^{a b} = (1 - \lambda_2)^2 + 3 \lambda_2^2 + (1 + 2 \lambda_2)^2,
  \qquad \text{for} \qquad \frac{1}{2} \geq \lambda_2 \geq -
  \frac{1}{3} .
\ee 
The maximum value occurs at $\lambda_2 = 1 / 2$, i.e., for
\be
  P^{a b} = \text{diag} \left( \frac{1}{2}, \frac{1}{2}, \frac{1}{2},
  \frac{1}{2}, - 2 \right) .
\ee 
We find $P^2_{\text{max}} = \frac{4}{4} + 4 = 5$, hence $\lambda_{\text{min}} =
\frac{1}{\sqrt{5}} = \frac{1}{\sqrt{d-2}}$, once again saturating our bound.

As before, there are BPS fundamental strings arising from M2 branes wrapping the various one-cycles of $T^4$, but now there are also BPS fundamental strings arising from M5 branes wrapping $T^4$; these are characterized by integer winding numbers $W^m$ and $W$, respectively. The two are intermixed by U-duality, so we define
\be
U^a = \begin{pmatrix} w^m + w C^m \\ w \end{pmatrix} \, \qquad \text{where $w^m = R^{-1/3} W^m$ and $w = R^{4/3} W$.}
\ee
Here the relative powers of $R$ and the term involving $C^m$ are both fixed by the fact that in an alternate basis upon which $SL(5,\mathbb{Z})$ naturally acts, $V_{a b} = \bigl(\begin{smallmatrix}
    \frac{1}{2} \varepsilon_{m n p q} W^{p q} & - N_m\\
    N_n & 0
  \end{smallmatrix}\bigr)$ and $U^a = \bigl(\begin{smallmatrix} W^m \\ W \end{smallmatrix}\bigr)$ both have integral components. Thus, U-duality together with a few easily-analyzed special cases fixes the tension formula for $1/2$ BPS strings
\be
T_{\text{string}} = \frac{1}{2\pi} \sqrt{\delta_{a b} U^a U^b} = \frac{\Vert g\Vert^{- \frac{1}{5}}}{2\pi} \sqrt{(w^m + w C^m)^2+\Vert g \Vert w^2} \,,
\ee
where in the second equality we return to an arbitrary basis where $g_{m n} \ne \delta_{m n}$, restoring the correct factors of $\Vert g \Vert$ by considering the special cases of wrapped M2 branes and of wrapped M5 branes with $C_{m n p} = 0$. A straightforward calculation then gives
\be
\zeta^{a b} = -\frac{1}{10} \delta^{a b} + \frac{1}{2} \frac{U^a U^b}{U^2} \,,
\ee
for the string oscillator modes. These are precisely the directions that saturated the bound $\lambda \ge 1/\sqrt{d-2}$ above, hence there is a string scale associated to each such direction as before.

\subsection{Six, Five, and Four Dimensions}

As seen above, U-duality plays an increasingly important role as we compactify further. In Appendix \ref{sec:dleq6} we use an approach that incorporates U-duality from the start to show that the formula $\lambda_\text{min}=1/\sqrt{d-2}$ persists for $d\in\{4,5,6\}$.

\section{Bottom-Up Evidence in Minimal Supergravity}\label{BOTTOMUP}

In the previous section, we saw that our proposed bound $\lambda \geq 1/\sqrt{d-2}$ is saturated in maximal supergravity in dimensions $d=\text{4 -- 10}$. In this section, we present further evidence that this bound is saturated in minimal supergravity in diverse dimensions. As the title of the section suggests, our analysis here proceeds by a bottom-up approach: with only a couple of exceptions, we will not study UV complete string/M-theory compactifications. Instead, following the approach of \cite{Gendler:2020dfp} in four dimensions, we study the scaling behavior of gauge couplings in infinite-distance limits in moduli space. Invoking the tower Weak Gravity Conjecture or the Weak Gravity Conjecture for strings then implies a tower of light charged particles/string oscillator modes in the limit of vanishing gauge couplings, and by working out the scaling of the gauge couplings with proper field distance, we may in turn determine the scaling of the particle masses in this limit. 

In the remainder of this section, we will find no counterexamples to the bound $\lambda \geq 1/\sqrt{d-2}$, and we will find many examples in which this bound is saturated by oscillator modes of a charged string. We will also find many examples of decompactification limits, in which one tower satisfies the Distance Conjecture with $\lambda = \sqrt{(d-1)/(d-2)}$ (as expected for a tower of Kaluza-Klein modes under dimensional reduction), while another satisfies the bound with $\lambda = 1/\sqrt{(d-1)(d-2)}$ (as expected for the Kaluza-Klein zero modes of a $D$-dimensional tower after dimensional reduction to $d=D-1$ dimensions). This provides support not only for our bound \eqref{prop}, but also for the Emergent String Conjecture, which holds than any infinite-distance limit is either a decompactification limit or an emergent string limit.

This approach relies on several important assumptions. First of all, it relies on the assumption of the tower Weak Gravity Conjecture and the Weak Gravity Conjecture for strings, but given the vast body of evidence in favor of these conjectures (see e.g. \cite{Heidenreich:2015nta, Heidenreich:2016aqi, Montero:2016tif, Andriolo:2018lvp, Lee:2018urn, Lee:2019tst, Klaewer:2020lfg, Alim:2021vhs}), this seems to be a relatively minor assumption.

Secondly, this approach relies on the assumption that the tensionless string which emerges in the weak coupling limit of a 2-form gauge field is a fundamental string, meaning that its core probes the deep ultraviolet. Otherwise, one would not expect an infinite tower of string oscillator modes, but merely a finite tower. This assumption follows from the Emergent String Conjecture \cite{Lee:2019wij} and the Distant Axionic String Conjecture \cite{Lanza:2021qsu}, and it is satisfied in many examples in string theory \cite{Lee:2018urn, Lee:2019xtm, Lee:2019xtm, Lanza:2021qsu, BPSStrings}, so it appears to be a valid assumption.

Finally, and perhaps most significantly, our discussion in this section will ignore questions of charge quantization, focusing instead on the scaling behavior of gauging couplings in the classical action in asymptotic limits of moduli space. Given a gauge kinetic matrix $a_{IJ}$, we may compute the eigenvalues of this matrix as a function of the moduli in the theory, and we may say that a gauge coupling vanishes when some eigenvalue of $a_{IJ}$ diverges. However, the eigenvector associated with the gauge coupling is generically an axion-dependent quantity, so the linear combination of gauge fields whose coupling vanishes in the infinite-distance limit is a function of these axions. The presence of these axions does not necessarily present a problem, since the axions appearing in such a linear combination may be fixed to particular values in an asymptotic limit. However, for certain values of the axions, there may not be any particles charged solely under the weakly coupled gauge field in question, due to charge quantization.

For instance, consider two 1-form gauge fields $A_1^1$, $A_1^2$ with field strengths $F_2^i = \rmd A_1^i$ and electric charges quantized in the $A^1$, $A^2$ basis as 
\be
 \frac{1}{g_1^2} \oint_{S^{d-2}}  \star F_2^1  ,\frac{1}{g_2^2} \oint_{S^{d-2}} \star F_2^2  \in \mathbb{Z}\,.
 \label{chargequant}
 \ee 
 We may then consider the linear combinations
\be
A_1^+ = \cos(\theta) A_1^1 + \sin(\theta) A_1^2\,,~~~A_1^- = - \sin(\theta) A_1^1 + \cos(\theta) A_1^2 \,,
\ee 
with associated gauge couplings 
\be
g_+ = \cos(\theta) g_1 + \sin (\theta) g_2\,,~~~g_- = -\sin(\theta) g_1 + \cos (\theta) g_2\,.
\ee 
Then, the limit $g_+ \rightarrow 0$ is a weak coupling limit for the gauge field $A_1^+$, and if $\tan(\theta)$ is rational, then the tower Weak Gravity Conjecture implies a tower of light particles, charged under $A_1^+$ but not $A_1^-$, whose masses vanish in the limit.

On the other hand, if $\tan(\theta)$ is irrational, then because of our charge quantization condition \eqref{chargequant}, there can be no particles charged under $A_1^+$ that are not also charged under $A_1^-$. As a result, the tower Weak Gravity Conjecture does not imply a tower of particles whose masses vanish in the limit $g_+ \rightarrow 0$, as long as $g_-$ remains finite in this limit.

In \cite{Alim:2021vhs, BPSStrings}, it was argued in the context of M-theory compactifications to $d=5$ dimensions that the $g_+ \rightarrow 0$ limit with $\tan(\theta)$ rational corresponds to an infinite-distance limit of moduli space, where the tower of light particles required by the tower Weak Gravity Conjecture also satisfies the Distance Conjecture. In contrast, the $g_+ \rightarrow 0$ limit with $\tan(\theta)$ irrational corresponds to a ``periodic boundary'': a boundary of moduli space in which a compact scalar field traverses its fundamental domain many times. Such a boundary is not at infinite distance in moduli space, so the Distance Conjecture is satisfied trivially despite the absence of a tower of light particles.

Guided by our understanding of 5d M-theory compactifications, we will assume throughout this section that the presence of a vanishing gauge coupling indicates either (a) an infinite-distance limit in moduli space in which the tower Weak Gravity Conjecture leads to a tower of massless particles charged solely under the weakly coupled gauge field or (b) a finite-distance, periodic boundary of moduli space, in which the Distance Conjecture is satisfied trivially, but there are no particles charged solely under the weakly coupled gauge field. Confirming this assumption would require us to go beyond low-energy supergravity and study these systems from a top-down approach in string/M-theory. However, the absence of any counterexamples to the Distance Conjecture in known string/M-theory compactifications offers solid justification for our assumption.

\subsection{Five Dimensions}

In \cite{Corvilain:2018lgw, Heidenreich:2020ptx}, it was argued that every infinite-distance point in vector multiplet moduli space is a point of vanishing gauge coupling, and vice versa. Thus, to place an upper bound on the Distance Conjecture coefficient $\lambda$ in five dimensions, we assume that the tower weak gravity conjecture holds, and we study the scaling of the gauge couplings at infinite distance in moduli space.

We begin by reviewing relevant aspects of supergravity in five dimensions, following \cite{Alim:2021vhs}. At a generic point in vector multiplet moduli space, the action for the bosonic fields in a gauge theory with $n$ vector multiplets is given by
\begin{align}
  S &= \frac{1}{2 \kappa_5^2}  \int d^5 x \sqrt{- g}  \left( \mathcal{R} -
  \frac{1}{2} \mathfrak{g}_{i j} (\phi) \partial \phi^i \cdot \partial \phi^j
  \right) - \frac{1}{2 g_5^2} \int a_{I J} (\phi) F^I \wedge
  \star F^J \nonumber \\
  &+ \frac{1}{6(2\pi)^2} \int C_{I J K} A^I \wedge F^J \wedge F^K,
  \label{eqn:5dsugra}
\end{align}
where $I = 0, \ldots, n$, $i = 1, \ldots, n$, and $g_5^2 = (2\pi)^{4/3} (2\kappa_5^2)^{1/3}$. The scalar metric
$\mathfrak{g}_{i j} (\phi)$, the gauge kinetic matrix $a_{I J} (\phi)$, and the
Chern-Simons couplings $C_{I J K}$ are all determined by a prepotential
$\mathcal{F} [Y]$, which is a cubic in $Y^I$. 
We define $\mathcal{F}_I \equiv \partial_I \mathcal{F}$, $\mathcal{F}_{I J}
\equiv \partial_I \partial_J \mathcal{F}$ and $\mathcal{F}_{I J K} \equiv
\partial_I \partial_J \partial_K \mathcal{F}$. The Chern-Simons couplings are determined by $C_{I J K} = \mathcal{F}_{I J K}$, and the gauge kinetic matrix is given by
\be
  a_{I J} (\phi) =\mathcal{F}_I \mathcal{F}_J 
  -\mathcal{F}_{I J}  \,.
  \label{aIJ}
\ee 

The vector multiplet moduli space corresponds to the slice $\mathcal{F} =1$. The metric on vector multiplet moduli space is the pullback of $a_{IJ}$ to this slice,
\be
\mathfrak{g}_{i j}  = a_{IJ} \partial_i Y^I \partial_j Y^J\,.
\ee 

It is useful to work in homogenous coordinates, invariant under $Y^I \rightarrow \lambda Y^I$. We may then drop the constraint $\mathcal{F}=1$ and instead set 
\be
  a_{I J} = \frac{\mathcal{F}_I \mathcal{F}_J }{\mathcal{F}^{4/3}} 
  - \frac{\mathcal{F}_{I J}}{ \mathcal{F}^{1/3} }  \,.
  \label{aIJhom}
\ee 
In homogenous coordinates, the metric on scalar field space may be written as \cite{Heidenreich:2020ptx}
\be
\mathfrak{g}_{IJ} = \frac{2}{3} \frac{\cF_I \cF_J}{\cF^2} - \frac{\cF_{IJ}}{\cF}\,.
\label{ghomeq}
\ee 
The distance of a path in moduli space, $\gamma= \gamma(s)$, $s \in [s_i, s_f]$, may then be written in homogenous coordinates as
\be
\ell = \frac{1}{\sqrt{2} \kappa_5} \int_{s_i}^{s_f}  \rmd s \sqrt{\mathfrak{g}_{IJ} \dot Y^I \dot Y^J} \,,
\label{distanceeq}
\ee 
where $\dot Y^I = \partial Y^I / \partial s$, and the factor $1/\sqrt{2} \kappa_5$ comes from the prefactor in the action \eqref{eqn:5dsugra}.

As noted in \cite{Heidenreich:2020ptx}, a path $\gamma(s)$ approaching an infinite-distance point $Y_0^i$ in homogenous coordinates may always be rescaled via $Y^I(s) \rightarrow \lambda Y^I(s)$ to ensure that $Y_0^I$ remains finite for all $I$, and $Y_0^I$ is nonzero for at least one $I$. The condition that $Y_0^i$ lies at infinite distance then requires that $\mathfrak{g}_{IJ}$ must diverge in the $Y^I \rightarrow Y_0^I$ limit. Since $Y^I$ was assumed finite, $\cF_I$ must also be finite, which by \eqref{ghomeq} means that $\mathcal{F}$ must vanish in the limit.

We assume that the path is a straight line in homogeneous coordinates:
\be
Y^I = Y_0^I + s Y_1^I \,, ~~~~s \in [0, 1]\,,
\ee 
such that $Y_0^I$ lies at infinite distance. Not all infinite-distance limits take this straight-line form, but such paths offer a useful starting point for analysis, and we will return to the more general case below.
Note that such straight lines necessarily remain within the moduli space due to the convexity of the vector multiplet moduli space (in homogeneous coordinates) for M-theory compactifications to five dimensions \cite{Alim:2021vhs}. We further assume that the path remains within a single K\"ahler cone, i.e., there are no flop transitions. This assumption can be justified in M-theory compactifications on Calabi-Yau threefolds if the strong birational cone conjecture of \cite{BPSStrings} holds true. (This conjecture is a strengthening of the birational cone conjecture of \cite{Morrison94}.)

By a suitable redefinition of coordinates, we may in fact set 
\be
Y_0^I = \delta_0^I, Y_1^I = \delta_1^I \,.
\ee 
We may then expand the prepotential along the path near $Y_0^I$ in powers of $s$:
\begin{align}
\cF &= \frac{1}{6} C_{IJK} Y^I Y^J Y^K \nonumber \\
&=
\frac{1}{6} C_{IJK} Y_0^I Y_0^J Y_0^K + \frac{1}{2} C_{IJK} Y_0^I Y_0^J Y_1^K  s + \frac{1}{2} C_{IJK} Y_0^I Y_1^J Y_1^K  s^2 +  \frac{1}{6} C_{IJK} Y_1^I Y_1^J Y_1^K  s^3 \label{Fexpansion} \\ 
& = \frac{1}{6} C_{000} + \frac{1}{2} C_{001}  s + \frac{1}{2} C_{011}  s^2 +  \frac{1}{6} C_{111} s^3 \,.\nonumber
\end{align}
By the argument above, $\cF$ must vanish as $s \rightarrow 0$ if we assume that the $Y^I$ have been rescaled homogeneously so that $Y_0^I$ is finite for all $I$. This implies $C_{IJK} Y_0^I Y_0^J Y_0^K  = 0$, which implies $\cF \sim s^m$, for $m=1$, $2$, or $3$. It turns out that $m=3$ is not at infinite distance, so we have only two options to consider: i) $\cF \sim s$ and ii) $\cF \sim s^2$. For reasons that will become clear shortly, we will refer to these as decompactification limits and emergent string limits, respectively.

\vspace{0.2cm}
\noindent
\underline{Decompactification limits: $\cF \sim s$}
\vspace{0.2cm} \\
\noindent
At an asymptotic boundary corresponding to a decompactification limit, we have $C_{000} = 0$ but $C_{001} \neq 0$, so $\cF = \frac{1}{2} C_{001} Y_0^I Y_0^J Y_1^K  s + O(s^2) =  \frac{1}{2} C_{001}  s + O(s^2)$. We then have
\begin{align}
\cF_I \dot Y^I  = \cF_I Y_1^I  = \frac{1}{2} C_{IJK} Y_0^I Y_0^J Y_1^K  = \frac{1}{s} \cF + O(s) \,, \\
\cF_{IJ} \dot Y^I   \dot Y^J = \cF_{IJ} Y_1^I  Y_1^J  = \frac{1}{2} C_{IJK} Y_0^I Y_1^J Y_1^K  =  O(s^0) \,.
\end{align}
Plugging these equations into \eqref{ghomeq} and using \eqref{distanceeq} gives in the limit $\epsilon \rightarrow 0$,
\be
\ell(\epsilon) = \frac{1}{\sqrt{2} \kappa_5} \int_\epsilon^1 \rmd s \sqrt{ \frac{2}{3 s^2} + O(1/s) } =  - \frac{1}{\sqrt{3} \kappa_5} \log(\epsilon) + O(\epsilon^0) \,,
\label{elleq1}
\ee 
so indeed, the point $Y_0^I$ is at infinite distance.

Meanwhile, from \eqref{Fexpansion}, we have
\begin{align}
\cF_0 =  C_{001} s  + O(s^2) \,,~~&~\cF_1 = \frac{1}{2} C_{001} + O(s) \,, \\
\cF_{00} = C_{001} s + O(s^2) \,,~~~ \cF_{01} = C_{001} &+ O(s)\,,  ~~~\cF_{11} = C_{011} + O(s) \,,
\end{align}
which by \eqref{aIJhom} gives
\be
a_{00} = ( \sqrt{2} C_{001})^{2/3} s^{2/3} + O(s^{5/3}) \,,~~~ a_{11} = ( C_{001} /2 )^{2/3} s^{-4/3}+ O(s^{5/3}) \,,~~~a_{01} = O(s^{2/3}) \, .
\label{amatdec}
\ee 
By positive-definiteness of $a_{IJ}$, the scaling of $a_{00}$ and $a_{11}$ with $s$ implies that one eigenvalue of $a_{IJ}$ must scale as $a_{\textrm{min}} \sim s^{2/3}$ while another scales at least as $a_{\textrm{max}} \sim s^{-4/3}$ in the limit $s \rightarrow 0$. The eigenvalues of $a_{IJ}$ are (up to normalization constants) simply the inverse-squares of the gauge couplings.
Thus, employing \eqref{elleq1}, we have in the limit $\epsilon \rightarrow 0$:
\be
g_{\text{min}} \sim a_{\textrm{max}}^{-1/2} \lesssim \epsilon^{2/3} \sim \exp \left( - \frac{2}{\sqrt{3}}  \kappa_5 \ell \right) \,.
\ee 
If the tower weak gravity conjecture is satisfied, we expect a tower of particles charged under this gauge field with mass scale $m \lesssim g_{\text{min}} $. The Distance Conjecture is satisfied in this case with a coefficient
\be
\lambda = \frac{2}{\sqrt{3}} \,.
\ee 
This coefficient matches the value $\lambda = \sqrt{(d-1)/(d-2)}$ expected for Kaluza Klein modes upon dimensional reduction from $D=6$ dimensions, hence justifying our use of the term ``decompactification limit.'' Further justification comes from noting that the vanishing eigenvalue $a_{\textrm{min}} \sim s^{2/3}$ from \eqref{amatdec} implies a diverging gauge coupling $g_{\textrm{max}} \sim s^{-1/3}$ in the limit $s\rightarrow 0$. The magnetic Weak Gravity Conjecture for this gauge field then implies a tensionless string with string oscillator modes beginning at the scale
\be
M_{\textrm{string}} = \sqrt{2 \pi T_{\textrm{string}} } \sim \epsilon^{1/6} \sim  \exp \left( - \frac{1}{2\sqrt{3}}  \kappa_5 \ell \right) \,.
\ee 
This coefficient $1/(2\sqrt{3})$ is precisely the value $1/\sqrt{(d-1)(d-2)}$ observed in \eqref{eq28}: it is the expected scaling for the Kaluza-Klein zero modes of a tower of particles in $D$ dimensions after reduction to $d$ dimensions. Thus, assuming the tower Weak Gravity Conjecture and the magnetic Weak Gravity Conjecture for strings in 5d, we find both of the towers expected in a Kaluza-Klein decompactification to six dimensions.

\vspace{0.2cm}
\noindent
\underline{Emergent string limit: $\cF \sim s^2$}
\vspace{0.2cm} \\
\noindent
Next, we turn our attention to the other type of boundaries, which have 
\be\cF = \frac{1}{2} C_{IJK} Y_0^I Y_1^J Y_1^K  s^2 + O(s^3) =  \frac{1}{2} C_{001}   s^2 + O(s^3) \,.
\ee  
We then have
\begin{align}
&\cF_I \dot Y^I = \cF_I Y_1^I   =  C_{IJK} Y_0^J   Y_1^K Y_1^I s + O(s^2)  =\frac{2}{s} \cF + O(s^2) 
\label{firsteq} \\
&\cF_{IJ} \dot Y^I   \dot Y^J   = \cF_{IJ} Y_1^I Y_1^J =  C_{IJK} Y_0^I Y_1^J Y_1^K  = \frac{2}{s^2} \cF + O(s) \,.
\end{align}
Plugging these equations into \eqref{ghomeq} and using \eqref{distanceeq} gives in the limit $\epsilon \rightarrow 0$,
\be
\ell(\epsilon) = \frac{1}{\sqrt{2} \kappa_5} \int_\epsilon^1 \rmd s \sqrt{ \frac{2}{3 s^2} +O(1/s) } =  - \frac{1}{\sqrt{3} \kappa_5} \log(\epsilon) + O(\epsilon^0) \,,
\label{elleq2}
\ee 
so indeed, the point $Y_0^I$ is at infinite distance along the path.

Meanwhile, we have 
\begin{align}
\cF_0 = \frac{1}{2} C_{011} s^2   \,,~&~~\cF_1 =  C_{011} + O(s^2) \,, \\
\cF_{00} = 0 \,,~~~ \cF_{01} = C_{011}  s &+ O(s^2)\,,  ~~~\cF_{11} = C_{011} + O(s) \,,
\end{align}
so by \eqref{aIJhom},
\be
a_{00} = ( C_{011} /2)^{2/3} s^{4/3} + O(s^{7/3}) \,,~~~ a_{11} = ( \sqrt{2} C_{011}  )^{2/3} s^{-2/3}+ O(s^{5/3}) \,,~~~a_{01} = O(s^{4/3}) \, .
\ee 
By positive-definiteness of $a_{IJ}$, the scaling of $a_{00}$ and $a_{11}$ with $s$ implies that one eigenvalue of $a_{IJ}$ must scale as $a_{\textrm{min}} \sim s^{4/3}$ while another scales at least as $a_{\textrm{max}} \sim s^{-2/3}$ in the limit $s \rightarrow 0$. The eigenvalues of $a_{IJ}$ are (up to normalization constants) simply the inverse-squares of the gauge couplings. Thus, employing \eqref{elleq1}, the smallest gauge coupling scales in the limit $\epsilon \rightarrow 0$ as:
\be
g_{\text{min}} \sim a_{\textrm{max}}^{-1/2} \lesssim \epsilon^{1/3} \sim \exp \left( - \frac{1}{\sqrt{3}}  \kappa_5 \ell \right) \,.
\ee 
If the tower weak gravity conjecture is satisfied, we expect a tower of particles charged under this gauge field with mass scale $m \lesssim g_{\text{min}} $. The Distance Conjecture is satisfied in this case with a coefficient
\be
\lambda = \frac{1}{\sqrt{3}} \,,
\ee 
thereby saturating our proposed bound $\lambda \geq 1 / \sqrt{d-2}$.

\begin{table}
\centering
\renewcommand{\arraystretch}{1.5}
\begin{tabular}{|c|c|c|c|} \hline
Geometry & Prepotential $\cF$ & Asymptotic Boundary & Type  \\ \hline
Symmetric Flop   & $ \frac{1}{3} X^3 + 2 X^2 Y $ & $X \rightarrow 0$  & ES \\ \hline
GMSV  & $ \frac{5}{6} X^3 + 2 X^2 Y$ & $X \rightarrow 0$  & ES\\ \hline
\multirow{ 2}{*}{$h^{1,1} = 3$ KMV  } &  {$ \frac{4}{3} X^3 + \frac{3}{2} X^2 Y + \frac{1}{2} X Y^2 $ }& $X \rightarrow 0$  & Decomp. \\ 
&$+ X^2 Z + XYZ$ & $X, Y \rightarrow 0$, $X/Y$ fixed  & ES \\\hline 
\end{tabular}
\caption{Types of asymptotic boundaries (emergent string or decompactification) for M-theory compactified on three examples of Calabi-Yau geometries (the symmetric flop geometry \cite{Alim:2021vhs}, the Greene-Morrison-Strominger-Vafa geometry \cite{Greene:1995hu,Greene:1996dh}, and the $h^{1,1}=3$ Klemm-Mayr-Vafa geometry \cite{Klemm:1996hh}). Further details of these examples can be found in Section 7 of \cite{Alim:2021vhs}.
}
\label{5dtable}
\end{table}

Evidence that this boundary corresponds to an emergent string limit can be seen by further analyzing the largest gauge coupling of the system, which by scales with the smallest gauge coupling as
\be
g_{\text{max}} \sim a_{\textrm{min}}^{-1/2}  \kappa_5 /g_{\text{min}}^{2}\,.
\ee 
In the $g_{\text{min}} \rightarrow 0$ limit, $g_{\text{max}}$ diverges, but the magnetic Weak Gravity Conjecture suggests that a string charged magnetically under this gauge field should have a tension bounded above as
\be
T_{\textrm{string}} \lesssim 1/ (\kappa_5 g_{\text{max}}) \sim \epsilon^{2/3} \sim \exp \left( - \frac{2}{\sqrt{3}}  \kappa_5 \ell \right)   \,,
\ee 
so we see that indeed, a tensionless string emerges in the limit $g_{\text{min}} \rightarrow 0$. Furthermore, this string will have a tower of string oscillator modes beginning at the mass scale $M_{\text{string}} = \sqrt{2 \pi T} \sim g_{\text{min}}/\kappa_5$, and it is natural to identify this with the tower required by tower Weak Gravity Conjecture for the gauge field with coupling $g_{\text{min}}$.

Indeed, the scaling $\cF \sim s^2$ implies a nonzero Chern-Simons coupling of the form $C_{0ij}$ for some $i, j > 0$, which by anomaly inflow \cite{Callan:1984sa} implies that a string charged magnetically under $A_0$ will carry electric charge under $A_i$ and $A_j$. This string is precisely the emergent string discussed above, which becomes tensionless in the $g_{\text{min}} \rightarrow 0$ limit \cite{BPSStrings}. For more details, see \cite{Heidenreich:2021yda, Kaya:2022edp} for the simple case of a theory with a single vector multiplet or \cite{BPSStrings} for the more general case. Several examples of asymptotic boundaries were discussed in Section 7 of \cite{Alim:2021vhs}. Table \ref{5dtable} classifies each of these boundaries by type.

Finally, we return to important point mentioned above: not every infinite-distance geodesic in moduli space takes the form of a straight line in homogeneous coordinates. However, our analysis of these straight line paths has revealed that they have precisely the towers expected for a decompactification limit and an emergent string limit, as discussed in \S\ref{DIMRED}. Thus, we expect that any system featuring two or more of these straight-line boundaries will essentially prove to be just a special case of the dilaton-radion system studied above. Just as the convex hull condition was satisfied for the system in \S\ref{CHC}, we expect that the convex hull condition will be satisfied here, so that towers satisfying our bound $\lambda \geq 1/\sqrt{d-2}$ will appear in any infinite-distance limit in field space, including the non-straight-line paths which approach an intermediate regime between an emergent string boundary and a decompactification boundary.

\begin{figure}
\centering
\includegraphics[width=70mm]{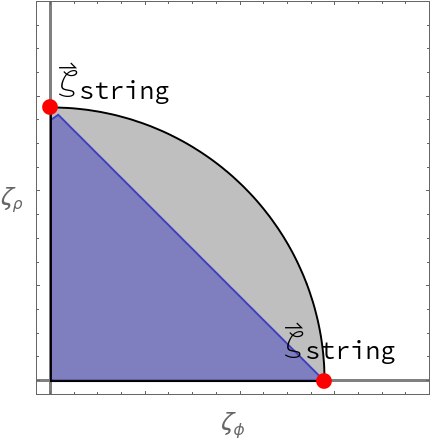}
\caption{A 5d supergravity theory with no decompactification limits. A theory with multiple emergent string boundaries but no decompactification limits will violate the convex hull condition for the Scalar Weak Gravity Conjecture in some directions in scalar field space, and it will violate our Distance Conjecture bound $\lambda \geq 1/\sqrt{d-2}$.}
\label{5dfig}
\end{figure}

One interesting corollary of our analysis from \S\ref{CHC} is that a single 5d supergravity theory cannot have multiple (straight-line) emergent string boundaries unless there is also a decompactification boundary: since the former saturate our bound \eqref{prop}, the convex hull condition will be violated in the intermediate regime between these two directions in field space, as shown in Figure \ref{5dfig}. It would be interesting to prove this statement, or at least to confirm it in examples of 5d supergravity theories arising from M-theory compactifications. More generally, such compactifications may offer a fertile testing ground for our proposed bound \eqref{prop}.

\subsection{Six Dimensions}

To begin, we review relevant aspects of 6d supergravity coupled to abelian gauge fields, following \cite{Riccioni:1999xq, Riccioni:2001bg}.  A generic 6d supergravity features one supergravity multiplet, $n_T$ tensor multiplets and $n_V$ vector multiplets. It may also include hypermultiplets, but we do not consider these in what follows. The supergravity multiplet includes the metric and
an anti-self-dual 2-form gauge field, but no scalar field. A tensor multiplet features
one self-dual 2-form gauge field and a scalar field. A vector multiplet features a 1-form
gauge field, but no scalar field.

As a result, a theory with $n_T$ tensor multiplets will have $n_T + 1$ 2-form gauge fields and $n_T$ scalar fields. These scalar fields parametrize the tensor multiplet moduli space, which is the coset space $SO(1, n_T )/SO(n_T )$. We may describe these in terms of
the $SO(1, n_T )$ matrix:
\be
V =\begin{pmatrix} v_r&\\
x_r^M \end{pmatrix}  
\ee 
where $r = 0, 1, ..., n_T$ and $M = 1, 2, ..., n_T$ . These are subject to the conditions
\be
v^r v_r=1  \,,~~~
v_r x^M_r=0 \,,~~~
v_r v_s - x^M_r x^M_s = \eta_{rs}\,,
\ee 
where here, repeated indices are summed, and $r$ and $s$ indices are raised and lowered
via the metric $\eta_{rs} = \text{diag}( -1, 1, 1, ..., 1)$. 

The gauge kinetic matrix for the tensor fields is then given by
\be
G= v_r v_s + x^M_r x^M_s  \,.
\ee 
and the relevant part of the action is given by
\be
S =  \frac{1}{2 \kappa_6^2}  \int   \biggl( \mathcal{R}  - \rmd v^r \wedge \star \rmd v_r  - \frac{1}{2} G_{rs} H_3^{r } \wedge \star H_{3}^s - \frac{1}{2} v_r c^{rab} F^a \wedge \star F^{b} + \frac{1}{2} c_r^{ab} B_2^r \wedge F_2^a  \wedge F_{2}^b  \biggr).
\label{6daction}
\ee 

To analyze the general case with $n_T$ tensor multiples, we begin by taking $V$ to be the identity matrix and perform boosts and rotations to get any matrix in $SO(1,n_T)$. Thus we have
\be
V= \Lambda \begin{pmatrix} v_{0,r}&\\
x_{0,r}^M \end{pmatrix}  
\ee 
where $
v_0=e_1=\begin{pmatrix}1&0&0&... \end{pmatrix} $, $
x^M_0=e_{M+1}
$,
and $\Lambda$ is a Lorentz transformation. Without loss of generality we can write $\Lambda$ as
  \be
 \Lambda= \biggl( B_{0}(\{\phi_{k}\} )\biggl) \biggl( \prod_{i,j\neq 0 } R_{ij}(\phi_{ij})\biggl)  \,,
 \ee 
 where $R_{ij}$  are rotations on the plane of $i$th and $j$th axis and $B_{0k}(\{\phi_{k}\} )$ are boosts along the $k$th direction.
 
The kinetic matrix for the 2-form gauge fields then takes the form
\begin{align}
G_{rs}=v_r v_s + x^M_r x^M_s &= \Lambda^k_{r} \Lambda^l_s \biggl((v_0)_k (v_0)_l + (x^M_0)_k (x^M_0)_l)\biggl) =  \Lambda^k_{r} \Lambda^l_s 
\delta_{kl}   = (\Lambda^T \Lambda)_{rs} \nonumber \\
 &= \biggl( \prod_{i,j\neq 0 } R_{ij}\biggl) ^T (B_0)^2 \biggl( \prod_{i,j\neq 0 } R_{ij}\biggl) \,.
\end{align}
since $B^T= B$ for any boost matrix. Since rotations do not change the eigenvalues of a matrix, the eigenvalues of $G$ are equal to the eigenvalues of $B_0^2$. 

To find the eigenvalues of the boost matrix $B_0$, we first write it in terms of the boost generators as 
\be
 B_0(\{\phi_k\})=e^{\phi_i K_i } = e^A,
 \ee 
 where $K_i$ is the standard  generator of boosts in the $i$th direction. The eigenvalues of $G (\{\lambda_i \} )$ are then given in terms of the eigenvalues of $A(\{a_i \} )$ as
 \begin{align}
 a_1= \sqrt{\sum_i (\phi_i)^2}= \phi \Rightarrow \lambda_1
  = e^{2 \phi}
\nonumber \\
 a_2= -\sqrt{\sum_i (\phi_i)^2}= -\phi \Rightarrow \lambda_1 
 = e^{- 2\phi}
 \\
 a_i =0   \Rightarrow \lambda_i=1 \text{   for    } i>2  \,, \nonumber
 \end{align}
 where we defined $\phi=\sqrt{\sum_i (\phi_i)^2} $ as the magnitude of the boost vector. In the limit $\phi \rightarrow \infty$, we see that one of the eigenvalues of $G_{rs}$ diverges as $e^{2 \phi}$, indicating a vanishing 2-form gauge coupling in this limit.

 The boosted vectors $v_r, x^M$ are given by

    \be
 v_r=Be_1=\begin{pmatrix}
 \cosh( \phi)   &\\
 \phi_1 \frac{\sinh(2 \phi) }{\phi}&\\
 \phi_2 \frac{\sinh(2 \phi) }{\phi}&\\
 ...&\\
 \phi_{N-1} \frac{\sinh(2 \phi) }{\phi}
 \end{pmatrix} \,,~~~~~
 (x^M)_r=Be_{M+1}=\begin{pmatrix}
 \phi_M \frac{\sinh(2 \phi) }{\phi}  &\\
 \phi_M \phi_1 \frac{\cosh( \phi)-1  }{\phi^2}\\
 ...&\\
 \frac{\sum_{i\neq M} (\phi_i)^2 + \phi_M^2 \cosh(\phi) }{\phi^2}&\\
 ...&\\
 \phi_M \phi_{N-1} \frac{\cosh( \phi)-1  }{\phi^2}
 \end{pmatrix} .
 \ee  
Defining angular variables $r_k= \frac{\phi_k}{\phi}$, we may rewrite $v_r$ as 
 \be
 v_r=
 \begin{pmatrix}
 \cosh(\phi) &\\
 r_1\sinh(\phi)  &\\ 
 r_2\sinh(\phi)  &\\ 
 ...&\\
 r_{N-1}\sinh(\phi)  &\\ 
 \end{pmatrix}  \,.
 \ee 
 Note that $\phi$ is independent of $r_i$, and $\sum_i r_i^2  = 1$. We keep $r_i$ constant and take the limit $\phi \rightarrow \infty$. With this, the scalar kinetic term in the Lagrangian \eqref{6daction} takes the form
  \be
  -\frac{1}{2 \kappa_6^2}\partial_\mu v^r \partial^\mu v_r  
  = -\frac{1}{2 \kappa_6^2}  \frac{\partial v^r}{\partial\phi} \frac{\partial v_r}{\partial\phi} \partial_\mu \phi \partial^\mu \phi +... \,,
  \ee 
 from which we can read off  
 \be
 g_{\phi\phi}=\frac{1}{\kappa_6^2}\frac{\partial v^r}{\partial\phi} \frac{\partial v_r}{\partial\phi} = \frac{1}{\kappa_6^2}\biggl(-\sinh^2(\phi)+ \sum_{i=1}^{N-1} r_i^2 \cosh^2(\phi) \biggl)= \frac{1}{\kappa_6^2}\,.
 \ee 
So, $\phi \rightarrow  \infty$ is indeed at infinite distance in moduli space, and the canonically normalized scalar field is given by $\hat \phi = \phi /\kappa_6$.

Assuming the Weak Gravity Conjecture is satisfied for the weakly coupled 2-form gauge field $B$ in the limit $\phi \rightarrow  \infty$, we expect a string with tension 
 \be
 T_{\textrm{string}} \lesssim   g_B \sim \exp (- \phi) \sim \exp ( - \kappa_6 \hat \phi)\,.
 \ee 
 The oscillator modes of this string will then give rise to a tower of light particles at the string scale
 \be
M_{\textrm{string}} = \sqrt{2\pi T_{\textrm{string}}} \lesssim  \sqrt{g_B} \sim  \exp(\kappa_6 \hat\phi /2 )\,,
 \ee 
 so the Distance Conjecture is satisfied with
  \be
\lambda = \frac{1}{2} = \frac{1}{\sqrt{d-2}}\biggl|_{d=6}\,,
 \ee 
 which saturates our proposed bound \eqref{prop}. An analogous computation applies in the $\phi \rightarrow - \infty$ limit as well. Note that each of these limits involve weakly coupled 2-form gauge fields, which by the Weak Gravity Conjecture imply emergent tensionless strings, so once again the scaling $\lambda = 1/\sqrt{d-2}$ is characteristic of an emergent string boundary, as we found in five dimensions above.

\subsection{Seven Dimensions}

Minimal 7d supergravity \cite{Bergshoeff:1985mr} features one supergravity multiplet and $n$ vector multiplets. The supergravity multiplet has a graviton, one scalar field, three abelian vector fields, and one 2-form gauge field, while each vector multiplet has three scalar fields and one vector field. 
This means that there are a total of $3n+1$ scalar fields: a dilaton $\sigma$, which comes from the gravity multiplet, and $3n$ scalars $\phi^\alpha$, which come from the vector multiplets and parametrize the coset $SO(3, n) / (SO(3) \times SO(n))$.

The $3n$ scalars of the vector multiplets may then be thought of as boosts $B_{ai}(\phi_{ai})$ in an ambient $\mathbb{R}^{3,n}$, with coordinates $\{t^{a}$, $x^i \}$, $a=1,2,3$, $i = 1, ..., n$. An infinite-distance path in the moduli space is a one-parameter family of boosts. As in the case of 6d supergravity above, we suppose that this path takes the simple linear form $\phi_{ai} = r_{ai} \phi$ for some constant $r_{ai}$. By an appropriate choice of coordinate axes, we may in fact further assume $r_{ai} = \delta_{a1} \delta_{i1}$, so the infinite-distance limit is simply the limit of an infinite boost $B_{11}(\phi \rightarrow \infty)$ in the $t^1$, $x^1$ plane of $\mathbb{R}^{3, n}$.

After this convenient choice of coordinates, the relevant part of the 7d supergravity action is given by \cite{Kaya:2022edp}:
\begin{align}
 S&= 
\frac{1}{2 \kappa_7^2}  \int \Big(  \mathcal{R} 
   - \frac{1}{2} e^\sigma  \left( F^1_2  \wedge \star F^1_2  + F^2_2 \wedge \star F^2_2  \right) - \frac{1}{2} e^\sigma  \left(e^{2 \phi} F^+_{2} \wedge \star F^{ +}_2  +e^{-2 \phi} F^-_{2} \wedge \star F^{ _2-}  \right) \nonumber \\
   &-\frac{1}2 e^{2\sigma} H_{3} \wedge \star H_3 
   - \frac{5}{4} \rmd \sigma \wedge \star \rmd \sigma
   -  \rmd \phi \wedge \star \rmd \phi  \Big)  \,,
    \end{align} 
    where
\be
H_{3}  =   \rmd B_{2} + \frac{1}{3 \sqrt{2}}   \left( A^1 \wedge F_{2}^1  +  A^2_{1} \wedge F_{2}^2  + \frac{1}{2} (A^+_{1} \wedge F_{2}^- +  A^-_{1}  \wedge F_{2}^+ )  \right) \,.
\ee 
The limit $\sigma$ is then a weak coupling limit for the 2-form $B_2$, and the Weak Gravity Conjecture implies a tower of string oscillator modes beginning at the string scale 
\be
M_{\textrm{string}} = \sqrt{2 \pi T_{\textrm{string}}} \sim \exp( - \sigma/2 )\,.
\label{stringscale}
\ee 
 After canonically normalizing the dilaton, this tower the Distance Conjecture with a coefficient of
\be
\lambda = \frac{ 1 }{ \sqrt{5 }} =  \frac{ 1 }{ \sqrt{d-2}} \,,
\ee 
which saturates our bound \eqref{prop}, as expected for a tower of string oscillator modes.

Meanwhile, the tower Weak Gravity Conjecture implies a tower of particles in the weak coupling limits $\sigma \pm 2 \phi \rightarrow \infty$ for the gauge fields $A^\pm$. These towers have
\be
\lambda = \sqrt{ \frac{6 }{5 } }\,,
\ee 
which is precisely what we expect for towers of Kaluza-Klein modes under dimensional reduction, from \eqref{KK}. 

Indeed, the parallels between this system and the one studied in \S\ref{DIMRED} become even clearer if we define canonically normalized scalar fields:
\be
\hat \rho \equiv \frac{1}{2  \kappa_7} \sqrt{\frac{5}{6}} \left( \sigma  + 2 \phi \right) \,,~~~\hat \sigma_8 = \frac{1}{2\kappa_7} \frac{1}{\sqrt{6}}  \left(- 5 \sigma + 2 \phi    \right) \,.
\ee 
In terms of these fields, the string scale \eqref{stringscale} is given by
\be
M_{\textrm{string}} \sim \exp\left( -  \frac{1  }{\sqrt{6}  } \kappa_7 \hat \sigma_8 +\frac{1}{\sqrt{30}}  \kappa_7 \hat \rho  \right)\,.
\ee 
This matches precisely with \eqref{eq28} upon setting $\lambda_D = 1 /\sqrt{D-2} $ and taking $\hat \rho$ to be the radion while $\hat \sigma_8$ is the 8d dilaton. Meanwhile, the Weak Gravity Conjecture tower for the gauge field $A^+$ begins at a scale \be
m \sim \exp \left( - \kappa_7 \sqrt{\frac{6}{5}} \hat \rho \right) \,.
\ee 
which matches precisely with \eqref{KK}. In other words, the towers required by the Weak Gravity Conjecture for $B$ and by the tower Weak Gravity Conjecture for $A^+$ have precisely the scaling behavior expected upon dimensional reduction of an 8d theory with a dilaton $\hat \sigma_8$ and a Distance Conjecture tower with $\lambda_8 = 1/ \sqrt{6}$. By a similar analysis for the gauge field $A^-$, taking $\phi \rightarrow - \phi$, the same is true for the limit $\phi \rightarrow - \infty$. As a result, the analysis of \S\ref{CHC} implies that the convex hull condition $\lambda \geq 1/\sqrt{5}$ will be satisfied in the limit $\phi \rightarrow \pm \infty$, $\sigma \rightarrow \infty$ for any value of the ratio $\phi /\sigma$. Minimal supergravity cannot tell us about the strongly coupled $\sigma \rightarrow - \infty$ limit, however, so we must invoke string dualities to cover this case.

Note that our analysis here also provides good evidence for the Emergent String Conjecture, as all of the infinite-distance limits of moduli space in seven dimensions we have introduce a light tower of  fields whose masses scale either as Kaluza-Klein modes or string oscillator modes.

\subsection{Eight Dimensions}

The case of 8d supergravity \cite{Awada:1985ag} is similar to 7d supergravity. It features one supergravity multiplet and $n$ vector multiplets. The supergravity multiplet has a graviton, one scalar field, two abelian vector fields, and one 2-form gauge field, while each vector multiplet has two scalar fields and one vector field. 
This means that there are a total of $2n+1$ scalar fields: a dilaton $\sigma$, which comes from the gravity multiplet, and $2n$ scalars $\phi^\alpha$, which come from the vector multiplets and parametrize the coset $SO(2, n) / (SO(2) \times SO(n))$.

Similar to 7d supergravity, we can think of the $2n$ scalar fields in the vector multiplets as boosts $B_{ai}(\phi_{ai})$ in an ambient $\mathbb{R}^{2,n}$, with coordinates $\{t^{a}$, $x^i \}$, $a=1,2$, $i = 1, ..., n$. An infinite-distance path in the moduli space is a one-parameter family of boosts. If we suppose that this path takes the simple linear form $\phi_{ai} = r_{ai} \phi$ for some constant $r_{ai}$, then by appropriate choice of axes we may set $r_{ai} = \delta_{a1} \delta_{i1}$, so the infinite-distance limit is simply the limit of an infinite boost $B_{11}(\phi \rightarrow \infty)$ in the $t^1$, $x^1$ plane of $\mathbb{R}^{2, n}$.

Under this assumption, we may ignore the angular directions in scalar field space (i.e., axions) and focus our attention on the scaling of gauge couplings with the dilaton $\sigma$ and the radial mode $\phi$. The relevant part of the supergravity action then takes the simple form
\begin{align}
 S&= 
\frac{1}{2 \kappa_8^2}  \int \Big(  \mathcal{R} 
   - \frac{1}{2} e^\sigma  \left( F^1_2  \wedge \star F^1_2   \right) - \frac{1}{2} e^\sigma  \left(e^{2 \phi} F^+_{2} \wedge \star F^{ +}_2  +e^{-2 \phi} F^-_{2} \wedge \star F^{ _2-}  \right) \nonumber \\
   &-\frac{1}{2} e^{2\sigma} H_{3} \wedge \star H_3 
   - \frac{3}{2} \rmd \sigma \wedge \star \rmd \sigma
   -  \rmd \phi \wedge \star \rmd \phi  \Big)  \,.
    \end{align} 
This action may be obtained simply by reducing pure supergravity in 9 dimensions and ignoring the axion $\theta$ associated with the holonomy of the 9d gauge field around the circle, setting the axion vev to zero. Consequently, this system takes precisely the form studied in \S\ref{DIMRED} above: the limit $\sigma \rightarrow \infty$ is an emergent string limit as the 2-form $B_2$ becomes weakly coupled, and the tower of string oscillator modes satisfy the Distance Conjecture with $\lambda = 1/\sqrt{6} = 1/\sqrt{d-2}$, saturating our proposed bound. Meanwhile, the limit $\sigma \pm 2 \phi \rightarrow 0$ is a decompactification limit, associated with a tower of light Kaluza-Klein modes with $\lambda = \sqrt{7/6}$ charged under the gauge fields $A^\pm$. The whole system is consistent with the Emergent String Conjecture, and by our analysis in \S\ref{DIMRED}, it is consistent with our proposed bound \eqref{prop} as well.

\subsection{Nine Dimensions}

9d supergravity \cite{Gates:1984kr} features one supergravity multiplet and $n$ vector multiplets. The supergravity multiplet has a graviton, a 2-form gauge field, a 1-form gauge field, and a scalar, while each vector multiplet has a vector field and a scalar field. 
This means that there are a total of $n+1$ scalar fields: a dilaton $\sigma$, which comes from the gravity multiplet, and $n$ scalars $\phi^\alpha$, which come from the vector multiplets and parametrize the coset $SO(1, n) / SO(n)$.

The relevant part of the supergravity action takes the form
\begin{equation}
S= 
  \int \frac{1}{2 \kappa_9^2} \Big(  \mathcal{R} - \frac{1}{2} e^{-2 \sigma} H_3 \wedge \star  H_3
-\frac{1}{2}  e^{- \sigma} a_{IJ} F_2^I \wedge \star  F^J_2 
  -  g_{\alpha \beta} \rmd \phi^\alpha \wedge \star \rmd \phi^\beta - \frac{7}{4} \rmd \sigma \wedge \star \rmd \sigma
  \Big)  \,.
\end{equation} 
Here, the metric on vector multiplet moduli space is identical to the metric on tensor multiplet moduli space in 6d supergravity considered above, as the moduli spaces in question are identical. Likewise, the gauge kinetic term $a_{IJ}$ is identical to the 2-form gauge kinetic term $G_{rs}$ in 6d supergravity. Thus, borrowing our results from our 6d analysis, we have that in an infinite-distance limit of the vector multiplet moduli space $\phi \rightarrow \infty$, we have eigenvalues of $a_{IJ}$ given by
\begin{equation}
\lambda_1 = e^{2 \phi} \,, ~~~~~  \lambda_2 = e^{-2 \phi}\,,~~~~~ \lambda_i = 1\,,~  i > 2\,,
\end{equation}
and the kinetic term for $g_{\phi\phi}$ is determined to be
\begin{equation}
g_{\phi\phi} = 1 \,.
\end{equation}
As expected, this system takes the form studied in \S\ref{DIMRED} above: the limit $\sigma \rightarrow \infty$ is an emergent string limit as the 2-form $B_2$ becomes weakly coupled, and the tower of string oscillator modes satisfy the Distance Conjecture with $\lambda = 1/\sqrt{7} = 1/\sqrt{d-2}$, saturating our proposed bound. Meanwhile, the limit $\sigma \pm 2 \phi \rightarrow 0$ is a decompactification limit, associated with a tower of light Kaluza-Klein modes with $\lambda = \sqrt{8/7} = \sqrt{(d-1)/(d-2)}$ charged under the gauge fields $A^\pm$ associated to the eigenvalues $\lambda_1$ and $\lambda_2$. The whole system is consistent with the Emergent String Conjecture, and by our analysis in \S\ref{DIMRED}, it is consistent with our proposed bound \eqref{prop} as well.

\section{ Discussion}\label{DISC}

In this paper, we have proposed and provided evidence for a bound on the minimal value of the coefficient $\lambda$ appearing in the Distance Conjecture. Whether or not this bound will stand up to further scrutiny is undoubtedly the most pressing question opened by our work. In the meantime, however, it is worth considering possible applications of our bound. In the remainder of this paper, we contemplate possible applications to various quantum gravity conjectures and cosmology.

\subsection{Scalar Field Potentials}\label{SFP}

If indeed our proposed bound $\lambda \geq 1/\sqrt{d-2}$ is correct, the next item of interest is possible applications of the bound. From a phenomenological perspective, the Distance Conjecture in its original formulation is not very interesting: the real world is not supersymmetric, and there are no massless scalar fields in the Standard Model. In order to connect the Distance Conjecture to observable physics, therefore, one must subscribe to its ``refined'' version \cite{Ooguri:2006in, Klaewer:2016kiy}, which holds that a tower of exponentially light charged particles should exist when \emph{any} scalar field is taken to infinite distance, not just a massless modulus. Before applying our bound to phenomenological context, it is therefore crucial to determine if the bound $\lambda \geq 1/\sqrt{d-2}$ applies to the refined version of the Distance Conjecture as well.

In the most optimistic scenario in which the refined Distance Conjecture is indeed satisfied with $\lambda \geq 1/\sqrt{d-2}$, our conjecture would have important consequences for scalar field potentials in asymptotic limits of scalar field space \cite{Hebecker:2018vxz}.
A scalar field displacement $\Delta \phi$ in $d$ dimensions implies a tower of light states beginning at the mass scale $m \lesssim \exp(- |\Delta \phi| \kappa_d/\sqrt{d-2} ) \Mpld$, which for $|\Delta \phi| \gg 1/\kappa_d$ in turn implies a large number of light species and a ``species bound'' cutoff \cite{Dvali:2007hz} no larger than
\be
\Lambda_{\textrm{UV}} \lesssim  \exp \left( - \frac{  \kappa_d |\Delta \phi|   }{ (d-1) \sqrt{d-2} }  \right) \Mpld \,.
\ee 
Demanding that this UV scale lies above the IR Hubble scale $H \sim \sqrt{V} \kappa_d \lesssim \LUV$ thus leads to a bound on the potential,
\be
V \lesssim \exp  \left( - \frac{ 2 \kappa_d |\Delta \phi|   }{ (d-1) \sqrt{d-2} }  \right) \Mpld^d\,.
\ee 
This suggests a bound 
\be
\frac{|\nabla V|}{V} \geq \frac{2}{(d-1) \sqrt{d-2}} \kappa_d\,
\ee 
in asymptotic regions of scalar field space. In other words, this simple argument suggests that the asymptotic de Sitter conjecture \cite{Obied:2018sgi, Hebecker:2018vxz} should be satisfied with a coefficient of $c = 2/((d-1) \sqrt{d-2})$.

However, our work above, combined with the Emergent String Conjecture, suggests that this result can be strengthened. This conjecture implies that every infinite-distance limit should be either an emergent string limit or a decompactification limit. Combined with our results above, this suggests that the Distance Conjecture should be satisfied in any infinite-distance limit by either a tower string oscillator modes with $\lambda = 1/\sqrt{d-2}$ or a tower of Kaluza-Klein modes with $\lambda \geq 1/\sqrt{d-2}$. Thus we have
 \be
  \textrm{min}(M_{\textrm{string}}, m_{\textrm{KK}}) \lesssim \exp \left(- \frac{ \kappa_d |\Delta \phi| }{ \sqrt{d-2}  } \right)  \Mpld \,.
  \label{cosmo1}
\ee  
At the string scale $M_{\textrm{string}}$, a Hagedorn density of states appears, and effective field theory breaks down. Consistency of the low-energy effective field theory therefore requires a bound on the Hubble scale, $H \lesssim M_{\textrm{string}}$. Similarly, the Hubble scale $H$ must lie below the Kaluza-Klein scale $m_{\textrm{KK}} = L^{-1}$, ensuring that the horizon $H^{-1}$ is larger than the size $L$ of the extra dimensions so that the system can be treated as a $d$-dimensional FRW cosmology. Thus we have a bound
 \be
H \lesssim \Lambda_{\textrm{UV}} \equiv  \textrm{min}(M_{\textrm{string}}, m_{\textrm{KK}})\, .
\label{cosmo2}
\ee  
A scalar field rolling in an exponential potential $V \sim \exp(- \lambda \kappa_d \phi) \Mpld^d$ gives rise at late times to a Hubble scale $H \sim \sqrt{V} \kappa_d$ \cite{Tsujikawa:2013fta}. Together with \eqref{cosmo1} and \eqref{cosmo2}, this implies
 \be
V \lesssim \exp \left( - \frac{ 2 \kappa_d |\Delta \phi|   }{  \sqrt{d-2} }  \right) \Mpld^d\,,
\label{strongpot}
\ee  
which gives
\be
\frac{|\nabla V|}{V} \geq \frac{2}{\sqrt{d-2}} \kappa_d\,.
\label{strongdSC}
\ee 
This bound is precisely the statement of the strong asymptotic de Sitter Conjecture \cite{Rudelius:2021oaz, Rudelius:2021azq} (see also \cite{Obied:2018sgi, Bedroya:2019snp}), which ensures that the strong energy condition is satisfied and forbids accelerated expansion in asymptotic regimes of scalar field space. This bound is saturated in a number of string theory examples in $d=4$ dimensions \cite{Andriot:2020lea}, as well as supercritical string theories in $d \geq 4$ spacetime dimensions \cite{Hellerman:2006nx}. This result also fits nicely with the results of \cite{Hellerman:2001yi, Rudelius:2021azq}, which proved that the bound \eqref{strongdSC} is satisfied for any equidimensional limit at which supersymmetry is restored.

It is important to note that this bound has been derived under the assumption of a $d$-dimensional FRW cosmology and therefore depends on the dynamics of the scalar field. In particular, the bound \eqref{strongdSC} may be violated at the expense of decompactification by relaxing the constraint $H^{-1} > L$. Furthermore, the bound \eqref{strongpot} assumes that the path in question is a gradient flow trajectory along which the scalar field $\phi$ rolls to infinity at late times. This means, as emphasized in \cite{Rudelius:2021oaz, Rudelius:2021azq}, that the strong asymptotic de Sitter Conjecture \eqref{strongdSC} bounds the gradient of the potential $| \nabla V|$ in asymptotic regions of field space rather than the derivative $|V'(\phi)|$ along an arbitrary path in field space. The two are equal for a gradient flow trajectory, but for a more general path the former will be larger than the latter, as it receives contributions from scalar fields orthogonal to the path. Indeed, the bound \eqref{strongdSC} is false in asymptotic regions of scalar field space if one replaces $\nabla V$ with the derivative $V'(\phi)$ along an arbitrary geodesic in field space \cite{Andriot:2020lea}.\footnote{For example, consider a 4d theory with two scalar fields $\phi$, $\rho$, with action $\mathcal{L} = -  \frac{1}{2} (\partial \phi)^2 -  \frac{1}{2} (\partial \rho)^2 -V(\phi)$, where $V(\phi) \sim V_0 \exp(- (  \sqrt{2} \phi+ \sqrt{2/3} \rho )/\Mpl )$ in asymptotic regions of scalar field space. Along the geodesic $\rho \rightarrow \infty$, we have $\Mpl |\partial_\rho V| /V = \sqrt{2/3}$, which violates the bound \eqref{strongpot}. However, this path is not a gradient descent trajectory, so \eqref{strongpot} need not apply to it. And indeed, \eqref{strongpot} is satisfied along a gradient descent trajectory because $\Mpl |\nabla V| / V = 2 \sqrt{2/3} \geq \sqrt{2}$, satisfying \eqref{strongdSC}. In fact, precisely this system appears in the dilaton-radion field space of heterotic string theory compactified to four dimensions \cite{Rudelius:2021oaz}. 
}

It is also important to note that our bound forbids indefinite periods of accelerated expansion in asymptotic regimes of scalar field space, but it does not restrict accelerated expansion of the universe in the interior of field space, nor does it forbid finite periods of accelerated expansion. For instance, certain compactifications of supercritical string theories support periods of accelerated expansion \cite{Dodelson:2013iba}, and even in a theory that satisfies \eqref{strongdSC}, one can attain a short burst of accelerated expansion by giving the scalar field a kick of kinetic energy towards the top of its potential.\footnote{We thank Thomas Van Riet for pointing this out to us.} Instead, our bound simply implies that any period of accelerated expansion in asymptotic regimes of string theory must end after a finite time. Indeed, it is widely believed that de Sitter vacua in string theory are at best metastable \cite{Dyson:2002pf, Goheer:2002vf, Bousso:2011aa}, so it seems quite plausible that any period of accelerated cosmological expansion in quantum gravity can persist for only a finite length of time.

This does not imply that eternal inflation is forbidden, however. In the standard picture of eternal inflation \cite{Starobinsky:1986fx, Vilenkin:1999pi, Linde:2007fr, Guth:2007ng}, a given observer will, with probability 1 \cite{Dyson:2002pf, Bousso:2011aa}, experience a phase transition after a finite period of time from a parent accelerating phase to a daughter decelerating/Minkowski/AdS phase. Inflation is nonetheless ``eternal'' in this scenario because bubbles of the daughter phase continue to inflate away from one another due to the accelerated expansion in the parent phase, and some nonzero fraction of the universe at comoving time $t$ remains in the parent accelerating phase even in the limit $t \rightarrow \infty$ (see e.g. \cite{Rudelius:2019cfh} for further explanation). This means that eternal inflation may occur even if every de Sitter vacuum is metastable and every period of accelerated expansion in quantum gravity can persist for only a finite length of time. Thus, while our results suggest that quintessence, like de Sitter, cannot exist indefinitely in quantum gravity, they do not necessarily point towards a ban on eternal inflation.

It is interesting that the coefficient $c_{\textrm{min}} = 2 /\sqrt{d-2}$ we have derived in \eqref{strongdSC} is twice the Distance Conjecture coefficient $\lambda_{\textrm{min}} = 1/\sqrt{ d-2 }$ we have proposed throughout this paper. Such a relationship between $c_{\textrm{min}}$ and $\lambda_{\textrm{min}}$ has been discussed previously \cite{Hebecker:2018vxz, Andriot:2020lea, Bedroya:2020rmd}, and it is encouraging here that the resulting value of $c_{\textrm{min}}$ matches precisely with a value that is already supported by several independent lines of evidence \cite{Rudelius:2021oaz, Rudelius:2021azq}.

\subsection{Large-Field Inflation}\label{LFI}

Specializing to four dimensions, the bounds developed here may lead to important consequences for large-field inflation. A scalar field displacement $\Delta \phi$ in four dimensions implies a tower of light states beginning at the mass scale $m \lesssim \exp(- |\Delta \phi| /\sqrt{2} )$ in Planck units, which for $\Delta \phi \gg \Mpl$ implies species bound cutoff no larger than
\be
\Lambda_{\textrm{UV}} \lesssim \exp \left( - \frac{|\Delta \phi| }{3 \sqrt{2} \Mpl} \right) \Mpl \,,
\ee 
assuming that the tower begins at or below the Planck scale $\Mpl$ when the field starts to roll.
 For a large-field model of inflation with $|\Delta \phi| = 10 \Mpl$, this leads to a bound $\Lambda_{\textrm{UV}} \lesssim 10^{-1} \Mpl$, which yields a narrow window for a hierarchy of scales between the IR cutoff $H \sim 10^{-4} \Mpl$ and the UV cutoff $\Lambda_{\textrm{UV}}$.

In fact, the window for large-field inflation may be even tighter than this. Invoking the Emergent String Conjecture once again, we assume that every infinite-distance limit in scalar field space is either an emergent string limit or a decompactification limit. This leads to a cutoff
 \be
  \Lambda_{\textrm{UV}} \equiv \min(M_{\textrm{string}}, m_{\textrm{KK}}) \lesssim \exp \left(- \frac{|\Delta \phi| }{ \sqrt{2} \Mpl } \right)  \Mpl \,.
\label{stringboundinf}
\ee  
Here, we have made the seemingly reasonable assumption \cite{Baumann:2014nda} that the string scale and the Kaluza-Klein are bounded above by the Planck scale when the field starts to roll. For $|\Delta \phi| = 10 \Mpl$, this yields a string scale cutoff $\Lambda_{\textrm{UV}} \sim 10^{-3} \Mpl$, and for $|\Delta \phi| = 15 \Mpl$, it yields a string scale cutoff $\Lambda_{\textrm{UV}} \sim 10^{-5} \Mpl$.

Thus, at best, the UV cutoff is only one order of magnitude above the IR cutoff $H \approx 10^{-4} \Mpl$ when $|\Delta \phi| \sim 10 \Mpl$, effectively ruling out any semblance of theoretical control over the effective field theory. Models with $|\Delta \phi| \gtrsim 15 \Mpl$, such as $m^2 \phi^2$ inflation, are excluded entirely for $H \approx 10^{-4} \Mpl$. A detailed investigation into our proposed bound and its applicability beyond the supersymmetric context could therefore have important consequences for the viability of large-field inflation (and relaxion models \cite{Graham:2015cka}) in string theory.

A couple of important caveats are in order here: first, we have assumed that string scale and the Kaluza-Klein scale begin below the Planck scale, so the tower of particles implied by the Distance Conjecture appears almost immediately as the scalar field starts to roll. Although the Distance Conjecture is, strictly speaking, a statement about asymptotic behavior of scalar field space, various works (see e.g. \cite{Klaewer:2016kiy, Erkinger:2019umg}) have found in examples that the exponential towers required by the Distance Conjecture appear rather quickly, when $|\Delta \phi| \lesssim 1 \, \Mpl$, so our assumption is likely a reasonable one.

Perhaps a more serious difficulty is the question of whether or not our argument here may be applied to axion models of inflation, which have historically represented the most popular models of large-field inflation \cite{Freese:1990rb, McAllister:2008hb,Silverstein:2008sg}. We will consider this question next.

\subsection{Axions and a Tower Scalar Weak Gravity Conjecture}\label{AXIONS}

As discussed in the introduction, the Distance Conjecture deals with infinite-distance limits of moduli space and so does not directly constrain axions, which have a compact field space. On the other hand, the Scalar Weak Gravity Conjecture does constrain couplings to axion fields, but it does not imply the existence of an infinite tower of light particles: a finite number of particles may be sufficient to satisfy the convex hull condition.

The evidence we have seen here for both the Distance Conjecture and the Scalar Weak Gravity Conjecture--along with the fact that neither conjecture is quite strong enough to imply the other--suggests a third, stronger conjecture which implies both of them. This led the authors of \cite{Calderon-Infante:2020dhm} to propose the Convex Hull Distance Conjecture, which we call the tower Scalar Weak Gravity Conjecture:
\vspace{.2cm}
           \begin{namedconjecture}[The Tower Scalar Weak Gravity Conjecture]
           Given a massless scalar field modulus $\phi$ in a quantum gravity theory in $d$ spacetime dimensions, there necessarily exists an infinite tower of particles of mass $m_n, n \in \mathbb{Z}$ satisfying
\be
\frac{(\partial_\phi m_n )^2}{g_{\phi\phi} m_n^2} \geq  {\lambda_{\text{min}}^2}{\kappa_d^2}  \equiv \frac{\kappa_d^2}{d-2}\,,~~~ \text{   with }  \partial_\phi m_n < 0 \,,
\label{TsWGC}
\ee 
where $g_{\phi\phi}$ is the $\phi\phi$ component of the metric on scalar moduli space.
            \end{namedconjecture}
    \vspace{.1cm}
\noindent
In other words, the Scalar Weak Gravity Conjecture should be satisfied by a whole tower of particles. When $\phi$ is a non-compact field, this tower is the usual Distance Conjecture tower, but this conjecture goes beyond the Distance Conjecture in that it also requires a tower if $\phi$ is a compact scalar field.

One way to justify the tower Scalar Weak Gravity Conjecture qualitatively (i.e., up to $O(1)$ factors) is by the emergence proposal \cite{Grimm:2018ohb, Heidenreich:2018kpg, Heidenreich:2017sim}. Large field traversals in quantum gravity generally come from weakly coupled scalar fields, which in turn come from integrating out towers of exponentially light particles. This is true even for axion fields \cite{Heidenreich:2018kpg}: a large-field model of natural inflation requires a large axion decay constant, which by the emergence proposal requires a tower of exponentially light particles, which points strongly towards the tower Scalar Weak Gravity Conjecture. This emergence argument only works, however, when the mass scale of the tower of particles in question is far below the quantum gravity scale $\Lambda_{\textrm{QG}}$, where gravity becomes strongly coupled. 

Thus, it seems likely that the arguments of the previous subsection do apply to large-field models of natural inflation as well, which may help explain why these models have so far resisted an embedding in quantum gravity \cite{Banks:2003sx, Rudelius:2015xta, delaFuente:2014aca, Heidenreich:2015wga, Montero:2015ofa, Brown:2015iha, Bachlechner:2015qja, Conlon:2016aea}. It is less clear, however, that the tower Scalar Weak Gravity Conjecture will place important constraints on axion monodromy models of inflation \cite{McAllister:2008hb,Silverstein:2008sg}, since these do not require a large axion decay constant, and towers of particles may remain heavy even as the axion traverses its fundamental domain many times (see however \cite{Baume:2016psm}). A more thorough investigation of axion monodromy in light of our results in this paper could be a worthwhile direction for future study.

\subsection{Black Holes and Repulsive Forces}\label{BHRF}

Consider a Einstein-Maxwell-dilaton action in $d$ dimensions for a 1-form gauge field:
\be
S = \frac{1}{2 \kappa_d^2} \int \rmd^d x \sqrt{-g} \left({\cal R} - \frac{1}{2} (\nabla \phi)^2\right) - \frac{1}{2 e^2} \int d^d x \sqrt{-g} \e^{-\alpha \phi} F_{2}^2 \,.
\ee
The extremality bound for a black hole of quantized charge $Q$ and mass $M$ takes the form
\be
\frac{e^2 Q^2}{M^2} \geq \left[   \frac{\alpha^2}{2} + \frac{d-3}{d-2} \right] \kappa_d^2 \,.
\label{WGCdil}
\ee 
For $\phi \rightarrow - \infty$, the gauge field becomes weakly coupled, and by the tower Weak Gravity Conjecture we expect a tower of particles whose mass scales as 
\be
m \sim \exp( -  \alpha  \phi/2  ) \sim \exp( -  \alpha \kappa_d \hat \phi/\sqrt{2}  ) \,,
\label{mscaling}
\ee 
where $\hat \phi = \phi \sqrt{2} \kappa_d$ is the canonically normalized scalar field. For large charge, we expect that this tower will approach the black hole extremality bound, so that \eqref{mscaling} applies to both quantum mechanical particles at small charge and semiclassical black holes at large charge and smoothly interpolates between them. 

Two copies of a given particle in this theory will repel each other if their gauge repulsion overwhelms the attraction due to the gravitation force and the scalar force--in other words, if
\be
\frac{e^2 Q^2}{M^2} \geq \left[  2 \frac{(\partial_\phi M)^2}{M^2}+ \frac{d-3}{d-2} \right] \kappa_d^2 \,.
\ee 
Plugging in $\partial_\phi M = \alpha M/2$ from \eqref{mscaling}, this is precisely the bound \eqref{WGCdil}: two copies of a particle in the tower will repel each other at long distances precisely if they are superextremal, so the Repulsive Force Conjecture is equivalent to the Weak Gravity Conjecture for this tower of particles.

So far, this discussion is simply a review (see e.g. \cite{Palti:2017elp, Lee:2018spm, Heidenreich:2019zkl, Heidenreich:2020upe}). The novelty comes when we impose our bound $\lambda \geq 1/\sqrt{d-2}$, which by \eqref{mscaling} implies $\alpha \geq \sqrt{2/(d-2)} $. As a result,  both the tower Weak Gravity Conjecture and the tower Repulsive Force Conjecture imply the bound
\be
\frac{e^2 Q^2}{M^2} \geq \kappa_d^2 
\ee 
for charged particles in the tower. Note that this bound only applies for exactly massless scalar fields, since massive scalar fields do not mediate long range forces. The consequences and validity of this bound are problems for future study.

\subsection{The Emergent String Conjecture and an Upper Bound on $\lambda$}\label{ESC}

The Emergent String Conjecture has played an important role throughout this paper. If every infinite-distance limit in scalar moduli space is either an emergent string limit or a decompactification limit, then our analysis in \S\ref{DIMRED} is not merely one illustrative example of what may happen in an asymptotic limit of scalar field space. Rather, it is more or less the \emph{only} example of what may happen. The Emergent String Conjecture therefore provides strong evidence that our proposed lower bound on $\lambda$ is correct. It also offers a compelling explanation for why towers with $\lambda=1/\sqrt{(d-1)(d-2)}$ are so often accompanied by other towers with $\lambda = \sqrt{(d-1)/(d-2)}$, since these values occur readily in dimensional reduction.

In turn, our analysis provides strong evidence for the Emergent String Conjecture. We have found in all of our examples that the limits which saturate our proposed bound $\lambda \geq 1/\sqrt{d-2}$ are emergent string limits, and the states which saturate this bound are oscillator modes of fundamental strings. The other towers we have encountered have the scaling properties expected of Kaluza-Klein towers, with $\lambda_{\textrm{KK}} > 1/\sqrt{d-2}$.

When our results are combined with the Emergent String Conjecture, further interesting results emerge. For one thing, if we assume that the \emph{lightest} tower of particles in a given infinite-distance limit is either a Kaluza-Klein tower or a tower of string oscillator modes, then we conclude that any generator of the convex hull in $\vec{\zeta}$-vector space must have length
\begin{equation}
|\vec{\zeta}^{(d)}| = \frac{1}{\sqrt{d-2}} ~~\text{ or } ~~ |\vec{\zeta}^{(d)}| = \sqrt{ \frac{ n+d-2 }{n (d-2)} }\,,
\end{equation}
for some $n \geq 1$. These values correspond to string oscillator modes and Kaluza-Klein towers, respectively.
As a corollary to this, we also deduce that the parameter $\lambda$ is upper-bounded as 
\begin{equation}
\lambda \leq \lambda_{\textrm{max}} = \sqrt{(d-1)/(d-2)}\,.
\end{equation}
This bound is satisfied in the examples we have encountered in this paper. Further exploration of this bound--as well as a more thorough investigation of the Emergent String Conjecture--is warranted in light of the results of this paper.

\section*{Acknowledgements}

We thank Matthew Reece for collaboration in the early stages of this work. We thank David Andriot, Matthew Reece, and Irene Valenzuela for comments on a previous version of this manscript. We thank David Andriot, Daniel Klaewer, Miguel Montero, Lisa Randall, Matthew Reece, Cumrun Vafa, and Irene Valenzuela for useful discussions. The work of SK and TR was supported in part by the Berkeley Center for Theoretical Physics; by the Department of Energy, Office of Science, Office of High Energy Physics under QuantISED Award DE-SC0019380 and under contract DE-AC02-05CH11231; and by the National Science Foundation under Award Number 2112880. The work of BH and ME was supported by NSF grant PHY-1914934. The work of YQ was supported by NSF grants PHY-1820675 and PHY-2112800.

\appendix
\section{Top Down Evidence in Maximal Supergravity for Six, Five, and Four dimensions \label{sec:dleq6}}
We now investigate the value $\lambda_\text{min}$ for the Scalar Weak Gravity Conjecture for maximal supergravity in six, five, and four dimensions, making extensive use of U-duality to simplify the calculation. In all cases, we obtain $\lambda_\text{min}=1/\sqrt{d-2}$.
\subsection{Symplectic Conventions}
The unitary symplectic group $\text{USp} (2 n)$ is the subgroup of $U (2 n)$
that preserves the symplectic form $\Omega^{a b}$, which is an antisymmetric
tensor with the property that
\be
  \Omega^{a c} \Omega^{\star}_{b c} = \delta^a_b . \label{eqn:symplecticCons}
\ee 
Note that an arbitrary antisymmetric tensor $A^{a b}$ of $U (2 n)$ can be put
in the block-diagonal form
\be
  A^{a b} = \text{diag} \left[ \left(\begin{array}{cc}
    0 & \lambda_1\\
    - \lambda_1 & 0
  \end{array}\right), \ldots, \left(\begin{array}{cc}
    0 & \lambda_n\\
    - \lambda_n & 0
  \end{array}\right) \right], \qquad \text{for $\lambda_i$ real and
  non-negative}
\ee 
by a unitary transformation $A \rightarrow U A U^{\top}$. In the case of the
symplectic form, the constraint \eqref{eqn:symplecticCons}
implies $\lambda_1 = \cdots = \lambda_n = 1$, hence all choices of $\Omega^{a
b}$ are equivalent up to a unitary transformation. Thus,
\be
  \Omega^{a b} = \left(\begin{array}{ccc}
    \mathbf{\varepsilon}_{2 \times 2} & \mathbf{0}_{2 \times 2} &
    \mathbf{0}_{2 \times 2}\\
    \mathbf{0}_{2 \times 2} & \ddots & \mathbf{0}_{2 \times 2}\\
    \mathbf{0}_{2 \times 2} & \mathbf{0}_{2 \times 2} &
    \mathbf{\varepsilon}_{2 \times 2}
  \end{array}\right), \qquad \text{or} \qquad \Omega^{a b} =
  \left(\begin{array}{cc}
    \mathbf{0}_{n \times n} & \mathbf{1}_{n \times n}\\
    -\mathbf{1}_{n \times n} & \mathbf{0}_{n \times n}
  \end{array}\right),
\ee 
in two closely related bases, where $\mathbf{\varepsilon}_{2 \times 2} = i
\sigma_2 = \left(\begin{array}{cc}
  0 & 1\\
  - 1 & 0
\end{array}\right)$. These two bases manifest the important subgroups
$\text{SU} (2)^n \subseteq \text{USp} (2 n)$ (with equality for $n = 1$) and
$U (n) \subset \text{USp} (2 n)$.

It is convenient to define $\Omega_{a b} = (\Omega^{a b})^{- 1} = (\Omega^{b
a})^{\star}$, so that $\Omega_{a b} \Omega^{b c} = \delta_a^c$. We raise and
lower indices with the symplectic form $\Omega_{a b}$ acting on the left, as
follows:
\be
  V^a = \Omega^{a b} V_b \qquad \Leftrightarrow \qquad V_a = \Omega_{a b} V^b
  .
\ee 
Note that this implies
\be
  V^a W_a = (\Omega^{a b} V_b) W_a = - V_b  (\Omega^{b a} W_a) = - V_b W^b,
\ee 
so it is important to keep track of index position.

\subsection{U-Duality Results}

Per \cite{Obers:1998fb}, the 1/2 BPS particles have a mass formula that is
quadratic in the central charge
\be
  M^2 = M_{I J} (\phi) Z^I Z^J,
\ee 
where $M_{I J} (\phi)$ is some moduli-dependent matrix. Moreover, the $1 / 2$
BPS condition takes the form
\be
  f^{\alpha} = S_{I J}^{\alpha} Z^I Z^J = 0,
\ee 
where $S_{I J}^{\alpha}$ is some $\phi$-independent, U-duality invariant such
that $\alpha$ sits in the string representation of the U-duality group.

We are interested in
\be
  \zeta_a = \frac{1}{M}  \frac{\partial M}{\partial \phi^a} = \frac{1}{2 M^2} 
  \frac{\partial M_{I J}}{\partial \phi^a} Z^I Z^J,
\ee 
where $\phi^a$ are the independent scalar fluctuations at a given point in the
moduli space, with accompanying metric $G^{a b} (\phi)$. Let us write this as
\be
  \zeta^a = G^{a b} \zeta_b = \zeta^a_{I J}  \hat{Z}^I  \hat{Z}^J, \qquad
  \hat{Z}^I \equiv \mathcal{N} \frac{Z^I}{\sqrt{M_{I J} Z^I Z^J}},
\ee 
where $\mathcal{N}$ is some normalization factor that depends on conventions.

The moduli space has a coset structure $\mathcal{M}= G / H$. Focusing on a
single point, the subgroup $H$ plays the role of a ``symmetry'' group in the
low-energy effective theory, albeit not an exact symmetry for the same reasons
that, e.g., $\text{SO} (2) \subset \text{SL} (2, \mathbb{R})$ is not an exact
symmetry of type IIB string theory: the charge lattice, massive spectrum,
etc., are not invariant, but the two-derivative effective action \emph{is}
invariant at the classical level.

This means that $M_{I J}$, $S_{I J}^{\alpha} $, and $\zeta^a_{I J}$ must be
$H$ invariants. Combined with representation theory, that is enough to fix the
formula for $\zeta^a$ in terms of the central charges (again, ignoring charge
quantization) up to a change of basis. In particular, to do so we need to know
what representation of $H$ the scalars $\phi^a$ transform under (call it
$R_{\text{mod}}$), as well as the representation $R_{\text{part}}$ under which
the central charges $Z^I$ transform and the representation $R_{\text{str}}$
under which the 1/2 BPS conditions $f^{\alpha} = 0$ transform. For $4
\leq d \leq 9$, we get:
\be
  \begin{array}{c|cccc}
    d & H & R_{\text{mod}} & R_{\text{part}} & R_{\text{str}}\\ \hline
    9 & \text{SO} (2) & \ydiagram{2} \oplus \mathbf{1} & \ydiagram1 \oplus
    \mathbf{1} & \ydiagram 1\\
    8 & \text{SO} (3) \times \text{SO} (2) & \left( \ydiagram{2},
    \mathbf{1} \right) \oplus \left( \mathbf{1}, \ydiagram{2} \right) &
    (\ydiagram 1, \ydiagram 1) & (\ydiagram 1, \mathbf{1})\\
    7 & \text{SO} (5) & \ydiagram{2} & \ydiagram{1,1} & \ydiagram 1\\
    6 & \text{Spin} (5) \times \text{Spin} (5) & (\ydiagram 1, \ydiagram 1) & (S, S) &
    (\ydiagram 1, 1) + (1, \ydiagram 1)\\
    5 & \text{USp} (8) & \ydiagram{1,1,1,1} & \ydiagram{1,1} & \ydiagram{1,1}\\\\
    4 & \text{SU} (8) & \ydiagram{1,1,1,1} & \ydiagram{1,1}_{\mathbb{C}} & \text{Adj} \oplus \ydiagram{1,1,1,1}
  \end{array} \label{tab:Ureps}
\ee 
where $S$ is the four-dimensional (pseudoreal) spinor of $\text{Spin} (5)$,
which is the $\ydiagram 1$ of $\text{USp} (4)$. All the reps are real except the
explicitly indicated $\ydiagram{1,1}_{\mathbb{C}}$ for particles in 4d, where the reality condition on
the vectors will relate $F$ and $\star F$, i.e., the complexity of the
representation is related to the presence of magnetic charge. Note that the 6d
line can also be written in terms of $H = \text{USp} (4) \times \text{USp}
(4)$, whence $R_{\text{mod}} = \left( \ydiagram{1,1}, \ydiagram{1,1} \right)$, $R_{\text{part}} = (\ydiagram 1, \ydiagram 1)$, and $R_{\text{str}} =
\left( \ydiagram{1,1}, 1 \right) \oplus \left( 1, \ydiagram{1,1} \right)$. This makes the embedding into $\text{USp} (8)$ in one
lower dimension more obvious.

To see where \eqref{tab:Ureps} comes from, note that $R_{\text{mod}}$ can be
fixed from the coset structure $\mathcal{M}= G / H$, where $G$ and $H$ are
well known (see, e.g., table B.3 of \cite{Polchinski:1998rr}). Moreover, $R_{\text{part}}$ and
$R_{\text{str}}$ descend from the irreps $R_{\text{part}}^{(G)}$ and
$R_{\text{str}}^{(G)}$ of $G$ that are listed in hep-th/9809039, so one only
needs to understand the $G \rightarrow H$ branching rules. In particular, we
have the following:
\be
  \begin{array}{c|cccc}
    d & G & H & R_{\text{part}}^{(G)} & R_{\text{str}}^{(G)}\\ \hline
    9 & \text{SL} (2, \mathbb{R}) \times \text{SO} (1, 1) & \text{SO} (2) &
    \ydiagram 1 \oplus \mathbf{1} & \ydiagram 1\\
    8 & \text{SL} (3, \mathbb{R}) \times \text{SL} (2, \mathbb{R}) & \text{SO}
    (3) \times \text{SO} (2) & (\ydiagram 1, \ydiagram 1) & (\ydiagram 1, \mathbf{1})\\
    7 & \text{SL} (5, \mathbb{R}) & \text{SO} (5) & \ydiagram{1,1} & \ydiagram 1\\
    6 & \text{Spin} (5, 5) & \text{Spin} (5) \times \text{Spin} (5) & S &
    \ydiagram 1\\
    5 & E_{6 (6)} & \text{USp} (8) & \mathbf{2}\mathbf{7} &
    \mathbf{2}\mathbf{7}'\\
    4 & E_{7 (7)} & \text{SU} (8) & \mathbf{5}\mathbf{6} & \text{Adj}
    =\mathbf{1}\mathbf{3} \mathbf{3}
  \end{array}
\ee 
Here $S$ denotes the minimal, 16-dimensional Majorana-Weyl spinor rep of
$\text{Spin} (5, 5)$, $E_{6 (6)}$ and $E_{7 (7)}$ denote the split real forms
of the $E_6$ and $E_7$ algebras (and their corresponding groups) and
$\mathbf{2}\mathbf{7}$ and $\mathbf{2}\mathbf{7}'$ are the two
different fundamental representations of $E_{6 (6)}$ related by the
$\mathbb{Z}_2$ outer automorphism of the Dynkin diagram. The non-trivial point
now is to understand the branching rules in the exceptional cases (which
depend only on the Lie algebras, not on the choice of real forms); these can be computed using SAGE~\cite{sagemath}.
One finds that
\begin{gather}
  E_6 \rightarrow C_4 : \qquad \mathbf{2}\mathbf{7},
  \mathbf{2}\mathbf{7}' \rightarrow \ydiagram{1,1}, \qquad \text{Adj} \rightarrow \text{Adj} \oplus
  \ydiagram{1,1,1,1},\\
  E_7 \rightarrow A_7 : \qquad \mathbf{5}\mathbf{6} \rightarrow
  \ydiagram{1,1} \oplus \overline{\ydiagram{1,1}}, \qquad \text{Adj} \rightarrow \text{Adj} \oplus
  \ydiagram{1,1,1,1} .
\end{gather}
Noting that $\text{Adj}_G \rightarrow \text{Adj}_H \oplus R_{\text{mod}}$,
this is enough to deal with the cases of interest in the table.

\subsection{Six Dimensions}

In 6d, $H = \text{USp} (4) \times \text{USp} (4)$ and the central charge $Z^{a
; b}$ transforms in the bivector representation $(\ydiagram 1, \ydiagram 1)$. While the rep
$\ydiagram 1$ of $\text{USp} (4)$ is pseudoreal, the $(\ydiagram 1, \ydiagram 1)$ rep is real.
Specifically, the reality condition
\be
  (Z^{a ; b})^{\star} = Z_{a ; b} = \Omega_{a c} \Omega_{b d} Z^{c ; d},
\ee 
is consistent, since
\be
  Z^{\star} = \Omega Z \Omega^{\top} \qquad \Rightarrow \qquad
  (Z^{\star})^{\star} = \Omega^{\star} Z^{\star} \Omega^{\dag} =
  \Omega^{\star} \Omega Z (\Omega^{\star} \Omega)^{\top} = Z
  \label{eqn:realCons}
\ee 
using $\Omega^{\star} \Omega = - 1$.

Noting that
\be
  (\ydiagram 1, \ydiagram 1) \otimes_S (\ydiagram 1, \ydiagram 1) = \left( \ydiagram{2}, \ydiagram{2}
  \right) \oplus \left( \ydiagram{1,1} \oplus \mathbf{1}, \ydiagram{1,1} \oplus \mathbf{1} \right),
\ee 
we see that the 1/2 $\text{BPS}$ shortening condition, in the $\left(
\ydiagram{1,1}, \mathbf{1} \right) \oplus \left( \mathbf{1}, \ydiagram{1,1} \right)$ rep, is uniquely fixed by $H$ invariance to be
\be
  Z^{a ; c} Z_{b ; c} = Z^{c ; a} Z_{c ; b} = \frac{1}{4} \delta^a_b Z^{c ; d}
  Z_{c ; d} .
\ee 
Likewise, due to $H$ invariance, the mass formula must take the form:
\be
  M^2 = Z^{a ; b} (Z^{a ; b})^{\star} = Z^{a ; b} Z_{a ; b},
\ee 
up to an overall normalization factor that we absorb into the definition of
$Z^{a ; b}$.

Last but not least, the scalar charge-to-mass ratio vector $\zeta^{a b ; c d}$
sits in the $\left( \ydiagram{1,1}, \ydiagram{1,1} \right)$ representation. Note that the representation
$\ydiagram{1,1}$ of $\text{USp} (4)$ (the $\ydiagram 1$ of $\text{SO} (5)$) is itself
real, i.e., the condition
\be
  (A^{a b})^{\star} = A_{a b} = \Omega_{a c} \Omega_{b d} A^{c d} 
\ee 
is consistent for much the same reason as in \eqref{eqn:realCons}. Likewise,
$\zeta$ satisfies
\be
  (\zeta^{a b ; c d})^{\star} = \zeta_{a b ; c d} .
\ee 
As there is a unique singlet in $\left( \ydiagram{1,1}, \ydiagram{1,1} \right) \otimes_S \left( \ydiagram{1,1}, \ydiagram{1,1} \right)$, i.e., written in terms of $\text{SO} (5) \times
\text{SO} (5)$ reps:
\be
  (\ydiagram 1, \ydiagram 1) \otimes_S (\ydiagram 1, \ydiagram 1) = \left( \ydiagram{2} \oplus
  \mathbf{1}, \ydiagram{2} \oplus \mathbf{1} \right) \oplus \left(
  \ydiagram{1,1}, \ydiagram{1,1} \right) =\mathbf{1} \oplus (\cdots),
\ee 
the metric on $\zeta$-space is also fixed by $H$ invariance to be
\be
  \zeta^2 = \zeta^{a b ; c d} (\zeta^{a b ; c d})^{\star} = \zeta^{a b ; c d}
  \zeta_{a b ; c d},
\ee 
up to an overall factor that we absorb into the definition of $\zeta^{a b ; c
d}$.

In the case of a $1 / 2$ BPS state, we should have an expression schematically
of the form
\be
  \zeta = \frac{1}{M^2} Z Z
\ee 
with some index structure. Because $(\ydiagram 1, \ydiagram 1) \otimes_S (\ydiagram 1, \ydiagram 1)$ has a
unique copy of $\left( \ydiagram{1,1}, \ydiagram{1,1} \right)$ in its decomposition, the index structure is again fixed
by $H$ invariance to be
\be
  \zeta^{a b ; c d} =\mathcal{N} \left( \hat{Z}^{a ; c}  \hat{Z}^{b ; d} -
  \hat{Z}^{a ; d}  \hat{Z}^{b ; c} - \frac{1}{8} \Omega^{a b} \Omega^{c d}
  \right), \qquad \hat{Z}^{a ; b} = \frac{Z^{a ; b}}{\sqrt{Z^{a ; b} Z_{a ; b}
  }},
\ee 
where the last term is required to ensure $\Omega_{a b} \zeta^{a b ; c d} = 0$
and $\zeta^{a b ; c d} \Omega_{c d} = 0$.

The normalization factor $\mathcal{N}$ can no longer be absorbed by
redefinitions. To determine it, first note that the unit-normalized central
charge $\hat{Z}^{a ; b}$ satisfies
\be
\hat{Z}^{a ; b}  \hat{Z}_{a ; c} = \frac{1}{4} \delta^b_c, \qquad \hat{Z}^{a
; b} \hat{Z}_{c ; b} = \frac{1}{4} \delta^a_c,
\ee
\be
\Rightarrow \qquad
\Omega_{a b} \hat{Z}^{a ; c} \hat{Z}^{b ; d} = - \frac{1}{4} \Omega^{c d},
\qquad \hat{Z}^{a ; c} \hat{Z}^{b ; d} \Omega_{c d} = - \frac{1}{4}
\Omega^{a b},
\ee
upon imposing the 1/2 BPS condition. Thus, for a $1 / 2$ BPS state
\be
\zeta^2 = 2\mathcal{N}^2  \left( \hat{Z}^{a ; c}  \hat{Z}^{b ; d} - \hat{Z}^{a ; d}  \hat{Z}^{b ; c} - \frac{1}{8} \Omega^{a b} \Omega^{c d} \right) \hat{Z}_{a ; c}  \hat{Z}_{b ; d} = 2\mathcal{N}^2 \left( 1 - \frac{1}{4} - \frac{1}{8} \right) = \frac{5}{4} \mathcal{N}^2 .
\ee 
However, we know that Kaluza Klein modes are $1 / 2$ BPS states, which moreover satisfy
$\zeta^2 = \frac{d - 1}{d - 2} = \frac{5}{4}$. Thefore, $\mathcal{N}= 1$.

Next, note that reality condition combined with the $1 / 2$ BPS condition
gives $\hat{Z}^{a ; b} (\hat{Z}^{c ; b})^{\star} = \frac{1}{4} \delta^a_c$,
i.e, $\hat{Z}  \hat{Z}^{\dag} = \frac{1}{4}$, so $2 \hat{Z}$ is a unitary
matrix. We can therefore fix $\hat{Z}^{a ;}\,_b = \frac{1}{2}
\delta^a_b$ by a unitary rotation on the right-hand factor of $\text{USp} (4)
\times \text{USp} (4)$. Thus, we obtain
\be
  \Omega^R_{b c}  \hat{Z}^{a ; c} = \frac{1}{2} \delta^a_b \qquad \Rightarrow
  \qquad \hat{Z}^{a ; b} = - \frac{1}{2} \Omega_R^{a b},
\ee 
where we are now careful to distinguish the symplectic form $\Omega_R$ for the
right-hand factor from the one $\Omega_L$ for the left-hand factor. However,
the reality condition on $\hat{Z}$ now gives:
\be
(\hat{Z}^{a ; b})^{\star} =\hat{Z}_{a ; b} = \Omega^L_{a c} \Omega^R_{b d} \hat{Z}^{c ; d}
\quad\Rightarrow\quad- \frac{1}{2} \Omega_{b a}^R =- \frac{1}{2} \Omega^L_{a c} \Omega^R_{b d} \Omega_R^{c d}
\quad\Rightarrow\quad\Omega_{a b}^R = \Omega^L_{a b},
\ee
using $(\Omega^{a b})^{\star} = \Omega_{b a}$, so they are in fact the same.

Thus, we obtain:
\be
  \zeta^{a b ; c d} = \frac{1}{4}  \left[ \Omega^{a c} \Omega^{b d} -
  \Omega^{a d} \Omega^{b c} - \frac{1}{2} \Omega^{a b} \Omega^{c d} \right],
\ee 
in this basis for a $1 / 2$ BPS state. As a check this satisfies $\Omega_{a b}
\zeta^{a b ; c d} = 0$, $\zeta^{a b ; c d} \Omega_{c d} = 0$, and
\be
  \zeta^2 = \frac{1}{8}  \left[ \Omega^{a c} \Omega^{b d} - \Omega^{a d}
  \Omega^{b c} - \frac{1}{2} \Omega^{a b} \Omega^{c d} \right] \Omega_{a c}
  \Omega_{b d} = \frac{1}{8}  [16 - 4 - 2] = \frac{5}{4} .
\ee 
Now we want to write $\zeta$ and $\zeta^{I ; J}$, a bivector of $\text{SO} (5)
\times \text{SO} (5)$ with metric
\be
  \zeta^2 = \zeta^{I ; J} \zeta_{I ; J} = \text{Tr} (\zeta \zeta^{\top}) .
\ee 
By choosing a $\hat{Z}^{a ; b}$ above, we picked a particular identification
between the two factors of the group, and having done so $\zeta$ could be
written purely in terms of invariants of the diagonal $\text{USp} (4)$ that
remained. Likewise, if we consider the diagonal $\text{SO} (5)$ there is a
unique invariant matrix, i.e., the identity matrix. Thus, we conclude that, in
this basis
\be
  \zeta_{I ; J} = \frac{1}{2} \delta_{I J},
\ee 
where the normalizing factor is chosen to ensure that
\be
  \zeta^2 = \frac{1}{4} \text{Tr} (1) = \frac{5}{4},
\ee 
as before.

Thus, in an arbitrary basis we have the $1 / 2$ BPS state $\zeta$ vectors
\be
  \zeta_{I ; J} = \frac{1}{2} O_{I J},
\ee 
where $O_{I J}$ is any element of $\text{SO} (5)$.

We want to understand the convex hull of these $\zeta$ vectors. Note that it
is actually $O (5) \times O (5)$ invariant. Consider some arbitrary bivector
$P_{I ; J}$. Using the singular value decomposition, we can set
\be
  P_{I ; J} = \text{diag} (\lambda_1, \ldots, \lambda_5), \qquad \lambda_I
  \geq 0
\ee 
after an $O (5) \times O (5)$ transformation. Thus, for arbitrary $\zeta_{I ;
J} = \frac{1}{2} O_{I J}$,
\be
  \text{Tr} (\zeta P^{\top}) = \frac{1}{2}  \sum_{I = 1}^5 \lambda_I O_{I I} .
\ee 
We have the constraints
\be
  \sum_I O_{I J} O_{I K} = \delta_{J K} .
\ee 
Clearly this requires $O_{I I} \leq 1$, and the constraint can be
saturated by taking $O_{I J} = \delta_{I J}$. Thus,
\be
  \text{Tr} (\zeta P^{\top}) \leq \frac{1}{2}  \sum_{I = 1}^5 \lambda_I =
  1,
\ee 
and we should take $\sum_{I = 1}^5 \lambda_I = 2$ to enfore $\zeta \cdot P
\leq 1$ with at least one $1 / 2$ BPS state saturating the constraint.

With this constraint, we seek to maximize
\be
  P \cdot P = \text{Tr} (P P^{\top}) = \sum_I \lambda_I^2,
\ee 
subject to the constraints $\sum_{I = 1}^5 \lambda_I = 2$ and $\lambda_I
\geq 0$. As before, for fixed $x + y = t$, $x^2 + y^2$ increases as the
difference between $x$ and $y$ increases. Thus, if more than one $\lambda_I$
is positive we can increase $P^2$ by reducing the value of the smaller one and
increasing the value of the larger one. As a result, we achieve the maximum
value when all but one vanishes, e.g., $\lambda_1 = 2$ and $\lambda_2 =
\lambda_3 = \lambda_4 = \lambda_5 = 0$. Thus,
\be
  P^2_{\text{max}} = \sum_I \lambda_I^2 = 4 .
\ee 
This gives $\lambda_{\min} = \frac{1}{P_{\text{max}}} = \frac{1}{\sqrt{4}} =
\frac{1}{2}$, which indeed verifies (and saturates) the bound
$\lambda_{\text{min}} \geq \frac{1}{\sqrt{d - 2}}$.

\subsection{Five Dimensions}\label{sec:5d}

In 5d, $H = \text{USp} (8)$ and the central charge $Z^{a b}$ transforms in the
traceless antisymmetric tensor representation $\ydiagram{1,1}$. This representation is real, with the reality condition
\be
  (Z^{a b})^{\star} = Z_{a b} \equiv \Omega_{a c} \Omega_{b d} Z^{c d} .
\ee 
Since $\ydiagram{1,1} \otimes_S \ydiagram{1,1} = \ydiagram{1,1,1,1} \oplus \ydiagram{2,2} \oplus \ydiagram{1,1} \oplus 1$, there is (1) a unique mass formula, coming from the $1$
component, (2) a unique $1 / 2$ BPS constraint, coming from the
$\ydiagram{1,1}$ component, and (3) a unique $\zeta \propto \hat{Z}^2$, coming
from the $\ydiagram{1,1,1,1}$ component. We examine these in turn. The mass formula is
\be
  M^2 = \frac{1}{2} Z^{a b}  (Z^{a b})^{\star} = \frac{1}{2} Z^{a b} Z_{a b},
\ee 
up to an overall normalization that we absorb into the definition of $Z^{a
b}$. The $1 / 2$ BPS constraint is
\be
  Z^{a c} Z_{b c} = \frac{1}{8} \delta^a_b Z^{c d} Z_{c d},
\ee 
transforming in the traceless antisymmetric tensor rep as required.

Likewise, the charge-to-mass ratio $\zeta^{a b c d}$ transforms in the
traceless four-form representation $\ydiagram{1,1,1,1}$. This representation is real, with the reality condition
\be
  (\zeta^{a b c d})^{\star} = \zeta_{a b c d} \equiv \Omega_{a e} \Omega_{b f}
  \Omega_{c g} \Omega_{d h} \zeta^{e f g h} . \label{eqn:5dzetareal}
\ee 
The metric is
\be
  \zeta^2 = \frac{1}{4!} \zeta^{a b c d}  (\zeta^{a b c d})^{\star} =
  \frac{1}{4!} \zeta^{a b c d} \zeta_{a b c d},
\ee 
up to an overall normalization absorbed into $\zeta^{a b c d}$, which is
unique because
\be
  \ydiagram{1,1,1,1} \otimes_S \ydiagram{1,1,1,1} = \ydiagram{2,2,2,2} \oplus \ydiagram{2,2} \oplus 1,
\ee 
contains only one singlet.

For $1 / 2$ BPS particles, we must have
\be
  \zeta^{a b c d} = 3\mathcal{N} \left[ \hat{Z}^{[a b  } 
  \hat{Z}^{  c d]} + \frac{1}{12} \Omega^{[a b  }
  \Omega^{  c d]} \right], \qquad \hat{Z}^{a b} = \frac{Z^{a
  b}}{\sqrt{\frac{1}{2} Z^{a b} Z_{a b}}},
\ee 
for a still-to-be determined normalization factor $\mathcal{N}$, where the
second factor is chosen to ensure tracelessness:

\begin{align}
\begin{aligned}
\Omega_{a b} \zeta^{a b c d} = & \mathcal{N} \Omega_{a b}  \left[ \hat{Z}^{a b}  \hat{Z}^{c d} + \hat{Z}^{a c}  \hat{Z}^{d b} + \hat{Z}^{a d}  \hat{Z}^{b c} + \frac{1}{12} \Omega^{a b} \Omega^{c d} + \frac{1}{12} \Omega^{a c} \Omega^{d b} + \frac{1}{12} \Omega^{a d} \Omega^{b c} \right]\\
= & \mathcal{N} \left[ 0 + \frac{1}{4} \Omega^{c d} + \frac{1}{4} \Omega^{c d} - \frac{2}{3} \Omega^{c d} + \frac{1}{12} \Omega^{c d} + \frac{1}{12} \Omega^{c d} \right] = 0 . 
\end{aligned}
\end{align}

We then find

\begin{align}
\begin{aligned}
\zeta^2 = & \frac{\mathcal{N}^2}{8}  \hat{Z}_{a b}  \hat{Z}_{c d}  \left[\hat{Z}^{a b}  \hat{Z}^{c d} + \hat{Z}^{a c}  \hat{Z}^{d b} + \hat{Z}^{a d} \hat{Z}^{b c} + \frac{1}{12} \Omega^{a b} \Omega^{c d} + \frac{1}{12}\Omega^{a c} \Omega^{d b} + \frac{1}{12} \Omega^{a d} \Omega^{b c} \right]\\
= & \frac{\mathcal{N}^2}{8}  \left[ 4 - \frac{1}{2} - \frac{1}{2} + 0 -\frac{1}{6} - \frac{1}{6} \right] = \frac{\mathcal{N}^2}{3} .  
\end{aligned}
\end{align}

Since Kaluza Klein modes are $1 / 2$ BPS and satisfy $\zeta^2 = \frac{d - 1}{d - 2} =
\frac{4}{3}$, we conclude that $\mathcal{N}= 2$, i.e.,
\begin{align}
\begin{aligned}
\zeta^{a b c d} = & 6 \left[ \hat{Z}^{[a b  }  \hat{Z}^{ c d]} + \frac{1}{12} \Omega^{[a b  } \Omega^{  c d]} \right]\\
= & 2 \left[ \hat{Z}^{a b}  \hat{Z}^{c d} + \hat{Z}^{a c}  \hat{Z}^{d b} + \hat{Z}^{a d}  \hat{Z}^{b c} + \frac{1}{12} \Omega^{a b} \Omega^{c d} + \frac{1}{12} \Omega^{a c} \Omega^{d b} + \frac{1}{12} \Omega^{a d} \Omega^{b c} \right] .   \label{eqn:5dzetaZ}
\end{aligned}
\end{align}

Now consider the conditions on $\hat{Z}^{a b}$. Written in terms of
$\hat{Z}^a\,_b \equiv \Omega_{b c}  \hat{Z}^{a c}$, we find
(recalling that $(\Omega_{a b})^{\star} = \Omega^{b a} = - \Omega^{a b}$):
\be
  \hat{Z}^a\,_a = 0, \qquad \left( \hat{Z}^a\,_b
  \right)^{\star} = \hat{Z}^b\,_a, \qquad
  \hat{Z}^a\,_c  \hat{Z}^c\,_b = \frac{1}{4}
  \delta^a_b, \label{eqn:5dZhatconds}
\ee 
i.e., $2 \hat{Z}^a\,_b$ is a traceless Hermitian matrix that
squares to $1$. Thus, after a unitary change of basis:
\be
  \hat{Z}^a\,_b = \frac{1}{2} \text{diag} (1, 1, 1, 1, - 1, -
  1, - 1, - 1) .
\ee 
In this basis, we observe that $\Omega^{a b}$ and $\hat{Z}^{a b} =
\hat{Z}^a\,_c \Omega^{b c}$ are both antisymmetric matrices.
This requires $\Omega^{a b}$ to be block diagonal in the $4 \times 4$ blocks
defined by $\hat{Z}^a\,_b$, i.e., the stabilizer subgroup of
$\hat{Z}$ is $\text{USp} (4) \times \text{USp} (4)$, where
\be
  \hat{Z}^{a b} = \frac{1}{2} (- \Omega_1^{a b} + \Omega_2^{a b}), \qquad
  \Omega^{a b} = \Omega_1^{a b} + \Omega_2^{a b}, \qquad \Longrightarrow
  \qquad \hat{Z}_{a b} = \frac{1}{2}  ([\Omega_1]_{a b} - [\Omega_2]_{a b}),
  \label{eqn:5dZhatCanon}
\ee 
in terms of the symplectic forms $\Omega_1^{a b}$ and $\Omega_2^{a b}$ for the
two $\text{USp} (4)$ blocks. Then:
\be
  \zeta^{a b c d} = 2 (\Omega_1^{[a b  } \Omega_1^{  c d]} +
  \Omega_2^{[a b  } \Omega_2^{  c d]} - \Omega_1^{[a b
   } \Omega_2^{  c d]}) . \label{eqn:5dzetaCanon}
\ee 
Now choose an arbitary traceless four-form $P^{a b c d}$ satisfying the
reality condition ($\ref{eqn:5dzetareal}$). We seek a 1/2 BPS central charge
$\hat{Z}^{a b}_{(0)}$ that globally maximizes $P^{a b c d} \zeta_{a b c
d}^{(0)}$ with $\zeta_{a b c d}$ given by \eqref{eqn:5dzetaZ}. Since all
$\hat{Z}^{a b}$'s are related by $\text{USp} (8)$ transformations, we can
rephrase this problem by going to the basis where $\hat{Z}^{a b}_{(0)}$ takes
the form \eqref{eqn:5dZhatCanon}, then requiring that $P^{a b c d}
\zeta^{(0)}_{a b c d}$ is maximized under $\text{USp} (8)$ rotations of $P^{a
b c d}$ with $\zeta^{(0)}_{a b c d}$ held fixed in its canonical form
\eqref{eqn:5dzetaCanon}. In other words, fixing $\Omega_1 = \Omega_2 =
\text{diag} \left[ \left(\begin{array}{cc}
  0 & 1\\
  - 1 & 0
\end{array}\right), \left(\begin{array}{cc}
  0 & 1\\
  - 1 & 0
\end{array}\right) \right]$ we maximize
\be
  P \cdot \zeta = \frac{1}{3}  (2 P^{1234} + 2 P^{5678} - P^{1256} - P^{1278}
  - P^{3456} - P^{3478}) .
\ee 
The traceless condition on $P^{a b c d}$ implies that

\begin{align}
\begin{aligned}
P^{1234} + P^{1256} + P^{1278} = 0, \qquad P^{1234} + P^{3456} + P^{3478} = 0,\\
P^{1256} + P^{3456} + P^{5678} = 0, \qquad P^{1278} + P^{3478} + P^{5678} = 0 .  
\end{aligned}
\end{align}

These equations can be rewritten as
\be
P^{5678} = P^{1234}, \qquad P^{3478} = P^{1256}, \qquad P^{3456} = P^{1278}, \qquad P^{1234} + P^{1256} + P^{1278} = 0,
\ee 
where the reality condition on $P^{a b c d}$ implies that all of these are
real. Thus,
\be
  P \cdot \zeta = P^{1234} + P^{5678} = 2 P^{1234} .
\ee 
Defining the four-form $\Pi_{a b c d}$ such that $\Pi_{1234} = + 1$ with other
components vanishing, we see $P \cdot \zeta = \frac{2}{4!} P^{a b c d} \Pi_{a
b c d}$ and the condition for a critical point is
\be
  \delta (P \cdot \zeta) \propto i T^a_e P^{e b c d} \Pi_{a b c d} = 0,
  \label{eqn:extCond}
\ee 
where $T^a_e$ is an arbitrary generator of $\text{USp} (8)$, i.e., a Hermitian
matrix satisfying $T^a_c \Omega^{c b} + T^b_c \Omega^{a c} = 0$. In
particular,
\be
  T^a\,_b = \left(\begin{array}{cc}
    T_1 & X\\
    \Omega_1^{- 1} X \Omega_2 & T_2
  \end{array}\right),
\ee 
for any $T_1, T_2$ generators of $\text{USp} (4)_{1, 2}$ and $X$ any $4 \times
4$ matrix. This means that \eqref{eqn:extCond} becomes
\be
  P^{e b c d} \Pi_{a b c d} = 0,
\ee 
i.e., $P^{a b c d}$ has no components with 3 legs on the first $4 \times 4$
block and 1 leg on the second. By an analogous argument $P^{a b c d}$ has no
components with 1 leg on the first block and 3 on the second either. Thus, we
can write
\be
  P^{a b c d} = \frac{3}{2} \lambda (\Omega_1^{[a b  }
  \Omega_1^{  c d]} + \Omega_2^{[a b  } \Omega_2^{ 
  c d]} - \Omega_1^{[a b  } \Omega_2^{  c d]}) + \hat{P}^{a
  b c d}, \label{eqn:Pdecomp}
\ee 
where $\lambda = 2 P^{1234}$ and $\hat{P}^{a b c d}$ only has components with
2 legs on each $4 \times 4$ block. Combined with the traceless condition, this
implies that $\hat{P}^{a b c d} (\Omega_1)_{c d} = \hat{P}^{a b c d}
(\Omega_2)_{c d} = 0$, and so $\hat{P}^{a b c d}$ transforms in the $\left(
\ydiagram{1,1}, \ydiagram{1,1} \right)$ irrep of $\text{USp} (4) \times \text{USp} (4)$, which is
the $(\ydiagram 1, \ydiagram 1)$ irrep of $\text{Spin} (5) \times \text{Spin} (5)$. Notice
that \eqref{eqn:Pdecomp} implies
\begin{align}
\begin{aligned}
\frac{1}{2} P^{a b c d}  \hat{Z}_{c d} = & \frac{3}{8} \lambda (\Omega_1^{[a b  } \Omega_1^{  c d]} + \Omega_2^{[a b  } \Omega_2^{  c d]} - \Omega_1^{[a b  } \Omega_2^{  c d]}) ([\Omega_1]_{c d} - [\Omega_2]_{c d})\\
= & \frac{\lambda}{2}  (- \Omega_1^{a b} + \Omega_2^{a b}) = \lambda \hat{Z}^{a b} .
\end{aligned}
\end{align}
using \eqref{eqn:5dZhatCanon}. We recognize this is an eigenvalue equation for the $28 \times 28$ matrix $P^{a b}_{\hspace{1.2em} c d}$
\be
  \frac{1}{2} P^{a b}_{\hspace{1.2em} c d}  \hat{Z}^{c d} = \lambda \hat{Z}^{a
  b} . \label{eqn:Peigen}
\ee 
We've shown that this is a necessary condition to maximize $P \cdot \zeta$. In
fact, it is also a sufficient condition, since
\be
P \cdot \zeta = \frac{1}{4}  \hat{Z}_{a b}  \hat{Z}_{c d} P^{a b c d},
\ee
\be
\Rightarrow \qquad \delta (P \cdot \zeta) \propto i T^a_e  \hat{Z}_{a b} \hat{Z}_{c d} P^{e b c d} = 2 \lambda i T^a_e  \hat{Z}_{a b}  \hat{Z}^{e b} = \frac{i \lambda}{2} T^a_a = 0,
\ee
since the generators of $\text{USp} (8)$ are traceless. The reformulation
\eqref{eqn:Peigen} of the condition for extremizing $P \cdot \zeta$ will be
useful again later.

So far, we have imposed the condition that $P \cdot \zeta$ is extremized, but
not that it is maximized, much less that it realizes its global maximum value.
To do so, we first use the fact that $\hat{P}^{a b c d}$ reduces to the
$(\ydiagram 1, \ydiagram 1)$ irrep of $\text{Spin} (5) \times \text{Spin} (5)$ to put it
into a canonical form. First, we need to determine the map between
$\ydiagram{1,1}$ of $\text{USp} (4)$ and $\ydiagram 1$ of $\text{Spin} (5)$, which is
related to the $\Gamma$ matrices for $\text{Spin} (5)$. In terms of the $2
\times 2$ Pauli matrices
\be
  \mathbf{\sigma}^1 = \left(\begin{array}{cc}
    0 & 1\\
    1 & 0
  \end{array}\right), \qquad \mathbf{\sigma}^2 = \left(\begin{array}{cc}
    0 & - i\\
    i & 0
  \end{array}\right), \qquad \mathbf{\sigma}^3 = \left(\begin{array}{cc}
    1 & 0\\
    0 & - 1
  \end{array}\right),
\ee 
we choose a basis where
\be
  (\Gamma^i)^a\,_b = \left(\begin{array}{cc}
    \mathbf{0} & - i\mathbf{\sigma}^i\\
    i\mathbf{\sigma}^i & \mathbf{0}
  \end{array}\right), \qquad (\Gamma^4)^a\,_b =
  \left(\begin{array}{cc}
    \mathbf{0} & \mathbf{1}\\
    \mathbf{1} & \mathbf{0}
  \end{array}\right), \qquad (\Gamma^5)^a\,_b =
  \left(\begin{array}{cc}
    \mathbf{1} & \mathbf{0}\\
    \mathbf{0} & -\mathbf{1}
  \end{array}\right) .
\ee 
Choosing the symplectic form
\be
  \Omega^{a b} = \left(\begin{array}{cc}
    \mathbf{\varepsilon} & \mathbf{0}\\
    \mathbf{0} & \mathbf{\varepsilon}
  \end{array}\right), \qquad \mathbf{\varepsilon}= i \sigma^2 =
  \left(\begin{array}{cc}
    0 & 1\\
    - 1 & 0
  \end{array}\right), \qquad \Rightarrow \qquad \Omega_{a b} =
  \left(\begin{array}{cc}
    -\mathbf{\varepsilon} & \mathbf{0}\\
    \mathbf{0} & -\mathbf{\varepsilon}
  \end{array}\right),
\ee 
we find that $\Gamma^m_{a b} \equiv \Omega_{a c}  (\Gamma^m)^c\,_b$ is given by
\be
  \Gamma^i_{a b} = \left(\begin{array}{cc}
    \mathbf{0} & i\mathbf{\varepsilon}\mathbf{\sigma}^i\\
    - i\mathbf{\varepsilon}\mathbf{\sigma}^i & \mathbf{0}
  \end{array}\right), \qquad \Gamma^4_{a b} = \left(\begin{array}{cc}
    \mathbf{0} & -\mathbf{\varepsilon}\\
    -\mathbf{\varepsilon} & \mathbf{0}
  \end{array}\right), \qquad \Gamma^5_{a b} = \left(\begin{array}{cc}
    -\mathbf{\varepsilon} & \mathbf{0}\\
    \mathbf{0} & \mathbf{\varepsilon}
  \end{array}\right) . \label{eqn:loweredGamma}
\ee 
Since $\mathbf{\varepsilon}^{\top} = -\mathbf{\varepsilon}$ and
\be
  i\mathbf{\varepsilon}\mathbf{\sigma}^1 = \left(\begin{array}{cc}
    i & 0\\
    0 & - i
  \end{array}\right), \qquad i\mathbf{\varepsilon}\mathbf{\sigma}^2 =
  \left(\begin{array}{cc}
    - 1 & 0\\
    0 & - 1
  \end{array}\right), \qquad i\mathbf{\varepsilon}\mathbf{\sigma}^3 =
  \left(\begin{array}{cc}
    0 & - i\\
    - i & 0
  \end{array}\right),
\ee 
are all symmetric, the $\Gamma^m_{a b}$ are all antisymmetric, with $\Omega^{a
b} \Gamma^m_{a b} = 0$ because the $(\Gamma^m)^a\,_b$ are
traceless. Thus, $(\Omega^{- 1} \Gamma^m)^{\top} = - \Omega^{- 1} \Gamma^m$,
or $\Gamma^m \Omega = \Omega (\Gamma^m)^{\top}$, from which one can readily
check that $\Omega$ is indeed a $\text{Spin} (5)$ invariant.

The $\Gamma_{a b}^m$ matrices \eqref{eqn:loweredGamma} provide the desired
dictionary for translating between the traceless antisymmetric tensor
representation of $\text{USp} (4)$ and the vector representation of
$\text{Spin} (5)$. Each $\Gamma_{a b}^m$ has non-zero components with indices
in disjoint pairs, i.e., $\Gamma_{13}^1 = - \Gamma_{24}^1 = i$, $\Gamma^2_{13}
= \Gamma_{24}^2 = - 1$, $\Gamma^3_{14} = \Gamma_{23}^3 = - i$, $\Gamma^4_{14}
= - \Gamma^4_{23} = - 1$ and $\Gamma^5_{12} = - \Gamma^5_{34} = - 1$, with all
other components vanishing except those related by antisymmetry. Thus, if we
write $\hat{P}^{a b c d}$ as a $(\ydiagram 1, \ydiagram 1)$ of $\text{Spin} (5) \times
\text{Spin} (5)$ and diagonalize it, then in 4-form langauge its non-vanishing
components are
\begin{align}
\begin{aligned}
\hat{P}^{1357} &= \hat{P}^{2468},\qquad \hat{P}^{1368} = \hat{P}^{2457}, \qquad \hat{P}^{1458} = \hat{P}^{2367}, \hat{P}^{1467} = \hat{P}^{2358},\\
\hat{P}^{1256} &= \hat{P}^{3478} = - \hat{P}^{1278} = - \hat{P}^{3456},
\end{aligned}
\end{align}
with all of them real. Working backwards, $P^{a b c d}$ therefore has real
components
\begin{align}
\begin{aligned}
P^{1357} &= P^{2468},\qquad P^{1368} = P^{2457},\qquad P^{1458} = P^{2367},\qquad P^{1467} =
P^{2358},\\
P^{1256} &= P^{3478},\qquad P^{1278} = P^{3456},\qquad P^{1234} = P^{5678},
\end{aligned}
\end{align}
subject to the tracelessness condition $P^{1234} + P^{1256} + P^{1278} = 0$.
We define
\begin{align}
\begin{aligned}
\lambda_1 &= P^{1234}, \quad \lambda_2 = P^{1256}, \quad \lambda_3 = P^{1278}, \quad \lambda_4 = P^{1357},\\
\lambda_5 &= P^{1368}, \quad \lambda_6 = P^{1458}, \quad \lambda_7 = P^{1467} .
\end{aligned} 
\end{align}
with $\lambda_1 + \lambda_2 + \lambda_3 = 0$. The matrix $P^{a
b}\,_{ c d}$ is now block diagonal. First, there is $4 \times 4$
block consisting of the $12, 34, 56$ and $78$ components:\footnote{Note that
raising the index pairs using the symplectic form takes $12 \rightarrow 12, 34
\rightarrow 34, 56 \rightarrow 56, 78 \rightarrow 78$.}
\be
  12, 34, 56, 78 : \left(\begin{array}{cccc}
    0 & \lambda_1 & \lambda_2 & \lambda_3\\
    \lambda_1 & 0 & \lambda_3 & \lambda_2\\
    \lambda_2 & \lambda_3 & 0 & \lambda_1\\
    \lambda_3 & \lambda_2 & \lambda_1 & 0
  \end{array}\right), \qquad \lambda_1 + \lambda_2 + \lambda_3 = 0.
\ee 
One can easily check that the non-zero eigenvalues $\lambda$ for this block
are $2 \lambda_1, 2 \lambda_2$ and $2 \lambda_3$ (the zero eigenvalue is
associated to the trace component and should be ignored). There are $6$
remaining $4 \times 4$ blocks, e.g.,\footnote{In this case, raising the index
pairs takes $13 \leftrightarrow 24$, $57 \leftrightarrow 68$, $14
\leftrightarrow - 23$, $58 \leftrightarrow - 67$.}
\be
  13, 24, 57, 68 : \left(\begin{array}{cccc}
    - \lambda_1 & 0 & \lambda_5 & \lambda_4\\
    0 & - \lambda_1 & \lambda_4 & \lambda_5\\
    \lambda_5 & \lambda_4 & - \lambda_1 & 0\\
    \lambda_4 & \lambda_5 & 0 & - \lambda_1
  \end{array}\right), \qquad 14, 23, 58, 67 : \left(\begin{array}{cccc}
    - \lambda_1 & 0 & - \lambda_7 & - \lambda_6\\
    0 & - \lambda_1 & - \lambda_6 & - \lambda_7\\
    - \lambda_7 & - \lambda_6 & - \lambda_1 & 0\\
    - \lambda_6 & - \lambda_7 & 0 & - \lambda_1
  \end{array}\right) .
\ee 
The eigenvalues in the first case are $\lambda = - \lambda_1 \pm_1 \lambda_4
\pm_2 \lambda_5$, and likewise $\lambda = - \lambda_1 \pm_1 \lambda_6 \pm_2
\lambda_7$ in the second. Similarly,\footnote{Now raising the index pairs
takes $15 \leftrightarrow 26$, $37 \leftrightarrow 48$, $16 \leftrightarrow -
25$, $38 \leftrightarrow - 47$, $17 \leftrightarrow 28$, $35 \leftrightarrow
46$, $18 \leftrightarrow - 27$, $36 \leftrightarrow - 45$.}
\be
  15, 26, 37, 48 : \left(\begin{array}{cccc}
    - \lambda_2 & 0 & - \lambda_6 & - \lambda_4\\
    0 & - \lambda_2 & - \lambda_4 & - \lambda_6\\
    - \lambda_6 & - \lambda_4 & - \lambda_2 & 0\\
    - \lambda_4 & - \lambda_6 & 0 & - \lambda_2
  \end{array}\right), \qquad 16, 25, 38, 47 : \left(\begin{array}{cccc}
    - \lambda_2 & 0 & \lambda_7 & \lambda_5\\
    0 & - \lambda_2 & \lambda_5 & \lambda_7\\
    \lambda_7 & \lambda_5 & - \lambda_2 & 0\\
    \lambda_5 & \lambda_7 & 0 & - \lambda_2
  \end{array}\right),
\ee 
with eigenvalues $\lambda = - \lambda_2 \pm_1 \lambda_4 \pm_2 \lambda_6$ and
$\lambda = - \lambda_2 \pm_1 \lambda_5 \pm_2 \lambda_7$, and finally
\be
  17, 28, 35, 46 : \left(\begin{array}{cccc}
    - \lambda_3 & 0 & \lambda_7 & \lambda_4\\
    0 & - \lambda_3 & \lambda_4 & \lambda_7\\
    \lambda_7 & \lambda_4 & - \lambda_3 & 0\\
    \lambda_4 & \lambda_7 & 0 & - \lambda_3
  \end{array}\right), \qquad 18, 27, 36, 45 : \left(\begin{array}{cccc}
    - \lambda_3 & 0 & - \lambda_6 & - \lambda_5\\
    0 & - \lambda_3 & - \lambda_5 & - \lambda_6\\
    - \lambda_6 & - \lambda_5 & - \lambda_3 & 0\\
    - \lambda_5 & - \lambda_6 & 0 & - \lambda_3
  \end{array}\right),
\ee 
with eigenvalues $\lambda = - \lambda_3 \pm_1 \lambda_4 \pm_2 \lambda_7$ and
$\lambda = - \lambda_3 \pm_1 \lambda_5 \pm_2 \lambda_6$.

Using \eqref{eqn:Peigen}, we see that the eigenvalues found above are
precisely the critical values of $P \cdot \zeta$ as we vary $\hat{Z}^{a b}$
with $P^{a b c d}$ held fixed.\footnote{We must be careful here, as we
have not yet imposed the conditions \eqref{eqn:5dZhatconds} on the $\hat{Z}^{a
b}$ eigenvectors. However, it is straightforward to check that they are
satisfied for all the eigenvectors in question, up to a (complex) overall
factor that can be freely chosen.} The critical value we found initially is $2
\lambda_1 = 2 P^{1234}$; now we can write down the necessary and sufficient
conditions for this to be a global maximum of $P \cdot \zeta$:

\begin{align}
\begin{aligned}
\lambda_1 \geq&\; \lambda_2, \quad \lambda_1 \geq \lambda_3, \quad 2 \lambda_1 \geq - \lambda_1 + | \lambda_4 | + | \lambda_5 |, \quad 2 \lambda_1 \geq - \lambda_1 + | \lambda_6 | + | \lambda_7 |,\\
 2 \lambda_1 \geq&\; - \lambda_2 + | \lambda_4 | + | \lambda_6 |, \quad 2 \lambda_1 \geq - \lambda_2 + | \lambda_5 | + | \lambda_7 |,\\
\quad 2 \lambda_1 \geq&\; - \lambda_3 + | \lambda_4 | + | \lambda_7 |, \quad 2 \lambda_1 \geq - \lambda_3 + | \lambda_5 | + | \lambda_6 | .
\end{aligned}
\end{align}

Using $\lambda_1 + \lambda_2 + \lambda_3 = 0$, the conditions on the second
line simply to
\begin{align}
\begin{aligned}
\lambda_1 - \lambda_3 \geq&\; | \lambda_4 | + | \lambda_6 |, \qquad \lambda_1 - \lambda_3 \geq | \lambda_5 | + | \lambda_7 |,\\
\lambda_1 - \lambda_2 \geq&\; | \lambda_4 | + | \lambda_7 |, \qquad \lambda_1 - \lambda_2 \geq | \lambda_5 | + | \lambda_6 |,
\end{aligned}
\end{align}
from which the conditions on the first line automatically follow. Thus,
setting $\lambda_1 = \frac{1}{2}$ (so that $P \cdot \zeta \leq 1$) we get
\begin{align}
\begin{aligned}
\lambda_3 + | \lambda_4 | + | \lambda_6 | \leq&\; \frac{1}{2}, \quad
\lambda_3 + | \lambda_5 | + | \lambda_7 | \leq \frac{1}{2}, \quad
\lambda_2 + | \lambda_4 | + | \lambda_7 | \leq \frac{1}{2},\\
\lambda_2 + | \lambda_5 | + | \lambda_6 | \leq&\; \frac{1}{2}, \quad
\lambda_2 + \lambda_3 = - \frac{1}{2} . \label{eqn:lambdaineq}
\end{aligned}
\end{align}
Subject to these constraints, we wish to maximize:
\be
  P^2 = 2 \lambda_1^2 + 2 \lambda_2^2 + 2 \lambda_3^2 + 2 \lambda_4^2 + 2
  \lambda_5^2 + 2 \lambda_6^2 + 2 \lambda_7^2 .
\ee 
Clearly we need to saturate at least two of the inequalities in
\eqref{eqn:lambdaineq} to do so, since otherwise we can increase at least one
of $\lambda_{4, 5, 6, 7}$ for free. In particular, we must either saturate the
first two or the second two, since if we only saturate one from the first pair
and one from the second then there is still one of $\lambda_{4, 5, 6, 7}$ that
can be increased for free. Because the problem is symmetric under the
simultaneous exchange $\lambda_2 \leftrightarrow \lambda_3$, $\lambda_4
\leftrightarrow \lambda_5$, we can assume that the saturated pair is
\be
  \lambda_2 + | \lambda_4 | + | \lambda_7 | = \lambda_2 + | \lambda_5 | + |
  \lambda_6 | = \frac{1}{2}, \label{eqn:lambdaSat}
\ee 
without loss of generality. Summing the inequalities for $\lambda_3$, this
implies that
\be
  2 \lambda_2 + | \lambda_4 | + | \lambda_5 | + | \lambda_6 | + | \lambda_7 |
  = 1 \geq 2 \lambda_3 + | \lambda_4 | + | \lambda_5 | + | \lambda_6 | +
  | \lambda_7 | \quad \Longrightarrow \quad \lambda_2 \geq \lambda_3 .
\ee 
Thus, for fixed $\lambda_2$ the sums $| \lambda_4 | + | \lambda_7 |$ and $|
\lambda_5 | + | \lambda_6 |$ are fixed. Maximizing $P^2$ with this constraint,
it is optimal to maximize the \emph{difference} between $| \lambda_4 |$
and $| \lambda_7 |$ and between $| \lambda_5 |$ and $| \lambda_6 |$, i.e., to
set one of each pair to zero. If, e.g., we set $| \lambda_6 | = | \lambda_7 |
= 0$, then the inequalities
\be
  \lambda_3 + | \lambda_4 | + | \lambda_6 | = \lambda_3 + | \lambda_4 |
  \leq \frac{1}{2}, \quad \lambda_3 + | \lambda_5 | + | \lambda_7 | =
  \lambda_3 + | \lambda_5 | \leq \frac{1}{2},
\ee 
follow from \eqref{eqn:lambdaSat} and $\lambda_2 \geq \lambda_3$, so we
can do so consistent with the $\lambda_3$ inequalities.\footnote{We cannot
necessarily set, e.g., $| \lambda_5 | = | \lambda_7 | = 0$, since the
$\lambda_3$ inequality $\lambda_3 + | \lambda_4 | + | \lambda_6 | \leq
\frac{1}{2}$ might then be violated; we could keep track of the point where we
first violate the $\lambda_3$ inequality in this direction, but the resulting
$P^2$ is submaximal, so we can just ignore this case altogether.} Thus, for
fixed $\lambda_2 \geq \lambda_3$, the maximum $P^2$ occurs when
\be
  | \lambda_4 | = | \lambda_5 | = \frac{1}{2} - \lambda_2, \qquad \lambda_6 =
  \lambda_7 = 0 .
\ee 
Specifically
\begin{align}
\begin{aligned}
P^2 &= 2 \left[ \frac{1}{4} + \lambda_2^2 + \left( - \frac{1}{2} - \lambda_2 \right)^2 + \left( \frac{1}{2} - \lambda_2 \right)^2 + \left( \frac{1}{2} - \lambda_2 \right)^2 + 0 + 0 \right],\\
\quad - \frac{1}{4} \leq&\;\lambda_2 \leq \frac{1}{2},
\end{aligned}
\end{align}
where the latter constraint comes from $\lambda_3 \leq \lambda_2
\leq \lambda_1$. Since this is a quadratic function of $\lambda_2$ with a
positive coefficient for $\lambda_2^2$, it is maximized at one boundary or the
other. In fact, the value is the \emph{same} at both boundaries:
\be
  P^2_{\max} = 3 .
\ee 
This indeed yields $\lambda_{\min} = \frac{1}{\sqrt{3}} = \frac{1}{\sqrt{d -
2}}$ as expected.

\subsection{Four Dimensions}

In 4d, $H = \text{SU} (8)$ and the central charge $Z^{a b}$ transforms in the
complex antisymmetric tensor representation $\ydiagram{1,1}$. Let us define $Z_{a b} \equiv (Z^{a b})^{\star}$. $Z$ can
alternately be thought of as living in the real representation
$\ydiagram{1,1} \oplus \overline{\ydiagram{1,1}}$ with twice the dimension, composed of components $Z^{a b}$ and
$Z_{a b}$ with the reality condition $Z_{a b} = (Z^{a b})^{\star}$.

Note that
\be
\left( \ydiagram{1,1} \oplus \overline{\ydiagram{1,1}} \right) \otimes_S \left( \ydiagram{1,1} \oplus \overline{\ydiagram{1,1}} \right) = \ydiagram{1,1,1,1} \oplus \ydiagram{2,2} \oplus \overline{\ydiagram{1,1,1,1}} \oplus \overline{\ydiagram{2,2}} \oplus \ydiagram{2,2,1,1,1,1} \oplus \text{Adj} \oplus 1.
\ee
This means that there is a unique
mass formula
\be
M^2 = \frac{1}{2} Z^{a b}  (Z^{a b})^{\star} = \frac{1}{2} Z^{a b} Z_{a b},
\ee 
up to an overall normalizing factor that we set to $1$ by redefining the
central charge. However, since $\ydiagram{1,1,1,1} \cong \overline{\ydiagram{1,1,1,1}}$ and $R_{\text{str}} = \text{Adj} \oplus \ydiagram{1,1,1,1}$ the half-BPS condition at first seems ambiguous. In particular,
\be
  X^{a b c d} = Z^{[a b  } Z^{  c d]},
\ee 
is complex, but the BPS condition should involve a \emph{real} four-form,
i.e., one satisfying
\be
  \omega^{\star}_{a b c d} = \frac{1}{4!} \varepsilon_{a b c d e f g h}
  \omega^{e f g h},
\ee 
which is a consistent condition since
\be
  [(\omega^{\star})^{\star}]^{a b c d} = \frac{1}{4!} \varepsilon^{a b c d e f
  g h} \omega_{e f g h}^{\star} = \frac{1}{4!^2} \varepsilon^{a b c d e f g h}
  \varepsilon_{e f g h i j k l} \omega^{i j k l} = \omega^{a b c d} .
\ee 
In general, the 1/2 BPS condition could involve a linear combination of the
real and imaginary parts of $X^{a b c d}$, i.e., schematically $\text{Im}
[e^{i \theta} Z Z] = 0$, but by redefining $Z^{a b}$ by a phase (leaving the
mass formula invariant) we can set $\theta = 0$. Thus, after this redefinition
the $1 / 2$ BPS conditions become
\be
  Z^{a c} Z_{b c} = \frac{1}{8} \delta^a_b Z^{c d} Z_{c d}, \qquad Z_{[a b
   } Z_{  c d]} = \frac{1}{4!} \varepsilon_{a b c d e f g h}
  Z^{e f} Z^{g h} . \label{eqn:4dhalfBPS}
\ee 
The scalar charge-to-mass ratio $\zeta^{a b c d}$ also sits in the
$\ydiagram{1,1,1,1}$ rep, satisfying the reality condition
\be
  \zeta^{\star}_{a b c d} = \frac{1}{4!} \varepsilon_{a b c d e f g h}
  \zeta^{e f g h} . \label{eqn:zetareal}
\ee 
Because
\be
  \ydiagram{1,1,1,1} \otimes_S \ydiagram{1,1,1,1} = \ydiagram{2,2,2,2} \oplus \ydiagram{2,2,1,1,1,1} \oplus 1,
\ee 
the metric is uniquely determined to be
\be
  \zeta^2 = \frac{1}{4!} \zeta^{a b c d}  (\zeta^{a b c d})^{\star} =
  \frac{1}{4!} \zeta^{a b c d} \zeta_{a b c d} = \frac{1}{4!^2} \varepsilon_{a
  b c d e f g h} \zeta^{a b c d} \zeta^{e f g h},
\ee 
up to an overall normalization that we can absorb into the definition of
$\zeta$, where $\zeta_{a b c d} \equiv (\zeta^{a b c d})^{\star}$.

Given the $1 / 2$ BPS condition, there is a unique $\ydiagram{1,1,1,1}$ component inside $\left( \ydiagram{1,1} \oplus \overline{\ydiagram{1,1}} \right) \otimes_S \left( \ydiagram{1,1} \oplus \overline{\ydiagram{1,1}} \right)$, so for $1 / 2$ BPS states we must have
\be
  \zeta^{a b c d} = 3\mathcal{N} \hat{Z}^{[a b  } 
  \hat{Z}^{  c d]} =\mathcal{N} (\hat{Z}^{a b}  \hat{Z}^{c d} +
  \hat{Z}^{a c}  \hat{Z}^{d b} + \hat{Z}^{a d}  \hat{Z}^{b c}), \qquad
  \hat{Z}^{a b} \equiv \frac{Z^{a b}}{\sqrt{\frac{1}{2} Z^{c d} Z_{c d}}},
\ee 
up to an unknown normalizing factor $\mathcal{N}$. To fix $\mathcal{N}$, we
compute $\zeta^2$ for a $1 / 2$ BPS state:
\be
  \zeta^2 = \frac{3}{4!} \mathcal{N}^2  (\hat{Z}^{a b}  \hat{Z}^{c d} +
  \hat{Z}^{a c}  \hat{Z}^{d b} + \hat{Z}^{a d}  \hat{Z}^{b c}) \hat{Z}_{a b} 
  \hat{Z}_{c d} = \frac{1}{8} \mathcal{N}^2  \left( 4 - \frac{1}{2} -
  \frac{1}{2} \right) = \frac{3\mathcal{N}^2}{8} .
\ee 
Since Kaluza Klein modes are $1 / 2$ BPS with $\zeta^2 = \frac{d - 1}{d - 2} =
\frac{3}{2}$, we conclude that $\mathcal{N}= 2$, i.e.,
\be
  \zeta^{a b c d} = 6 \hat{Z}^{[a b  }  \hat{Z}^{  c d]} = 2
  (\hat{Z}^{a b}  \hat{Z}^{c d} + \hat{Z}^{a c}  \hat{Z}^{d b} + \hat{Z}^{a d}
  \hat{Z}^{b c}), \qquad \hat{Z}^{a b} \equiv \frac{Z^{a
  b}}{\sqrt{\frac{1}{2} Z^{c d} Z_{c d}}} . \label{eqn:zetaZhat}
\ee 
Note that we can solve the $1 / 2$ BPS conditions
\be
  \hat{Z}^{a c}  \hat{Z}_{b c} = \frac{1}{4} \delta^a_b, \qquad \hat{Z}_{[a b
   }  \hat{Z}_{  c d]} = \frac{1}{4!} \varepsilon_{a b c d e
  f g h}  \hat{Z}^{e f}  \hat{Z}^{g h}, \label{eqn:ZhatBPS}
\ee 
as follows. After an $\text{SU} (8)$ transformation we can fix
\be
  \hat{Z}^{a b} = e^{i \phi} \text{diag} \left[ \left(\begin{array}{cc}
    0 & \lambda_1\\
    - \lambda_1 & 0
  \end{array}\right), \ldots, \left(\begin{array}{cc}
    0 & \lambda_4\\
    - \lambda_4 & 0
  \end{array}\right) \right],
\ee 
where $\lambda_1, \ldots, \lambda_4$ are real and non-negative and the overall
phase factor is needed because we can only perform a \emph{special
unitary} change of basis, not a general unitary change of basis. The first
equation of \eqref{eqn:ZhatBPS} then implies that $\lambda_1 = \lambda_2 =
\lambda_3 = \lambda_4 = \frac{1}{2}$. The second one implies that
\begin{align}
\begin{aligned}
\zeta^2 &= \frac{3}{2} = \frac{3}{2}  \hat{Z}^{[a b  }  \hat{Z}^{  c d]}  \hat{Z}_{[a b  }  \hat{Z}_{  c d]} = \frac{3}{2} \cdot \frac{1}{4!} \varepsilon_{a b c d e f g h}  \hat{Z}^{a b}  \hat{Z}^{c d} \hat{Z}^{e f}  \hat{Z}^{g h} = \frac{3}{2} \text{Pf} (2 \hat{Z}^{a b})\\
\Rightarrow\qquad& \text{Pf} (2 \hat{Z}^{a b}) = 1,
\end{aligned}
\end{align}
where
\be
  \text{Pf} (M^{a b}) = \frac{1}{2^n n!} \varepsilon_{a_1 \ldots a_n} M^{a_1
  a_2} \cdots M^{a_{n - 1} a_n}
\ee 
is the Pfaffian. Thus, we obtain $\text{Pf} (2 \hat{Z}^{a b}) = e^{4 i \phi} =
1$, so $e^{i \phi}$ is a fourth root of unity. However, the $\text{SU} (8)$
transformation $\text{diag} (e^{\pi i / 4}, \ldots, e^{\pi i / 4})$ imparts an
overall phase $e^{i \pi / 2}$ to $\hat{Z}^{a b}$, so we can set $e^{i \phi} =
1$ by such transformations, leaving
\be
  \hat{Z}^{a b} = \text{diag} \left[ \left(\begin{array}{cc}
    0 & 1\\
    - 1 & 0
  \end{array}\right), \ldots, \left(\begin{array}{cc}
    0 & 1\\
    - 1 & 0
  \end{array}\right) \right], \label{eqn:canonZhat}
\ee 
in this basis. In other words, all the $1 / 2$ BPS choices of $\hat{Z}^{a b}$
are equivalent up to $\text{SU} (8)$ transformations, and a particular choice
of $\hat{Z}^{a b}$ is left invariant by a $\text{USp} (8) \subset \text{SU}
(8)$ subgroup.

Now choose an arbitary four-form $P^{a b c d}$ satisfying the reality
condition ($\ref{eqn:zetareal}$). We seek a 1/2 BPS central charge $\hat{Z}^{a
b}_{(0)}$ that globally maximizes $P^{a b c d} \zeta_{a b c d}^{(0)}$ with
$\zeta_{a b c d}$ given by \eqref{eqn:zetaZhat}. Since all $\hat{Z}^{a b}$'s
are related by $\text{SU} (8)$ transformations, we can rephrase this problem
by going to the basis where $\hat{Z}^{a b}_{(0)}$ takes the form
\eqref{eqn:canonZhat}, then requiring that $P^{a b c d} \zeta^{(0)}_{a b c d}$
is maximized under $\text{SU} (8)$ rotations of $P^{a b c d}$ with
$\zeta^{(0)}_{a b c d}$ held fixed in its canonical form
\eqref{eqn:canonZhat}. In other words, we maximize
\be
  P^{1234} + P^{1256} + P^{1278} + P^{3456} + P^{3478} + P^{5678} = 2
  \text{Re} [P^{1234} + P^{1256} + P^{1278}],
\ee 
under $\text{SU} (8)$ rotations, where $P^{5678} = (P^{1234})^{\star}$,
$P^{3478} = (P^{1256})^{\star}$ and $P^{3456} = (P^{1278})^{\star}$ by the
reality condition ($\ref{eqn:zetareal}$).

Maximizing with respect to the Cartan subgroup, it is evident that $P^{1234}$,
$P^{1256}$ and $P^{1278}$ must be real and non-negative. More generally, the
condition to obtain an extremum of $P \cdot \zeta$ is
\be
  \delta (P \cdot \zeta) = \frac{1}{6} i T^a_e P^{e b c d} \zeta_{a b c d} = 0
\ee 
for any generator $T^a_b$ of $\text{SU} (8)$, i.e., for any traceless
Hermitian matrix $T^a_b$. In fact, since the condition is $\mathbb{C}$-linear
and any traceless matrix can be written as $T_1 + i T_2$ for traceless
Hermitian matrices $T_1$ and $T_2$, the condition holds for \emph{any}
traceless matrix $T^a_b$, so that
\be
  \frac{1}{3} P^{e b c d} \zeta_{a b c d} = \delta^e_a  (P \cdot \zeta) .
\ee 
In terms of $\hat{Z}_{a b}$, this becomes:
\be
  2 P^{e b c d}  \hat{Z}_{a b}  \hat{Z}_{c d} = \delta^e_a  (P \cdot \zeta)
  \qquad \Leftrightarrow \qquad \frac{1}{2} P^{a b c d}  \hat{Z}_{c d} =
  \hat{Z}^{a b} (P \cdot \zeta),
\ee 
using \eqref{eqn:ZhatBPS}. Thus, decomposing
\be
  P^{a b c d} = 4 \lambda \hat{Z}^{[a b  }  \hat{Z}^{  c d]}
  + \hat{P}^{a b c d}, \qquad \lambda = P \cdot \zeta,
\ee 
we find that $\hat{P}^{a b c d}  \hat{Z}_{c d} = 0$. In other words
$\hat{P}^{a b c d}$ transforms in the traceless 4-form representation
$\ydiagram{1,1,1,1}$ of the residual $\text{USp} (8)$ preserving $\hat{Z}^{a b}$.
Moreover, because any two top-forms are proportional
\be
  \frac{1}{2^4 \cdot 4!}  \hat{Z}_{[a b  }  \hat{Z}_{c d}  \hat{Z}_{e
  f}  \hat{Z}_{  g h]} = \frac{1}{8!} \varepsilon_{a b c d e f g h} 
  (\text{Pf} \hat{Z}) = \frac{1}{2^4 \cdot 8!} \varepsilon_{a b c d e f g h},
\ee 
where the overall normalization can be fixed by contracting with
$\varepsilon^{a b c d e f g h}$. Thus, the reality condition
\eqref{eqn:zetareal} for $P^{a b c d}$ becomes
\be
  \hat{P}^{\star}_{a b c d} = \frac{8!}{4!^2}  \hat{Z}_{[a b  } 
  \hat{Z}_{c d}  \hat{Z}_{e f}  \hat{Z}_{  g h]} \hat{P}^{e f g h} =
  2^4  \hat{Z}_{a e}  \hat{Z}_{b f}  \hat{Z}_{c g}  \hat{Z}_{d h}  \hat{P}^{e
  f g h},
\ee 
i.e., $\hat{P}^{a b c d}$ is real in the sense of \eqref{eqn:5dzetareal} with
symplectic form $\Omega_{a b} = 2 \hat{Z}_{a b}$. Thus, per the discussion in
{\textsection}\ref{sec:5d}, we can put $\hat{P}^{a b c d}$ into canonical form
with non-vanishing real components
\begin{align}
\begin{aligned}
\hat{P}^{1357} &= \hat{P}^{2468},\quad \hat{P}^{1368} = \hat{P}^{2457},\quad \hat{P}^{1458} = \hat{P}^{2367},\quad \hat{P}^{1467} = \hat{P}^{2358},\\
\hat{P}^{1256} &= \hat{P}^{3478},\quad \hat{P}^{1278} = \hat{P}^{3456},\quad \hat{P}^{1234} = \hat{P}^{5678},
\end{aligned}
\end{align}
subject to the tracelessness condition $\hat{P}^{1234} + \hat{P}^{1256} +
\hat{P}^{1278} = 0$. Working backwards, we see that $P^{a b c d}$ has real
components
\begin{align}
\begin{aligned}
P^{1357} &= P^{2468}, \quad P^{1368} = P^{2457},\quad P^{1458} = P^{2367},\quad P^{1467} =
P^{2358},\\
P^{1256} &= P^{3478},\quad P^{1278} = P^{3456},\quad P^{1234} = P^{5678}, \label{eqn:FanoComponents}
\end{aligned}
\end{align}
satisfying no further constraints.

To see that \eqref{eqn:FanoComponents} is not, in fact, some random collection
of directions, note that exactly one in each pair has a leg along any given
direction, e.g., the $1$ direction. Stripping off this leg, we find the
combinations
\be
234,\quad 256,\quad 278,\quad 357,\quad 368,\quad 458,\quad 467.
\ee 
These are nothing but the lines on a Fano plane with points $2, \ldots, 8$ (up
to convention-dependent relabling). To present it more symmetrically,
represent each number with 3 digit binary code, in this case the binary digits
of that number minus 1:
\be
2 \rightarrow 001,\quad 3 \rightarrow 010,\quad 4 \rightarrow 011,\quad 5 \rightarrow 100,\quad
  6 \rightarrow 101,\quad 7 \rightarrow 110,\quad 8 \rightarrow 111,\quad
\ee 
where the codes are chosen such that 3 points are colinear iff the sums of
their digits $\text{mod} 2$ are all zero (e.g., $001 + 011 + 010 = 000$ with
\emph{no carrying}, so $001 - 011 - 010$ forms the line 234). The
symmetries of the Fano plane are then $\text{GL} (3 ; \mathbb{Z}_2)$ acting on
these digits. Note that this group is 2-transitive---any pair of points can be
mapped to any other pair of points---and the stabilizer subgroup fixing any
pair of points is $\mathbb{Z}_2 \times \mathbb{Z}_2$; e.g., if the fixed
points are $100$ and $010$ then one $\mathbb{Z}_2$ is generated by adding the
third digit to the first ($001 \leftrightarrow 101, 011 \leftrightarrow 111$)
and the other by adding the third digit to the second ($001 \leftrightarrow
011, 101 \leftrightarrow 111$).

Using these facts, one can also show that $\text{GL} (3 ; \mathbb{Z}_2)$ acts
2-transitively on the set of lines, with the stabilizer subgroup fixing a pair
of lines equal to $\mathbb{Z}_2 \times \mathbb{Z}_2$; e.g., if the fixed lines
are $001 - 010 - 011$ and $001 - 100 - 101$ then one $\mathbb{Z}_2$ is
generated by adding the first digit to the third ($100 \leftrightarrow 101,
110 \leftrightarrow 111$) and the other is generated by adding the second
digit to the third ($010 \leftrightarrow 011, 110 \leftrightarrow 111$). In
fact, the Fano plane is self-dual under the exchange of lines and points, so
this similarity between the action on points and lines is no accident.
Explicitly, on such point-line duality is
\be
  (001, 010, 011) \rightarrow 100, \quad (100, 011, 111) \rightarrow 011,
  \quad (110, 101, 011) \rightarrow 111,
\ee 
with the other elements of the dictionary given by permuting the three digits.
With this in mind, we assign labels:

\begin{align}
\begin{aligned}
\lambda_{001} &= P^{1357} = P^{2468},\quad \lambda_{010} = P^{1256} =
P^{3478},\quad \lambda_{011} = P^{1458} = P^{2367},\\
\lambda_{100} &= P^{1234} = P^{5678}, \quad \lambda_{101} = P^{1368} = P^{2457},\quad \lambda_{110} = P^{1278} = P^{3456},\\
\quad \lambda_{111} &= P^{1467} = P^{2358} .
\end{aligned}
\end{align}

Now we study the ``eigenvalue'' problem:
\be
  \frac{1}{2} P^{a b c d}  \hat{Z}_{c d} = \lambda (\hat{Z}_{a b})^{\star},
\ee 
which is a necessary and sufficient condition for a given $\hat{Z}_{a b}$
satisfying \eqref{eqn:ZhatBPS} to extremize $P \cdot \zeta$, where the
eigenvalue $\lambda \in \mathbb{R}$ is equal to $P \cdot \zeta$. In general,
\be
  M v = \lambda v^{\star} \qquad \Leftrightarrow \qquad
  \left(\begin{array}{cc}
    0 & M\\
    M^{\star} & 0
  \end{array}\right) \left(\begin{array}{c}
    v\\
    v^{\star}
  \end{array}\right) = \lambda \left(\begin{array}{c}
    v\\
    v^{\star}
  \end{array}\right),
\ee 
so this becomes a standard eigenvalue problem for a matrix twice as large,
supplemented by a reality condition on the eigenvectors
\be
  \left(\begin{array}{cc}
    0 & 1\\
    1 & 0
  \end{array}\right) \left(\begin{array}{c}
    v\\
    v^{\star}
  \end{array}\right) = \left(\begin{array}{c}
    v\\
    v^{\star}
  \end{array}\right)^{\star} .
\ee 
In particular, if $M$ is real and symmetric then it has a basis of real
eigenvectors with real eigenvalues
\be
  M v_i = \lambda_i v_i, \qquad v_i = v_i^{\star} .
\ee 
From each such eigenvector we obtain a \emph{pair} of
$\left(\begin{array}{cc}
  0 & M\\
  M & 0
\end{array}\right)$ eigenvectors satisfying the reality condition:
\be
  \left(\begin{array}{cc}
    0 & M\\
    M & 0
  \end{array}\right) \left(\begin{array}{c}
    v_i\\
    v_i
  \end{array}\right) = \lambda_i  \left(\begin{array}{c}
    v_i\\
    v_i
  \end{array}\right), \qquad \left(\begin{array}{cc}
    0 & M\\
    M & 0
  \end{array}\right) \left(\begin{array}{c}
    i v_i\\
    - i v_i
  \end{array}\right) = - \lambda_i  \left(\begin{array}{c}
    i v_i\\
    - i v_i
  \end{array}\right) .
\ee 
Thus the eigenvalues in the equation $M v = \lambda v^{\star}$ are those of
$M$ together with those of $- M$.

Fortunately, the real symmetric matrix $P^{a b c d}$ (viewed as a $28 \times
28$ matrix of index pairs) is block diagonal. For example, in the $12, 34, 56,
78$ block we find:
\be
  M = \left(\begin{array}{cccc}
    0 & \lambda_{100} & \lambda_{010} & \lambda_{110}\\
    \lambda_{100} & 0 & \lambda_{110} & \lambda_{010}\\
    \lambda_{010} & \lambda_{110} & 0 & \lambda_{100}\\
    \lambda_{110} & \lambda_{010} & \lambda_{100} & 0
  \end{array}\right) .
\ee 
The eigenvalues of this matrix are $\lambda_{100} + \lambda_{010} +
\lambda_{110}$, $\lambda_{100} - \lambda_{010} - \lambda_{110}$, $-
\lambda_{100} + \lambda_{010} - \lambda_{110}$ and $- \lambda_{100} -
\lambda_{010} + \lambda_{110}$, so in combination with those of $- M$, we
obtain $\pm_1 \lambda_{100} \pm_2 \lambda_{010} \pm_3 \lambda_{110}$, the
largest eigenvalue of which is
\be
  \lambda_{\max} = | \lambda_{100} | + | \lambda_{010} | + | \lambda_{110} | .
\ee 
It is straightforward to check that the resulting eigenvectors $\hat{Z}_{a b}$
satisfy \eqref{eqn:ZhatBPS} after fixing their overall normalization as
required.

Fortunately, due to the symmetries of the Fano plane, there is no need to
consider the other six $4 \times 4$ blocks of $P^{a b c d}$, which are related
by permutations on the indices $1, \ldots, 8$ that correspond to these
symmetries. Thus, the requirement $P \cdot \zeta \leq 1$ for all half-BPS
central charges $\hat{Z}_{a b}$ translates to the seven conditions,
\be
  | \lambda_{i j k} | + | \lambda_{i' j' k'} | + | \lambda_{(i + i') (j + j')
  (k + k')} | \leq 1, \label{eqn:FanoCons}
\ee 
one for each line on the Fano plane with points $100$, $010$, etc. Meanwhile,
\be
  P^2 = 2 \sum_{i, j, k} | \lambda_{i j k} |^2 .
\ee 
We seek to maximize $P^2$ while satisfying all the constraints. To do so,
first note that we can take $\lambda_{i j k} \geq 0$ without loss of
generality. If any given $\lambda_{i j k}$ does not show up in a constraint
\eqref{eqn:FanoCons} that is saturated then we can increase it at no cost,
increasing $P^2$ at no cost. Thus, the lines corresponding to the saturated
constraints must intersect every point on the Fano plane. Therefore, there
must be a least one point for which \emph{all} the lines intersecting it
are saturated constraints.\footnote{Otherwise, since any two lines on the
place intersect at a point, we can form a chain of saturated lines. Once we
have three such lines we will have a loop of saturated lines but this only
covers six of the points, leaving one out.} Choose this to be $\lambda_{111}$,
so that
\be
  \lambda_{111} + \lambda_{101} + \lambda_{010} = \lambda_{111} +
  \lambda_{100} + \lambda_{011} = \lambda_{111} + \lambda_{001} +
  \lambda_{110} = 1 . \label{eqn:3lines}
\ee 
If there were no other constraints, then for fixed $\lambda_{111}$ the maximum
value of $P^2$ would occur when one of each of the other pairs vanishes, e.g.,
$\lambda_{101} = \lambda_{110} = \lambda_{011} = 0$, so that
\be
  \lambda_{100} = \lambda_{010} = \lambda_{001} = 1 - \lambda_{111} .
\ee 
This satisfies the remaining constraints provided that $\lambda_{111}
\geq \frac{1}{2}$, and since
\be
  P^2 = 2 (\lambda_{111}^2 + 3 (1 - \lambda_{111})^2),
\ee 
the maximum values are achieved at the boundaries $\lambda_{111} = 1$
(implying $\lambda_{100} = \lambda_{010} = \lambda_{001} = 0$) and
$\lambda_{111} = \frac{1}{2}$ (implying $\lambda_{100} = \lambda_{010} =
\lambda_{001} = \frac{1}{2}$), with $P^2 = 2$ in both cases.

Thus, by process of elimination we consider the case $\lambda_{111} <
\frac{1}{2}$. Now after maximizing $P^2$ the constraint associated to at least
one other line in the Fano plane must be saturated. Using the symmetries of
the Fano plane, we can fix this to be
\be
  \lambda_{110} + \lambda_{101} + \lambda_{011} = 1, \label{eqn:4thline}
\ee 
without loss of generality. Ignore the other constraints for the time being.
Applying the symmetries of the Fano plane, we may assume that $\lambda_{110}
\geq \lambda_{101}$ and $\lambda_{110} \geq \lambda_{011}$ (without loss of generality).
Then taking
\be
  \lambda_{110} \rightarrow \lambda_{110} + \Delta, \qquad \lambda_{101}
  \rightarrow \lambda_{101} - \Delta, \qquad \lambda_{001} \rightarrow
  \lambda_{001} - \Delta, \qquad \lambda_{010} \rightarrow \lambda_{010} +
  \Delta,
\ee 
preserves the constraints \eqref{eqn:3lines} and \eqref{eqn:4thline}, whereas
\be
  P^2 \rightarrow P^{\prime 2} = P^2 + 4 (\lambda_{110} + \lambda_{010} -
  \lambda_{101} - \lambda_{001}) \Delta + 8 \Delta^2 = P^2 + 8 (\lambda_{110}
  - \lambda_{101}) \Delta + 8 \Delta^2 > P^2,
\ee 
since $\lambda_{110} \geq \lambda_{101}$ by assumption, so $P^2$
increases. This can be continued until either $\lambda_{101} = 0$ or
$\lambda_{001} = 0$. If the former occurs, then we exchange the roles of
$\lambda_{101}$ and $\lambda_{011}$ and carry out the same process until
either $\lambda_{011}$ also vanishes or $\lambda_{001} = 0$. If the former,
then \eqref{eqn:4thline} implies that $\lambda_{110} = 1$, but then
\eqref{eqn:3lines} implies that $\lambda_{001} = 0$, so we ultimately obtain
$\lambda_{001} = 0$ regardless. Next, we again apply the symmetries of the
Fano plane to fix $\lambda_{101} \geq \lambda_{011}$, and then carry out
an analogous process to increase $\lambda_{101}$ at the expense of
$\lambda_{011}$ while steadily increasing $P^2$. This terminates when either
$\lambda_{011} = 0$ or $\lambda_{010} = 0$. If the former occurs, then our
Fano plane now looks like
\be
  \begin{array}{ccccc}
    &  & \lambda_{111} &  & \\
    & \lambda_{001} & \lambda_{010} & \lambda_{100} & \\
    \lambda_{110} &  & \lambda_{101} &  & \lambda_{011}
  \end{array} = \begin{array}{ccccc}
    &  & x &  & \\
    & 0 & 1 - 2 x & 1 - x & \\
    1 - x &  & x &  & 0
  \end{array},
\ee 
but then $\lambda_{110} + \lambda_{010} + \lambda_{100} = 1 - x + 1 - 2 x + 1
- x = 3 - 4 x > 1$ since $x = \lambda_{111} < \frac{1}{2}$, so we reach a
contradiction. If the latter, the our Fano plane now looks like:
\be
  \begin{array}{ccccc}
    &  & \lambda_{111} &  & \\
    & \lambda_{001} & \lambda_{010} & \lambda_{100} & \\
    \lambda_{110} &  & \lambda_{101} &  & \lambda_{011}
  \end{array} = \begin{array}{ccccc}
    &  & x &  & \\
    & 0 & 0 & 2 - 3 x & \\
    1 - x &  & 1 - x &  & 2 x - 1
  \end{array},
\ee 
but now since $x = \lambda_{111} < \frac{1}{2}$, $\lambda_{011} = 2 x - 1 < 0$
and we again reach a contradiction.

Thus, we conclude that another line must be saturated. Using the symmetries of
the Fano plane, we can fix it to be
\be
  \lambda_{001} + \lambda_{010} + \lambda_{011} = 1 . \label{eqn:5thline}
\ee 
Imposing all the constraints \eqref{eqn:3lines}, \eqref{eqn:4thline},
\eqref{eqn:5thline}, the Fano plane looks like:
\be
  \begin{array}{ccccc}
    &  & \lambda_{111} &  & \\
    & \lambda_{001} & \lambda_{010} & \lambda_{100} & \\
    \lambda_{110} &  & \lambda_{101} &  & \lambda_{011}
  \end{array} = \begin{array}{ccccc}
    &  & x &  & \\
    & y & 1 - x - y & 1 - 2 x & \\
    1 - x - y &  & y &  & x
  \end{array},
\ee 
and the remaining two lines impose the constraints
\be
  1 - 2 x \leq y \leq x .
\ee 
As usual, the maximum value of $P^2$ will be found on a boundary, either where
$y = x$ or $y = 1 - 2 x$, where in either case we obtain $P^2 = 2 (4 x^2 + 3
(1 - 2 x)^2)$. Since $x \geq \frac{1}{3}$ to satisfy the previous
inequality and $x \leq \frac{1}{2}$ by prior assumption, we obtain $P^2 =
2$ for $x = \frac{1}{2}$ (as before) and $P^2 = \frac{14}{9}$ for $x =
\frac{1}{3}$. The former is larger of course, so we finally obtain
\be
  P^2_{\text{max}} = 2,
\ee 
which corresponds to $\lambda_{\text{min}} = \frac{1}{\sqrt{d - 2}} =
\frac{1}{\sqrt{2}}$ precisely as expected.

\bibliographystyle{utphys}
\bibliography{ref}
\end{document}